\documentclass[%
 reprint,
 superscriptaddress,
 amsmath,amssymb,
 aps,
floatfix,
]{revtex4-2}

\usepackage{xcolor}
\usepackage{graphicx}
\usepackage{dcolumn}
\usepackage{bm}
\usepackage{hyperref}
\usepackage{siunitx} % added by Boris
\usepackage{soul}
\usepackage{rotating}
\usepackage{subfigure}
\usepackage{url}
\usepackage{comment}

\begin{document}

\preprint{APS/123-QED}

\title{Extensive search for axion dark matter over 1\,GHz with CAPP's Main Axion eXperiment}

\author{Saebyeok Ahn}
\affiliation{Center for Axion and Precision Physics Research, Institute for Basic Science (IBS), Daejeon 34051, Republic of Korea}

% Specific author order will be discussed later on
\author{JinMyeong Kim}
\affiliation{Department of Physics, Korea Advanced Institute of Science and Technology (KAIST), Daejeon 34141, Republic of Korea}
\affiliation{Center for Axion and Precision Physics Research, Institute for Basic Science (IBS), Daejeon 34051, Republic of Korea}

\author{Boris I. Ivanov}
\affiliation{Center for Axion and Precision Physics Research, Institute for Basic Science (IBS), Daejeon 34051, Republic of Korea}

\author{Ohjoon Kwon}
\affiliation{Center for Axion and Precision Physics Research, Institute for Basic Science (IBS), Daejeon 34051, Republic of Korea}

\author{HeeSu Byun}
\affiliation{Center for Axion and Precision Physics Research, Institute for Basic Science (IBS), Daejeon 34051, Republic of Korea}

\author{Arjan F. van Loo}
\affiliation{RIKEN Center for Quantum Computing (RQC), Wako, Saitama 351-0198, Japan}
\affiliation{Department of Applied Physics, Graduate School of Engineering, The University of Tokyo, Bunkyo-ku, Tokyo 113-8656, Japan}

\author{SeongTae Park}
\affiliation{Center for Axion and Precision Physics Research, Institute for Basic Science (IBS), Daejeon 34051, Republic of Korea}

\author{Junu Jeong}
\affiliation{Center for Axion and Precision Physics Research, Institute for Basic Science (IBS), Daejeon 34051, Republic of Korea}

\author{Soohyung Lee}
\affiliation{Center for Axion and Precision Physics Research, Institute for Basic Science (IBS), Daejeon 34051, Republic of Korea}

\author{Jinsu Kim}
\affiliation{Center for Axion and Precision Physics Research, Institute for Basic Science (IBS), Daejeon 34051, Republic of Korea}

\author{Çağlar Kutlu}
\affiliation{Center for Axion and Precision Physics Research, Institute for Basic Science (IBS), Daejeon 34051, Republic of Korea}

\author{Andrew K. Yi}
\affiliation{Department of Physics, Korea Advanced Institute of Science and Technology (KAIST), Daejeon 34141, Republic of Korea}
\affiliation{Center for Axion and Precision Physics Research, Institute for Basic Science (IBS), Daejeon 34051, Republic of Korea}

\author{Yasunobu Nakamura}
\affiliation{RIKEN Center for Quantum Computing (RQC), Wako, Saitama 351-0198, Japan}
\affiliation{Department of Applied Physics, Graduate School of Engineering, The University of Tokyo, Bunkyo-ku, Tokyo 113-8656, Japan}

\author{Seonjeong Oh}
\affiliation{Center for Axion and Precision Physics Research, Institute for Basic Science (IBS), Daejeon 34051, Republic of Korea}

\author{Danho Ahn}
\affiliation{Center for Axion and Precision Physics Research, Institute for Basic Science (IBS), Daejeon 34051, Republic of Korea}

\author{SungJae Bae}
\affiliation{Department of Physics, Korea Advanced Institute of Science and Technology (KAIST), Daejeon 34141, Republic of Korea}
\affiliation{Center for Axion and Precision Physics Research, Institute for Basic Science (IBS), Daejeon 34051, Republic of Korea}

\author{Hyoungsoon Choi}
\affiliation{Department of Physics, Korea Advanced Institute of Science and Technology (KAIST), Daejeon 34141, Republic of Korea}

\author{Jihoon Choi}
\thanks{Present address: Korea Astronomy and Space Science Institute, Daejeon 34055, Republic of Korea.}
\affiliation{Center for Axion and Precision Physics Research, Institute for Basic Science (IBS), Daejeon 34051, Republic of Korea}

\author{Yonuk Chong}
\affiliation{SKKU Advancd Institute of Nano Technology (SAINT) and Department of Nano Engineering
Sung Kyun Kwan University (SKKU), Suwon 16419, Republic of Korea}

\author{Woohyun Chung}
\thanks{corresponding author}
\email{gnuhcw@ibs.re.kr}
\affiliation{Center for Axion and Precision Physics Research, Institute for Basic Science (IBS), Daejeon 34051, Republic of Korea}

\author{Violeta Gkika}
\affiliation{Center for Axion and Precision Physics Research, Institute for Basic Science (IBS), Daejeon 34051, Republic of Korea}

\author{Jihn E. Kim}
\affiliation{Department of Physics, Seoul National University, 1 Gwanak-Ro, Seoul 08826, Republic of Korea}

\author{Younggeun Kim}
\affiliation{Center for Axion and Precision Physics Research, Institute for Basic Science (IBS), Daejeon 34051, Republic of Korea}

\author{Byeong Rok Ko}
\thanks{corresponding author}
\email{brko@ibs.re.kr}
\affiliation{Center for Axion and Precision Physics Research, Institute for Basic Science (IBS), Daejeon 34051, Republic of Korea}

\author{Lino Miceli}
\affiliation{Center for Axion and Precision Physics Research, Institute for Basic Science (IBS), Daejeon 34051, Republic of Korea}

\author{Doyu Lee}
\thanks{Present address: Samsung Electronics, Gyeonggi-do 16677, Republic of Korea.}
\affiliation{Center for Axion and Precision Physics Research, Institute for Basic Science (IBS), Daejeon 34051, Republic of Korea}

\author{Jiwon Lee}
\affiliation{Department of Physics, Korea Advanced Institute of Science and Technology (KAIST), Daejeon 34141, Republic of Korea}
\affiliation{Center for Axion and Precision Physics Research, Institute for Basic Science (IBS), Daejeon 34051, Republic of Korea}

\author{Ki Woong Lee}
\affiliation{Center for Axion and Precision Physics Research, Institute for Basic Science (IBS), Daejeon 34051, Republic of Korea}

\author{MyeongJae Lee}
\thanks{Present address: Department of Physics, Sungkyunkwan University, Suwon 16419, Republic of Korea.}
\affiliation{Center for Axion and Precision Physics Research, Institute for Basic Science (IBS), Daejeon 34051, Republic of Korea}

\author{Andrei Matlashov}
\affiliation{Center for Axion and Precision Physics Research, Institute for Basic Science (IBS), Daejeon 34051, Republic of Korea}

\author{Pallavi Parashar}
\affiliation{Department of Physics, Korea Advanced Institute of Science and Technology (KAIST), Daejeon 34141, Republic of Korea}
\affiliation{Center for Axion and Precision Physics Research, Institute for Basic Science (IBS), Daejeon 34051, Republic of Korea}

\author{Taehyeon Seong}
\affiliation{Center for Axion and Precision Physics Research, Institute for Basic Science (IBS), Daejeon 34051, Republic of Korea}

\author{Yun Chang Shin}
\affiliation{Center for Axion and Precision Physics Research, Institute for Basic Science (IBS), Daejeon 34051, Republic of Korea}

\author{Sergey V. Uchaikin}
\thanks{corresponding author}
\email{uchaikin@ibs.re.kr}
\affiliation{Center for Axion and Precision Physics Research, Institute for Basic Science (IBS), Daejeon 34051, Republic of Korea}

\author{SungWoo Youn}
\thanks{corresponding author}
\email{swyoun@ibs.re.kr}
\affiliation{Center for Axion and Precision Physics Research, Institute for Basic Science (IBS), Daejeon 34051, Republic of Korea}

\author{Yannis K. Semertzidis}
\affiliation{Center for Axion and Precision Physics Research, Institute for Basic Science (IBS), Daejeon 34051, Republic of Korea}
\affiliation{Department of Physics, Korea Advanced Institute of Science and Technology (KAIST), Daejeon 34141, Republic of Korea}

\begin{abstract}
We report an extensive high-sensitivity search for axion dark matter above 1\,GHz at the Center for Axion and Precision Physics Research (CAPP).
The cavity resonant search, exploiting the coupling between axions and photons, explored the frequency (mass) range of 1.025\,GHz (4.24\,$\mu$eV) to 1.185\,GHz (4.91\,$\mu$eV).
We have introduced a number of innovations in this field, demonstrating the practical approach of optimizing all the relevant parameters of axion haloscopes, extending presently available technology.
The CAPP 12\,T magnet with an aperture of 320\,mm made of Nb$_3$Sn and NbTi superconductors surrounding a 37-liter ultralight-weight copper cavity is expected to convert DFSZ axions into approximately $10^2$ microwave photons per second. 
A powerful dilution refrigerator, capable of keeping the core system below 40\,mK, combined with quantum-noise limited readout electronics, achieved a total system noise of about 200\,mK or below, which corresponds to a background of roughly $4\times 10^3$ photons per second within the axion bandwidth.
The combination of all those improvements provides unprecedented search performance, imposing the most stringent exclusion limits on axion--photon coupling in this frequency range to date.
These results also suggest an experimental capability suitable for highly-sensitive searches for axion dark matter above 1\,GHz.

\end{abstract}


\maketitle

\section{Introduction}
Dark matter, which is evident in many astronomical observations, is generally accepted to be one of the major substances in our universe, making up approximately 85\% of  matter and strongly influencing the formation and evolution of galaxies~\cite{article:dark_matter, article:Planck}.
However, the mysterious hypothetical substance has eluded decades of dedicated searches. It remains invisible and its identity unknown.
The possible candidates for dark matter include individual heavy particles,\ such as weakly interacting massive particles~\cite{article:WIMP1,article:WIMP2}, coherently oscillating waves such as axions (in generic, axion-like particles)~\cite{article:axion1, article:axion2}, or astronomical objects such as primordial black holes~\cite{article:PBH1, article:PBH2}.
Among these, axions are theoretically promising particles that were originally proposed to solve a fundamental problem in strong interactions,  known as the strong-CP problem.
The Charge-Parity (CP) symmetry that is expected to be violated according to quantum chromodynamics (QCD) of the Standard Model (SM), appears to be conserved in nature, as indicated by the absence of an electric dipole moment (EDM) for the neutron~\cite{article:nEDM} and proton~\cite{article:pEDM}.
%An elegant solution to this naturalness problem was suggested by R. Peccei and H. Quinn.
%The breakthrough is to promote the CP violating term in the QCD Lagrangian, referred to as $\bar{\theta}$, to a scalar field by introducing a new U(1) global symmetry. 
An elegant solution to this naturalness problem is the Peccei-Quinn (PQ) mechanism, where the CP-violating term in the QCD Lagrangian, referred to as $\bar{\theta}$, is promoted to a scalar field by introducing a new U(1) global symmetry as a minimal extension of the SM~\cite{article:PQ77}.
The symmetry becomes spontaneously broken at some energy scale, dynamically relaxing the $\bar{\theta}$ parameter to zero, thus resolving the strong-CP problem.
The Goldstone boson associated with the spontaneous symmetry breaking is the axion~\cite{article:Weinberg78, article:Wilczek78}.
The mass and coupling (strength of axion--photon interaction) are inversely proportional to the symmetry breaking scale $f_a$, also known as the decay constant.
Since searches for the standard axion at $f_a\sim v_{\rm EW}$ (electroweak scale) have yielded null results, attention has shifted to detecting ``invisible" axions with very large $f_a$, i.e., with very small mass and very weak coupling to ordinary matter.

The unique properties of the QCD axion turn out to provide a suitable explanation for the missing matter of the universe with a cosmological constraint on the decay constant $f_a \lesssim 10^{12}$\,GeV, which imposes a boundary of $\gtrsim 1\,\mu{\rm eV}$ on the mass ($\gtrsim 0.25\,{\rm GHz}$ on the frequency)~\cite{article:axion_cosmology1, article:axion_cosmology2, article:axion_cosmology3}.
Depending on the production and evolution of the cosmic axion field, there are two possible scenarios for axion cosmology.
In the pre-inflationary scenario, where cosmic inflation selects a single patch of the universe free of topological defects, the PQ symmetry breaking results in a homogeneous initial misalignment value of the axion field throughout the observable universe and determines the axion relic (present) density with a mass within a wide range as low as 1~peV/$c^2$, where $c$ is the speed of light, accounting for the present abundance of dark matter ~\cite{article:pre_infl1, article:pre_infl2}.
Alternatively, the post-inflationary axion field embraces many patches and thus takes on multiple initial values, resulting in topological defects that additionally contribute to axion production.
Using a spatially averaged initial misalignment value, recent QCD lattice calculations and simulations have been able to predict the axion mass within a relatively narrow range from $\mathcal{O}(10^1)$ to $\mathcal{O}(10^2)~\mu$eV~\cite{article:post_infl1, article:post_infl2, article:post_infl3}.

At present, dark-matter axions are assumed to be in a virialized state (a sort of thermalization process without energy loss from the total system) with their velocity following the Maxwell--Boltzmann distribution. 
This results in a dispersion of $\sim10^{-3}c$,  which in turn yields an average quality factor of $1.0\times10^6$ for an observer on earth~\cite{article:Turner}.

Although feeble, the QCD axion can interact with SM particles through anomalous couplings to gluons. 
Depending on the type of particles contributing to the anomalous loop, there are two general classes of invisible axion models: Kim-Shifman-Vainshtein-Zakharov (KSVZ) with a newly introduced heavy quark carrying the PQ charge~\cite{article:KSVZ1, article:KSVZ2} and Dine-Fischler-Srednicki-Zhitnitsky (DFSZ) with ordinary quarks and leptons carrying the charge~\cite{article:DFSZ1, article:DFSZ2}.
The electromagnetic interaction has been widely accepted for practical searches, since it mediates axion--photon coupling, providing an experimental signature, i.e., photons, which can be detected with existing well-developed detection technologies.

The detection principle, first suggested by Sikivie~\cite{article:Sikivie83}, relies on the macroscopic version of the (inverse) Primakoff effect, where a classical magnetic field serves as a sea of virtual photons with which axions interact and convert into real photons. The photon frequency corresponds to the total energy of the detected axion.
There are several types of search strategies depending on the source of the axion. These include: (i) haloscope searches for dark matter axions in the galactic halo (CAPP~\cite{article:CAPP-8TB, article:CAPP-9T, article:CAPP-PACE,article:CAPP-PACE-JPA, 12TB-PRL}, ADMX~\cite{article:ADMX-1, article:ADMX-2, article:ADMX-3}, HAYSTAC~\cite{article:HAYSTAC-1, article:HAYSTAC-2, article:HAYSTAC-3}, QUAX~\cite{article:QUAX-1, article:QUAX-2}, ORGAN~\cite{article:ORGAN-1, article:ORGAN-2}, MADMAX~\cite{article:MADMAX}, DM Radio~\cite{article:DMRadio}, CAST-CAPP~\cite{article:CAST-CAPP}, etc.), (ii) helioscopes pointing at the Sun to look for solar axions (CAST~\cite{article:CAST} and IAXO~\cite{article:IAXO}) and (iii) light-shining-through-the-wall schemes, which are configured to generate and detect axions in the lab (OSQAR~\cite{article:OSQAR} and ALPS (II)~\cite{article:APLS, article:ALPSII}).
Among these, the cavity haloscope method provides the most sensitive approach, particularly in the microwave region. 

A cavity haloscope consists of a microwave resonator immersed in a strong magnetic field where axions induce photons which couple with one of the cavity resonant modes, e.g., the TM$_{010}$ mode for a cylindrical geometry.
The expected power of the axion--photon conversion is given by
\begin{widetext}
\begin{equation}
    P_{a\gamma\gamma} = 8.7 \times 10^{-23}\,{\rm W} \left(\frac{g_{\gamma}}{0.36}\right)^2 
    \left(\frac{\rho_a}{0.45\, {\rm GeV/cm^3}}\right) 
    \left(\frac{\nu_a}{1.1\,{\rm GHz}}\right) 
    \left(\frac{B_{\textrm{rms}}}{10.3\,{\rm T}}\right)^2 
    \left(\frac{V}{37\,{\rm L}}\right) 
    \left(\frac{G}{0.6}\right) 
    \left(\frac{Q_c}{10^5}\right),
    \label{eq:conv_power}
\end{equation}
\end{widetext}
where $g_{\gamma}$ is the model-dependent coupling coefficient with a value of $-$0.97 and 0.36 for the KSVZ and DFSZ models, respectively, $\rho_{a}$ is the axion density in the local dark-matter halo, $\nu_a$ is the axion Compton frequency, and $B_{\textrm{rms}}$ is the externally applied magnetic field. $V$ and $Q_c$ are the volume and quality factor of the cavity, and $G$ is the form factor associated with the resonant mode, providing a measure of how well the external magnetic field is matched with the electric field associated with the relevant cavity mode. 
Assuming that local dark matter is composed solely of DFSZ axions, a cavity haloscope with the experimental parameters used in Eq.~(\ref{eq:conv_power}) would expect about 120 photons/s to be converted in the cavity and about 30~photons/s to be picked up by a critically coupled receiver antenna.
A measure of experimental sensitivity is given by the signal-to-noise ratio (SNR),
\begin{equation}
    {\rm SNR} \equiv \frac{P_{a\gamma\gamma}}{\delta P_{\rm sys}} = \frac{P_{a\gamma\gamma}}{k_B T_{\rm sys}}\sqrt{\frac{\Delta t}{\Delta \nu}},
    \label{eq:snr}
\end{equation}
where $\delta P_{\rm sys}(=P_{\rm sys}/\sqrt{\Delta \nu \Delta t}$) denotes fluctuations in system noise power over a bandwidth $\Delta \nu$ during an integration time $\Delta t$, and $P_{\rm sys}(=k_B T_{\rm sys}\Delta\nu$) describes the Johnson-Nyquist noise with the Boltzmann constant $k_B$ and the equivalent noise temperature $T_{\rm sys}$.
Since the axion mass is unknown a priori and the allowed mass range is vast, a fast search speed with a given sensitivity becomes the figure of merit for the experimental design.
The relevant quantity is obtained by inserting Eq.~(\ref{eq:conv_power}) into Eq.~(\ref{eq:snr}) as
\begin{widetext}
\begin{equation}
    \frac{d\nu}{dt} = 1.2\,{\rm \frac{GHz}{year}} \left(\frac{0.6}{\eta}\right)
    \left(\frac{5}{snr}\right)^2 \left(\frac{0.2\,{\rm K}}{T_{\rm sys}}\right)^2
    \left(\frac{P_{a\gamma\gamma}}{8.7\times10^{-23}\,{\rm W}}\right)^2
    \left(\frac{10^5}{Q_c}\right),
    \label{eq:scan_rate}
\end{equation}
\end{widetext}
where $\eta$ is the data acquisition (DAQ) efficiency and $snr$ is the target SNR value.
Major R\&D efforts have been made to increase the search rate by maximizing the figure of merit ($\mathcal{FOM} \equiv B_{\textrm{rms}}^4V^2G^2Q_c/T^2_{\rm sys}$), the metric that defines the strength of the experiment for a given frequency.

The Center for Axion and Precision Physics Research (CAPP) of the Institute for Basic Science (IBS) was established in October 2013 at the Korea Advanced Institute of Science and Technology (KAIST) in Daejeon, Republic of Korea. IBS-CAPP's main focus has been cavity haloscope searches utilizing multiple experimental setups designed for different mass regions.
Among them, CAPP-MAX (Main Axion eXperiment), formerly known as CAPP-12TB, is the most sensitive (flagship) experiment at CAPP, taking advantage of available cutting-edge technologies learned from smaller experiments running in parallel.
The main components of the system include a superconducting solenoid with a center field of 12\,T, based on a $\rm Nb_3 Sn$ superconducting magnet (inner) and a regular $\rm NbTi$ superconducting magnet (outer), built by Oxford Instruments~\cite{OI-PAPER}. 
A cryogenic dilution refrigerator (DR) with a measured cooling power of 1\,mW at 90\,mK is capable of reaching a base temperature of 5.6\,mK.
The detector comprises a 37-liter lightweight copper cavity connected to a readout chain consisting of flux-driven Josephson parametric amplifiers~\cite{article:Kutlu21} (JPA), followed by a series of linear semiconductor amplifiers.
This experiment has recently reported a search using unprecedented sensitivity to probe DFSZ axion physics above 1\,GHz~\cite{12TB-PRL, Sagitarius} of about 20\,MHz.
In this paper, we report new results from an extended search for invisible axion dark matter with CAPP-MAX, the first search at or near DFSZ sensitivity over a large frequency range of more than 100\,MHz above 1~GHz.

The remainder of this paper is organized as follows. 
In Sec.~II, we introduce the major experimental equipment including the 12\,T superconducting magnet and the wet-type cryogenic dilution refrigerator system.
Their manufacturing specifications and the actual measurements of the relevant parameters such as magnetic field and base temperature are provided.
Section III describes the detection system consisting of the microwave cavity and readout electronics, featuring quantum-noise-limited amplifiers (QNLA) in the first stage of the chain.
The characteristics and quality of the individual components are presented in sufficient detail, with particular focus on noise calibration and cavity performance.
Section IV is dedicated to the automated data acquisition and monitoring system, specifically designed for continuous data flow and high-efficiency real-time monitoring of the entire system.
Section V presents the analysis procedure and the results. 
The analysis follows the standard procedure commonly adopted in the community and is similar to the analysis reported in our previous searches~\cite{12TB-PRL, Sagitarius, article:CAPP-8TB, article:CAPP-9T, article:CAPP-PACE, article:CAPP-PACE-JPA}. 
Finally, we highlight the importance of the experimental results and provide a perspective for our near-future work in Sec.~VI.

\section{Major equipment}
\subsection{Magnet} \label{Magnet}
\begin{figure}[ht]
\begin{center}
\includegraphics[width=\linewidth]{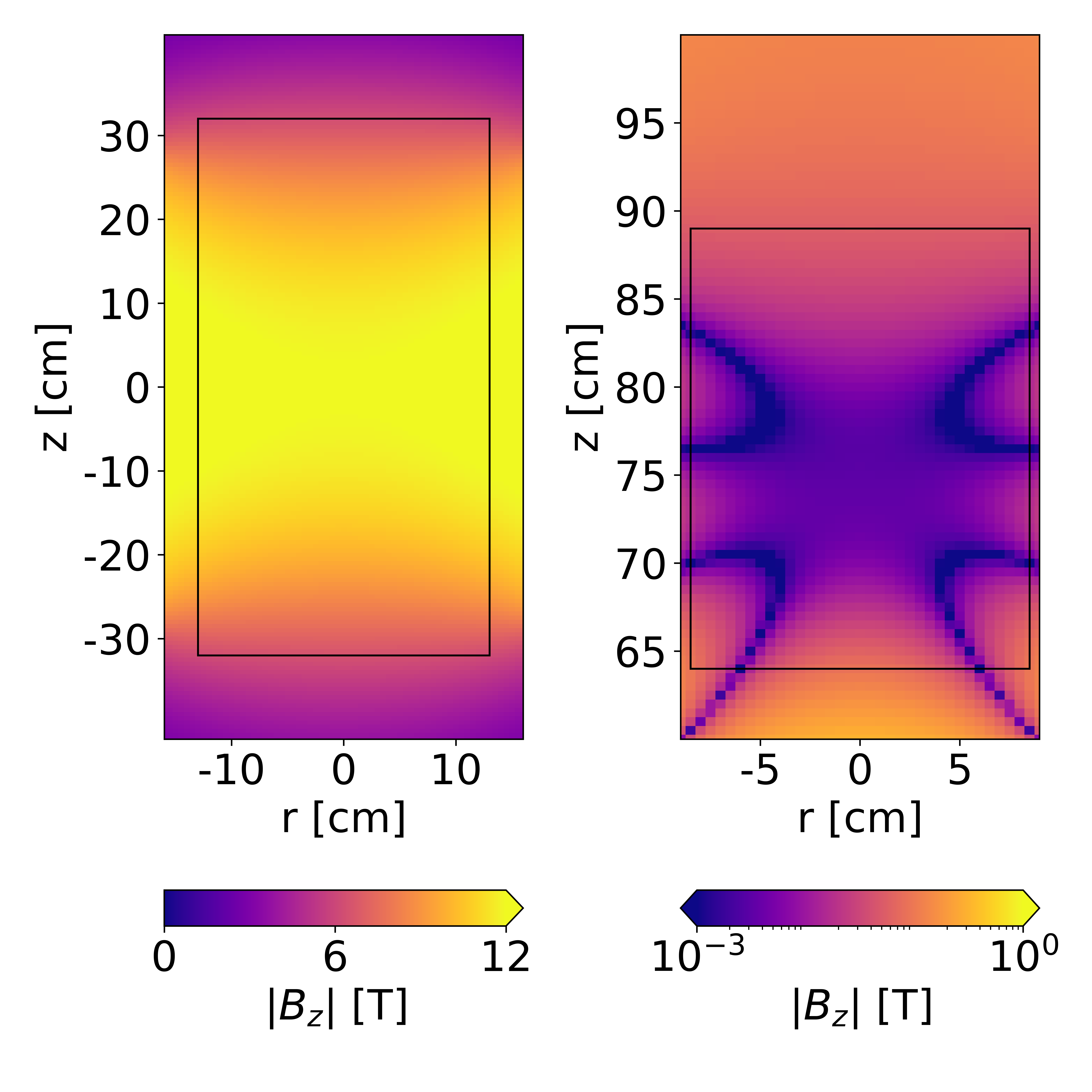}
\caption{\label{fig:field_maps} Magnetic-field distributions in the cavity region (left) and cancellation region (right), respectively. The coordinates $z$ and $r$ refer to the longitudinal and radial axes of the solenoid, correspondingly. The black outline on the left corresponds to the location of the cavity, while on the right it corresponds to the location of the JPA assembly. The magnet dimensions are given in the text.}
\end{center}
\end{figure}

The $\mathcal{FOM}$ in axion haloscope searches is dominated by the strength and volume of the magnet, making it the key factor in CAPP-MAX. Our superconducting magnet, classified as ``wet'', necessitates direct contact with liquid helium (LHe) to keep the coils in the superconducting state.
 It has three main characteristics: 
\begin{enumerate}
\item 
The cylindrical magnet used in our experiment, manufactured by Oxford Instruments~\cite{OI-PAPER}, can operate at a central magnetic field of 12\,\si{\tesla} when cooled to a temperature of 4.2\,\si{\kelvin} using LHe. In Fig.~\ref{fig:field_maps}, the magnetic field map in the cavity region is shown in the plot on the left. The total energy content of the magnet, when fully energized to 12\,\si{\tesla}, is 5.562\,MJ. 
\item 
The solenoidal magnet used in CAPP-MAX has a cold bore diameter of 320\,\si{\milli\meter}. The magnet is composed of two different low-temperature superconductors: Nb$_3$Sn for the inner coil and NbTi for the outer one. These coils are nested together to form a concentric solenoid. The inner coil has a length of 560\,\si{\milli\meter}, while the outer  has a length of 640\,\si{\milli\meter}. Taking into account the dimensions of the coil and the radiation shields of the dilution refrigerator insert (see Section~\ref{sec:detection}), the experiment can accommodate a cylindrical cavity with a volume of 37\,liters inside the solenoidal magnet's bore. The average root mean square (rms) magnetic field along the magnet axis over this volume measures 10.55\,\si{\tesla}.

\item 
The field cancellation region is located 750\,\si{\milli\meter} above the center of the magnet, and has  both diameter and length dimensions of 100\,\si{\milli\meter} in cylindrical coordinates. The cancellation coil is connected in series with the main solenoid, sharing the same power supply and superconducting short circuit. Within this region, the generated magnetic fields are weaker than $10^{-2}$\,\si{\tesla}, as illustrated on the right of Fig.~\ref{fig:field_maps}. The presence of this cancellation region is of utmost importance to ensure the proper operation of the JPAs, which serve as the initial-stage RF amplifier, as well as for the required microwave circulators and other RF components in the setup.
\end{enumerate}

The combination of three characteristics discussed above are unparalleled worldwide, allowing us to conduct axion dark-matter searches with increased sensitivity~\cite{12TB-PRL,Sagitarius}. These distinctive features not only enabled our current research but also pave the way for continued investigations into axion dark matter at higher frequencies.

The magnet can be safely ramped up at a rate of 0.13\,\si{\tesla}/min without any load inside the magnet bore. In addition, the magnet system also incorporates a crucial feature known as persistent-mode operation, which allows it to operate without the need of an external power supply once it has been ramped up. 

This persistent-mode operation significantly reduces the evaporation rate of LHe to 30\,liters/day, compared to 40\,liters/day during driven-mode operation. The decay rate of persistent mode operation is approximately 30\,parts per million per day, which has no practical impact on a typical axion-haloscope search schedule.

The magnet system underwent a quench test to evaluate its response before delivery by Oxford Instruments~\cite{OI, OI-PAPER}. 
It has been successfully installed inside a vapor shield cryostat, with a height of approximately 3.3\,\si{\meter}. 
The cryostat provides a useful volume of 478\,liters for containing the LHe. 
The magnet system was delivered to the CAPP experimental hall in March 2020. 
However, due to the impact of the COVID-19 pandemic, the commissioning run was delayed until August of the same year and finally completed in December 2020. 
The magnet was powered up in spring 2021 and has been maintained at a cold temperature, with only one instance of raising its temperature above 100\,\si{\kelvin} up to summer 2023.

Our magnet weighs approximately 1200\,\si{\kg}, and it is most convenient to keep it on a Low Vibration Pad of our experimental hall during operation and during the period between experiments. 
This necessity requires us to move the DR up and down during the installation and warming-up procedures. 

\subsection{Cryogenics}
One of the critical factors in Eq.~(\ref{eq:snr}) is the system noise temperature, which is comprised of various components, including the cavity, the initial amplifier stages, passive RF components, cable and connector losses, among others.
To minimize the impact of these effects, it is essential to reduce their physical temperature.
Furthermore, certain components, such as QNLA, operate only under cryogenic conditions, and their added noise depends on their physical temperatures.
High-electron-mobility-transistor (HEMT) amplifiers, while functional over a wide temperature range, exhibit their best noise properties below 10\,\si{\kelvin}~\cite{Shleeh2012, Ivanov2020}.
Therefore, having cryogenic equipment capable of cooling these crucial components in the readout circuit is essential to achieve the best performance of the system.

The determination of how low we need to cool our critical components can be derived from our target frequency range, which, for CAPP-MAX at this stage, falls between 1 and 2\,GHz.
According to the standard quantum limit, the noise temperature ($T_{\rm SQL}=\hbar\omega /k_B$) for a frequency of 1\,GHz is 48\,mK,  one component of which is the added amplifier noise of 24\,mK for 1\,GHz ($T_{\rm add,min}=\hbar\omega /2k_B$).
This provides us with guidance on the cooling temperatures required for our components to achieve optimal performance.
Temperatures below 30\,mK can be achieved with a standard dilution refrigerator.

Another part of our system that requires cryogenic temperatures is the superconducting magnet, manufactured by Oxford Instruments and featuring a complex design (see Sect.~\ref{Magnet}). 
During a warm-up phase of the magnet to 100\,K, in order to remove solidified nitrogen and oxygen from the main dewar, the outer vacuum chamber (OVC) was pumped to improve the isolation and remove any penetrating gas. 
This process improved vacuum isolation and reduced the helium gas pressure in the main dewar by a factor of 6.

%The description of the cryostat was provided earlier in Sec.~\ref{Magnet}. 
%The superconducting magnet has been maintained at LHe temperatures, except for a brief period when it was warmed up above 90\,\si{\kelvin}. During this warm-up phase, the outer vacuum chamber (OVC) was pumped to remove any penetrating gas. This process improves vacuum isolation and reduces the LHe evaporation rate by a factor of 6.

\begin{figure*}[ht]
    \begin{center}
    \includegraphics[width=1\linewidth]{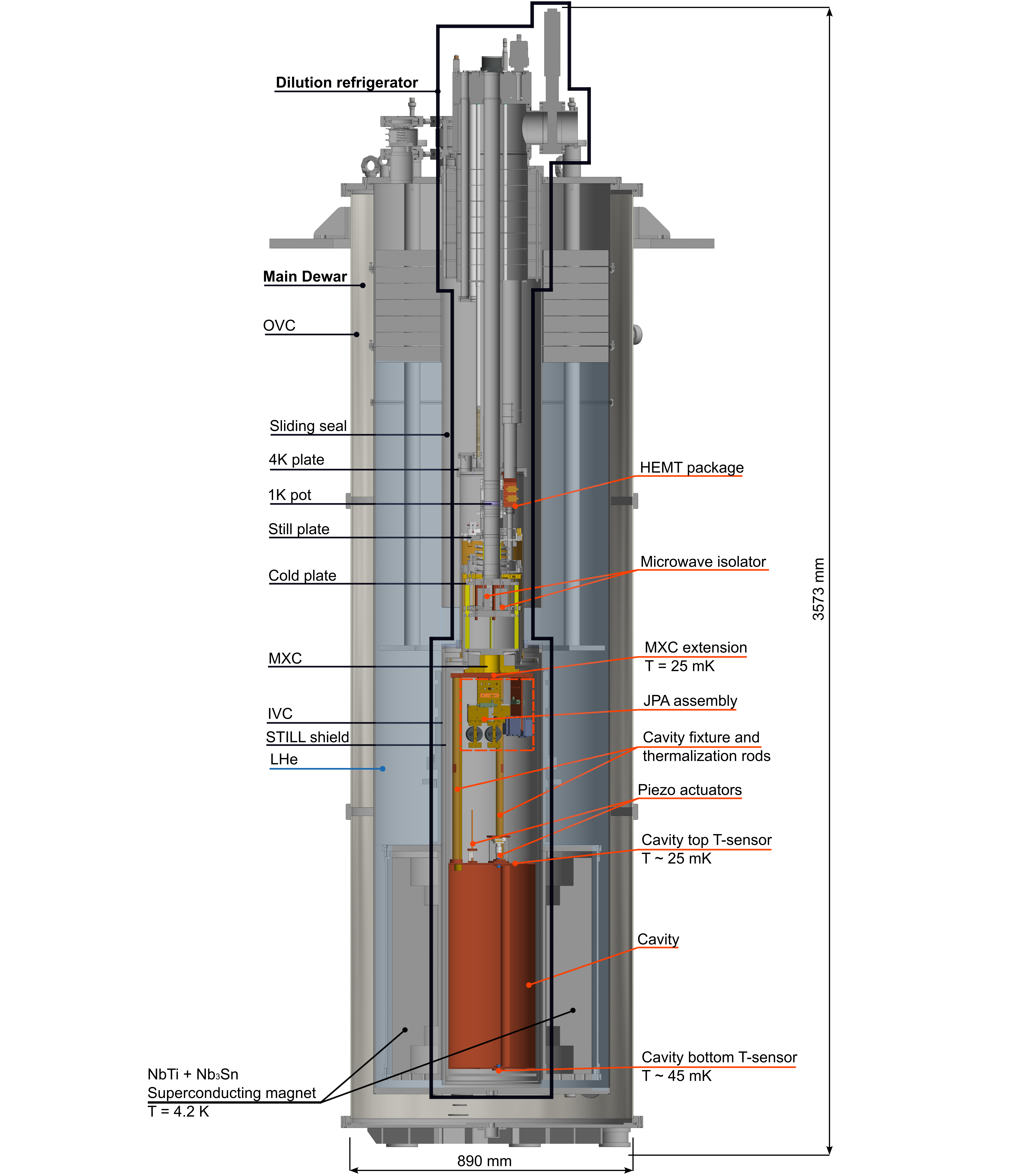} 
    \caption{\label{fig:main view} Comprehensive view of the detector setup within the cryostat. It includes a cross-sectional view of the main Dewar containing LHe and the inserted dilution refrigerator (outlined in black). The HEMT package is located on the 4\,K plate, while the microwave isolator is located on the Cold plate. 
    The cavity and MXC assembly are affixed to the gold-plated copper extension of the mixing chamber via four gold-plated copper thermalization rods. The Still radiation shield is linked to the Still extension. The cavity has two temperature sensors to monitor top and bottom temperatures and two piezo actuators to adjust the cavity resonance and coupling coefficient. 
    }
    \end{center}
\end{figure*}

We utilize a Leiden Cryogenics DRS1000M~\cite{DRS1000M} dilution refrigerator, which offers a nominal cooling power of approximately 1.3\,mW at a temperature of 120\,m\si{\kelvin}, (marked as Dilution refrigerator on Fig.~\ref{fig:main view}). 
The DR is equipped with three clear 25-m\si{\meter}-diameter tubes extending from the top of the DR down to the 4\,\si{\kelvin} stage, which is used for wiring purposes. Three 24-pin Fischer connectors~\cite{Fischer} facilitate connections to the mixing chamber (MXC), while one of the connectors is used for wiring to the 4\,\si{\kelvin} plate. 
The wiring is constructed of phosphor-bronze twisted pairs. Two differential lines are used to control the piezo actuators, and they are made of copper wire that runs from the room-temperature (RT) flange down to the 4\,\si{\kelvin} plate.
From the 4\,\si{\kelvin} plate down to the MXC, four and five differential pairs made out of NbTi wires are connected in parallel to enable the supply of high-current pulses, with amplitudes reaching a few amperes, to the piezo actuators.

All RF connections that run from the RT flange to the 4\,\si{\kelvin} plate are constructed using 0.034-inch-diameter CuNi cables. 
One of these cables is kept as spare and kept terminated inside the DR. 
The next four cables that connect the system inputs are made of  0.047-inch-diameter BeCu cables, while the output cable is constructed from superconducting 0.034-inch-diameter NbTi cables. 
All interconnections between cables are made using gold-plated SMA connectors and bulkheads are constructed of brass and BeCu.

The DR is equipped with five standard temperature sensors from Leiden Cryogenics, which are used to monitor the temperature and control it through a proportional-integral-differential (PID) controller.
We used an additional 10 channels to accurately measure the following temperatures: 4\,\si{\kelvin} plate, two HEMT amplifiers, two microwave circulators thermally anchored to the cold plate, the top and bottom of the cavity, and the extension of the mixing chamber. 
One of the thermometers was attached to the piezo rotator within Run 4 and later on it was placed at the cold plate within Run 5.

The inner vacuum chamber (IVC) shield maintains a temperature at the level of LHe, while inside, there is an immersed gold-plated radiation shield with a temperature ranging from approximately 550 to 800\,m\si{\kelvin} during the refrigerator operation. Due to space constraints, we have opted not to use the radiation shield that is typically thermally connected to the refrigerator's cold plate. This is one of the reasons why the base temperature is higher than the one specified in the manual.

\begin{figure*}[ht]
    \begin{center}
    \includegraphics[width=\linewidth]{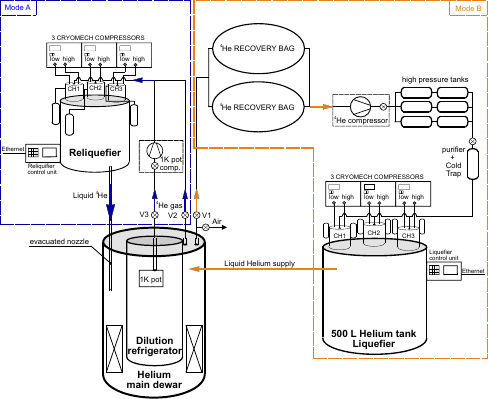}
    \caption{\label{fig:He_recovey_diagram} Schematic view of the helium gas recovery system (GRS) operating in two modes  shown with blue and orange dashed outlines: mode A -- Data taking and mode B -- DR inserting and main dewar maintenance.
    In the first configuration mode, the main recovery line is followed by the manual valve (V1), the two He gas recovery bags and the Cryomech He compressor~(model HRCP), the six high-pressure tanks, the purifier~(model  HRSMP-Purifier), the three cold heads (CH1--CH3) (model  PT415) charged by the Cryomech He compressors~(model  CP1110) placed on the top of the 500-liter non-magnetic He liquefier~(model LHeP60), and the liquid-cryogen level monitor~(model LM-510) with PID control.    
    In the second mode, the He closed-cycle operation is used. The main dewar is followed by the manual valve (V2), the reliquefier~(RL)~(model HeRL60), the three Cold Heads (CH1--CH3)~(model  PT420-RM), charged by the 3 Cryomech compressors~(model  CPA1114). The pressure and temperature control unit with PID from Stanford research systems~(model CTC100) is used. The output of the reliquefier is followed by the evacuated-jacket metal nozzle.
    The 1~\si{\kelvin} pot is followed by the manual valve (V3), by the Cryomech 1\,K pot compressor~(model  PHRS1E350006), and is connected to the input of the RL.
    }
    \end{center}
\end{figure*}

The DR is immersed in a LHe cryostat, which also houses a superconducting magnet. The IVC is filled with helium exchange gas, which aids in precooling the DR insert to LHe temperature.
Before immersing the DR, the 1\,\si{\kelvin} pot needs to be pressurized, typically to a pressure of around 1.3\,\si{\bar}. 
This pressurization is essential to prevent non purified helium gas, which may include slight nitrogen contamination, from entering these capillaries connecting the 4\,\si{\kelvin} pot to the cryostat. 
If helium gas mixed with nitrogen were to penetrate these capillaries, it could lead to the anti-sublimation of nitrogen on the capillary walls. 
This, in turn, could block the capillaries, rendering the operation of the 1\,\si{\kelvin} pot ineffective or even impossible.

Once the DR insert reaches approximately liquid-helium temperature, we initiate the process of pumping out the exchange gas from the IVC using dry rotary and turbo-molecular drag pumps, reducing the pressure below $5 \times 10^{-5}$\,\si{\milli\bar}. 
This process typically takes about 18\,hours to complete and requires running the sorption pump heater.
The primary cooling process takes place predominantly within the 1\,\si{\kelvin} pot. 
LHe is delivered from the main dewar through two tubes connected to the 4\,\si{\kelvin} plate. 
From there, it flows through thin capillaries, filling the 1\,\si{\kelvin} pot with LHe. 
To facilitate cooling, the 1\,\si{\kelvin} pot is pumped using an external compressor, shown as ``1\,\si{\kelvin} pot comp.'' in Fig.~\ref{fig:He_recovey_diagram}. 
As pumping progresses, the temperature of the 1\,\si{\kelvin} pot drops to the range of 1.4--1.8\,\si{\kelvin}, and it also pre-cools the insert part of the DR. 

Once the 1\,\si{\kelvin} pot is sufficiently cold, the $^3$He--$^4$He mixture can be introduced into the fridge by the condensing process.
When the mixture is filled into the MXC, circulation can begin. 
This circulation causes the still and mixing chamber to cool to below 1\,\si{\kelvin}. Once the mixing chamber has reached its base temperature (for fully loaded DR it is currently about 25\,m\si{\kelvin}), the Still heater can be activated, typically requiring a couple of m\si{\watt} of power.
This way we are able to control the flow rate of the mixture.
Generally, with this level of heating power, the pressure in the Still line stabilizes at around $8.2\times 10^{-2}$\,m\si{\bar}. 

The cooling power and time of each stage in a dilution refrigerator depend on several factors, including the design of the refrigerator, the heat load imposed on it, and the heat capacitance of the components integrated into the custom system. 
Achieving a good thermal connection between components is a crucial practical consideration. When two surfaces touch, the actual contact occurs only at the points where there are protrusions or asperities. To ensure effective connection, it is necessary to increase the contact area by plastic deformation by applying pressure during the connection process. Additionally, to compensate for thermal expansion and contraction in linear dimensions when using threaded links, spring washers should be employed. To eliminate the effect of metal oxide on the surface during thermalization, copper parts are gold-plated.

To increase the surface area of contact between dielectrics or between dielectrics and metals, we apply Apiezon~N grease~\cite{Apiezon} between them.
In general, we aim to minimize the use of dielectric components or reduce their size on the \si{\milli\kelvin} stages to expedite the cooling process and limit their impact.

The RF cables were thermally connected to the bulkheads on the 4\,\si{\kelvin} stage and the MXC. They were also thermally linked to the heat exchangers of the DR in between.
We used superconducting cables at temperatures below 4\,\si{\kelvin}.

The main evaporation of the He gas comes from the main dewar and the 1\,\si{\kelvin} pot of the DR.
In order to avoid any helium losses during the maintenance of the main dewar, the insertion of the DR and the data taking process we use He recovery lines (controlled by V1, V2, V3, see Fig.~\ref{fig:He_recovey_diagram}).

The initial He recovery line operates during the insertion of the DR into the main dewar or while warming up or cleaning the main dewar of contamination (Mode B in Fig.~\ref{fig:He_recovey_diagram}). For this operation, we employ a LHeP60 helium liquefier plant~\cite{LHeP60}, with the ability to liquefy up to 60\,liters helium per day and equipped with a storage capacity of 60\,liters of helium gas in gas tanks.

This line consists of  two He recovery bags, a helium compressor, high-pressure tanks, two purifiers for water absorption, a liquid-nitrogen cold trap, and a 500-liter helium tank (referred to as \text{Liquefier} in Fig.~\ref{fig:He_recovey_diagram}). Three cold heads, marked CH1--CH3 in Fig.~\ref{fig:He_recovey_diagram}, run the condensing process of the incoming He gas flow. The PID system, based on the cryogenic level monitor controls  the temperatures of the cold heads and the pressure within the liquefier's dewar.

After the He liquefaction, the transfer of liquid He to the main dewar is accomplished by directly connecting the main dewar and the liquefier dewar using an evacuated siphon. The liquefier dewar is constructed from non-magnetic metals, allowing its placement in close proximity to the main dewar, even within the high magnetic field area. The Cryomech He compressor is complemented by closed-cycle distilled water chillers, positioned outside the laboratory building for better cooling efficiency.

The second He recovery line, identified in the \emph{Mode A} in Fig.~\ref{fig:He_recovey_diagram}, operates in a closed He loop and does not require a high-volume He reservoir. The He gas evaporates from both the He main dewar, through valve~2, and from the 1~\si{K} pot, through valve~3, shown with arrows in Fig.~\ref{fig:He_recovey_diagram} to the liquefier tank and condenses at the three cold heads.

The liquid helium is returned to the main dewar via a vacuum thermo-isolated nozzle. The vertical position of the reliquefier (RL) HeRL60~\cite{HeRL60}, is adjustable. Varying the height of the RL maintains high operational efficiency since it keeps the end of the nozzle above the liquid helium level. The incorporation of an evacuated jacket around the metal nozzle has effectively reduced internal helium evaporation from the main dewar, resulting in a substantial decrease in the main dewar pressure from 2.4\,psi to nearly 0\,psi.
In this way, we are able to set the PID system to support 0.3\,psi of the input pressure using internal heaters inside of the RL.
During the data taking, the liquid He flows continuously, closing the loop between the RL and the main dewar.

\section{\label{sec:detection}Detection system}
\subsection{Microwave cavity} 
The axion haloscope employs a tunable cavity with a high quality factor, placed in a strong magnetic field, to detect signals generated when the axion frequency falls within the cavity resonance. As seen in Eq.~(\ref{eq:conv_power}), the larger the magnetic field and cavity volume, the more of the axion dark matter field is converted into microwave photons. A higher quality factor results in converted microwave photons surviving inside the cavity for longer periods, amplifying the signal. 

The CAPP-MAX relies on a powerful 12\,\si{\tesla} magnetic field generated within a large 320\,\si{\milli\meter} bore. This requires the design of a large cavity that allows high $B^2 V$. To address this, CAPP designed an ultralight cavity (ULC) that weighs less than 5\,kg including the tuning rod and occupies a volume of 37\,L, while maintaining a Q-factor of approximately $10^5$. This enables axion searches within the range of 1.02$-$1.185\,GHz.

The size of the cavity for the CAPP-MAX is influenced by the bore size of the magnet and the dimensions of its various internal radiation shields. 
The innermost 50\,\si{\milli\kelvin} shield is positioned within the still radiation shield, with an inner diameter of 281\,\si{\milli\meter}. Based on this fact, the outer diameter of the cavity was determined to be 274\,\si{\milli\meter}.
In the described axion experiments, the 50\,\si{\milli\kelvin} shield was not utilized. Without the 50\,\si{\milli\kelvin} shield, the additional heat load generated from the still shield, when calculated according to the Stefan-Boltzmann law, is negligible, amounting to less than 0.1\,\si{\micro\watt}. This decision has shortened the system setup time and provided additional space, enabling the use of larger cavities in subsequent experiments, thereby enhancing sensitivity.

\begin{figure*}[ht]
    \begin{center}
        \includegraphics[width=\linewidth]{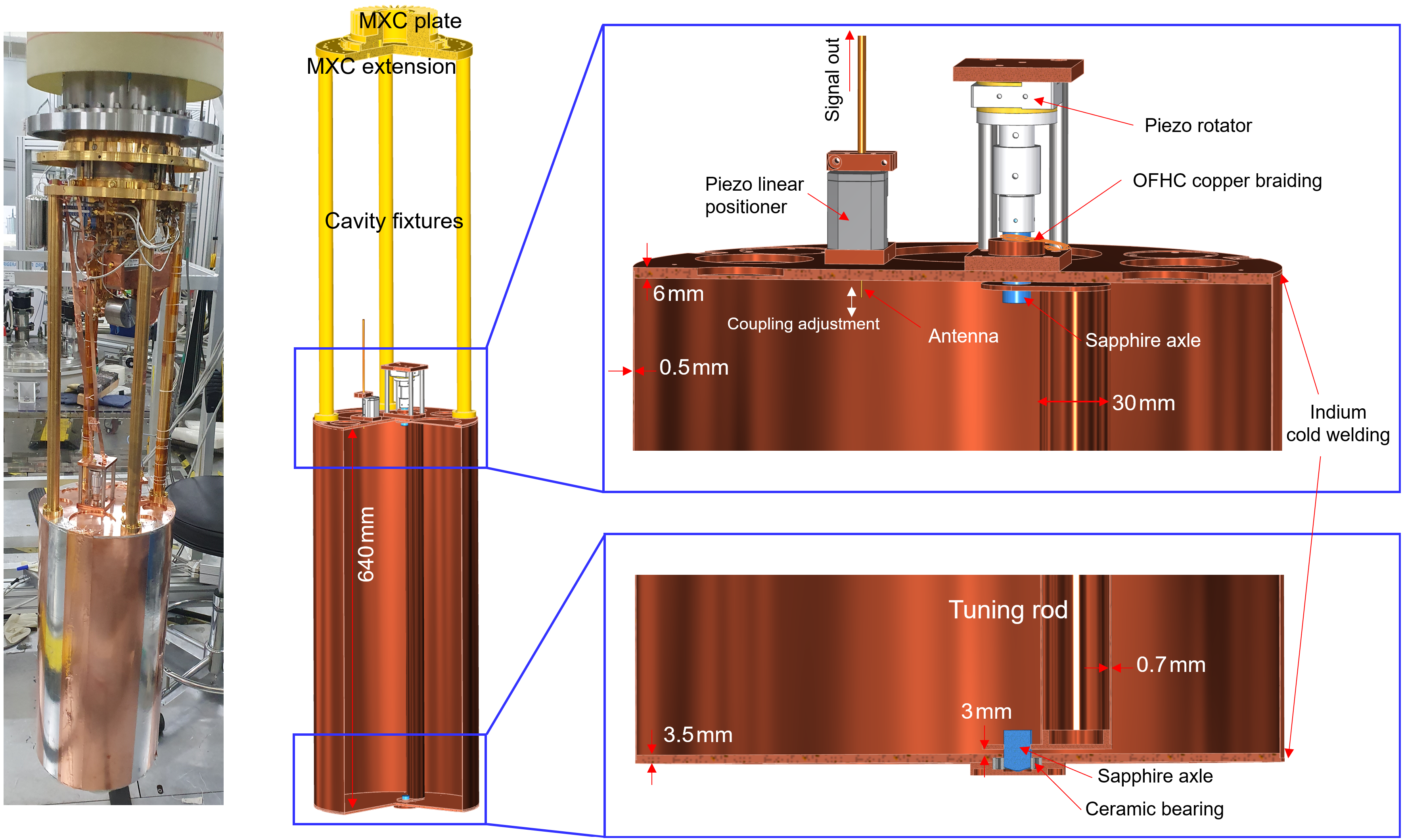}
        \caption{\label{fig:ULC_design} Photo of the ULC (left), a 3-dimensional cross-sectional view of the mechanical design (middle), and an enlarged view of the top and bottom (right). The four cavity fixtures and thermalization rods made of gold-plated copper thermally anchor the cavity which is affixed to the MXC extension. The piezo linear positioner adjusts the antenna coupling between the readout chain and the cavity. The piezo rotator steers the copper tuning rod with the sapphire axle, which is thermalized via OFHC copper braiding. Indium cold welding connects the cavity top and bottom to the side walls.} 
    \end{center}
\end{figure*}

To minimize space loss due to the thickness of the cavity walls and to reduce its weight, thin metal sheets were rolled to create the cavity's side walls. Oxygen-free high-thermal-conductivity (OFHC) copper, known for its high thermal and electrical conductivity, was used as the material of choice, allowing for a high Q-factor while facilitating effective cooling.
Figure~\ref{fig:ULC_design} depicts the structure of the first CAPP-MAX cavity. The top and bottom plates of the cavity are 6\,\si{\milli\meter}  and   3.5\,\si{\milli\meter} thick, respectively. The height of the cavity is 640\,\si{\milli\meter} with the thickness of its side walls  0.5\,\si{\milli\meter}. Excluding the tuning rod, the ULC weighs only about 4\,kg.

To adjust the resonance frequency of the cavity, we rotated the position of an empty cylinder with a diameter of 30\,\si{\milli\meter}, a height of 636\,\si{\milli\meter}, a thickness of 0.7\,\si{\milli\meter} within the cavity, and the rotation axis located 25\,\si{\milli\meter} away from the center of the cavity. This rotation had a 50-\si{\milli\meter} diameter that can be used to tune the TM$_{010}$-like mode resonance frequency in the range of 1.02$-$1.185\,GHz. 
Simulations show~\cite{COMSOL} that the tuning rod's relatively small size compared to the cavity diameter and its restricted movement within 50\,\si{\milli\meter} from the cavity center help mitigate the mode localization often observed when using a metal tuning rod in cavities\cite{article:HAYSTAC-3}.
We attached a sapphire axle to the center of rotation of the tuning rod, allowing precise tuning with a resolution of less than 1\,millidegree using the piezo rotator from Attocube Systems AG~\cite{ANC350}. This setup allowed for a tuning resolution of less than 1\,kHz. 
Additionally, the alignment of the upper and lower sapphire axles was adjusted within a 1~\si{\milli\meter} margin of error to keep the maximum tilting angle of the tuning rod below 0.2\,degrees, ensuring that the form factor remains high on average.

\begin{figure}[ht]
    \subfigure[]{
        \includegraphics[width=.48\textwidth]{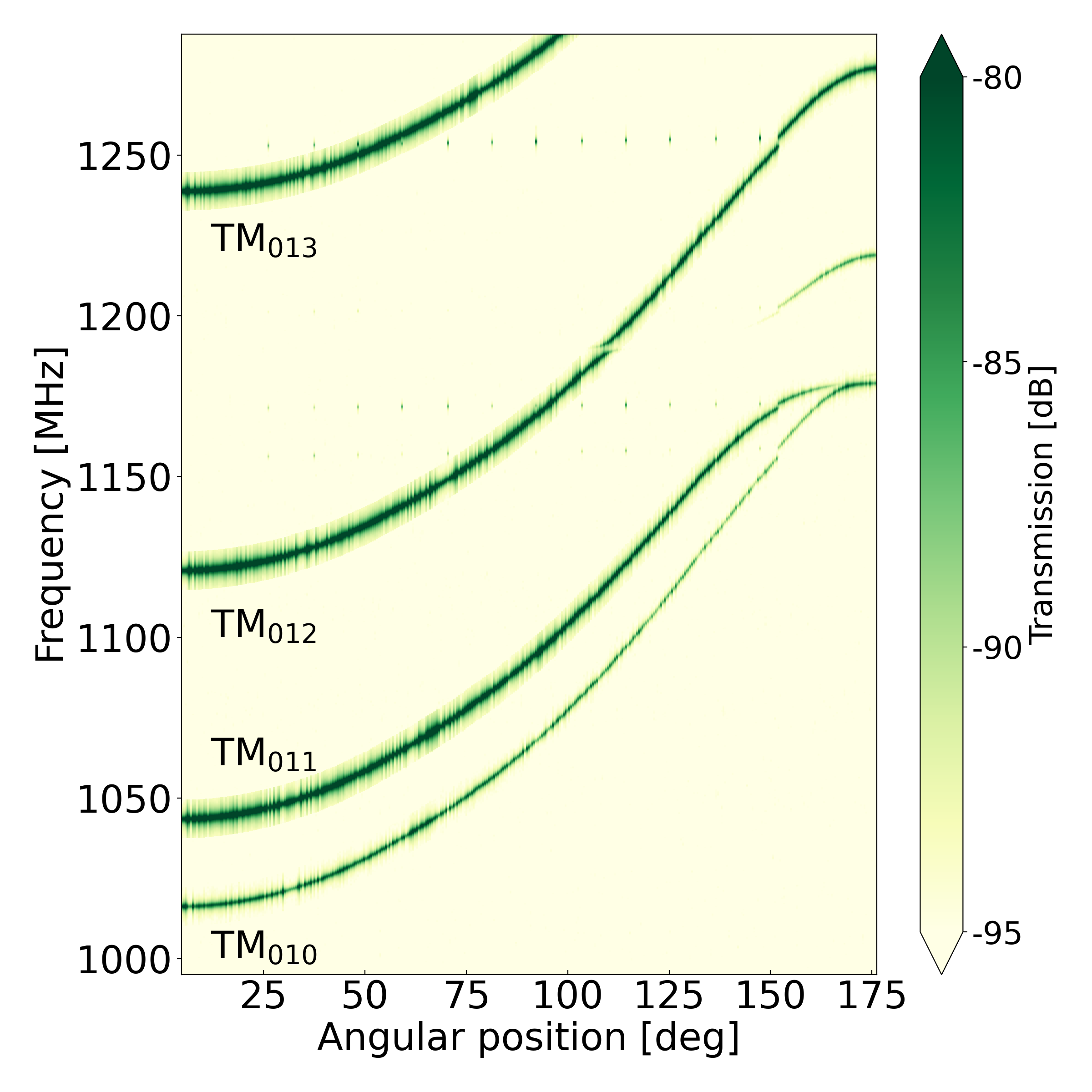}
    }
    \subfigure[]{    
        \includegraphics[width=.48\textwidth]{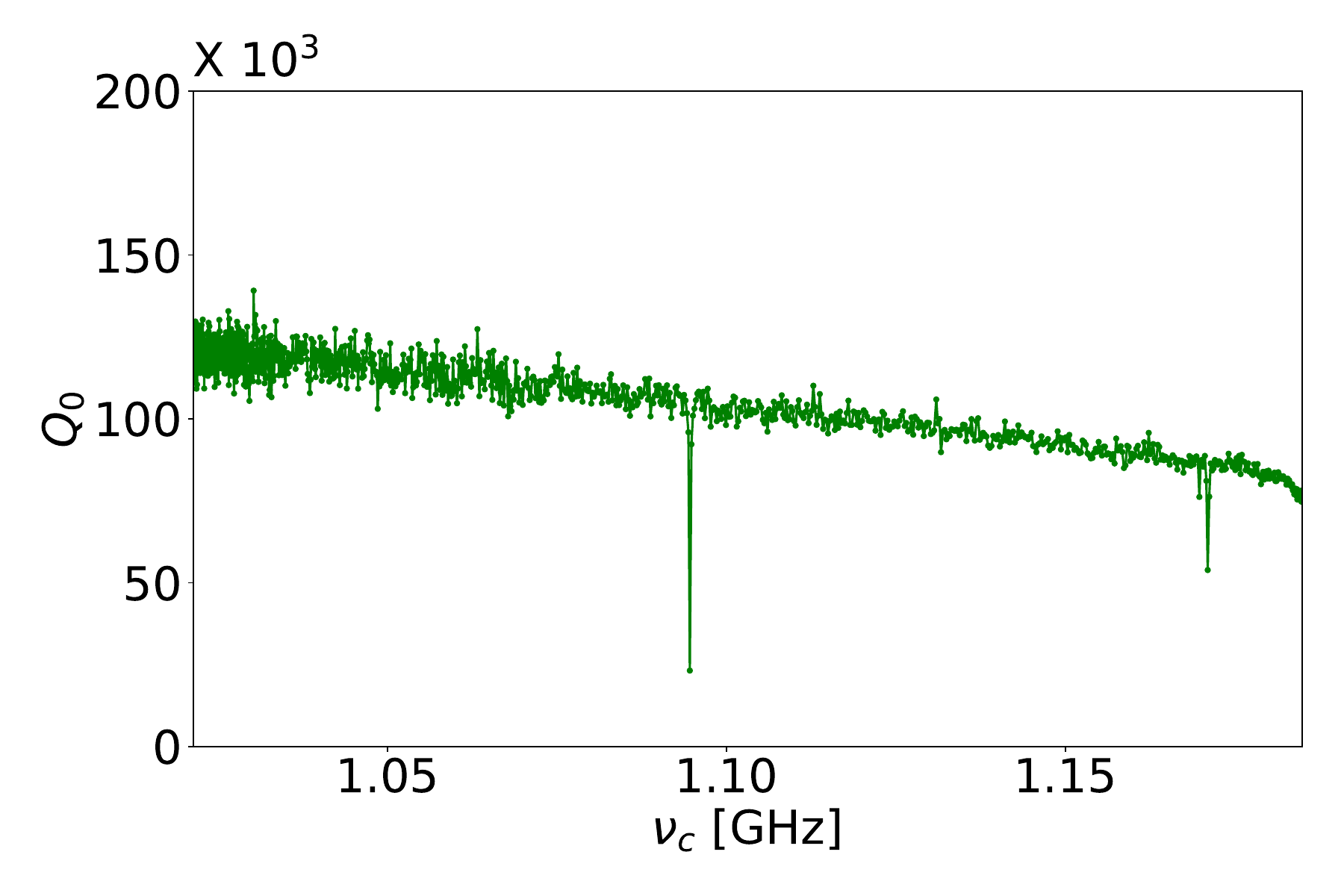}
    }    
    \caption{\label{fig:ULC_modemap} Cavity parameters obtained at 4.2\,K. (a) Mode map: resonant frequency vs. angular position of the tuning rod. The intensity of the lines correspond to the forward transmission coefficient (S21) in logarithmic scale. The individual modes are identified from simulation. (b) Unloaded quality factor as a function of frequency. The two vertical dips correspond to mode crossings.}
\end{figure}

Transmission scattering parameters were measured using a vector network analyzer (VNA) to create a mode map and the Q-factor versus the resonance frequency of the cavity for the 1$-$1.2\,GHz frequency range, as shown in Fig.~\ref{fig:ULC_modemap}.
The mint-colored line at the bottom of the mode map represents the resonance frequency range of the TM$_{010}$ mode. Although minor mode crossings occurred at some frequencies throughout the tuning range, there were no significant mode mixings, consistent with the initial design.
The unloaded cavity Q-factor was greater than $1.2\times 10^5$ when the tuning rod was farthest from the center (50\,mm) and remained at $0.9\times 10^5$ when positioned at the center, maintaining over $10^5$ throughout most of the tunable frequency range.

Effective thermal management is essential to minimize the physical noise from the cavity and the tuning rod. The key steps involve optimizing thermal conductivity, reducing the number of thermal interfaces, and enhancing efficiency at inevitable interfaces \cite{Ekin_cryo_book}.

Initially, we secured four OFHC copper rods with gold plating to the upper part of the cavity using bolts. This approach facilitated efficient thermalization. Interfaces between the upper and lower cavity sections were bonded through indium cold welding. Additionally, OFHC copper strips, secured with bolts, formed a direct connection between the upper and lower cavity plates, further augmenting thermalization.

The task of cooling the tuning rod posed a multifaceted challenge. Because the tuning rod must rotate smoothly and cannot be rigidly attached to other components, conventional cooling methods were rendered impractical. Furthermore, the substantial heat contribution of the tuning rod relative to its surface area potentially creates the so-called ``hot-rod problem''~\cite{article:HAYSTAC-1}. To address this issue, we employed varnish and glue to create robust thermal interfaces between the upper cap of the tuning rod and the sapphire axle, resulting in improved thermal conductivity. To further enhance the thermal connection, copper braiding was employed to link the sapphire axle and the cavity's top plate. This braiding was secured with bolts and enveloped with varnish. 
%This strategy ensured highly-effective heat transfer. 
To additionally stabilize the resonance frequency and diminish heat generated by vibrations, the sapphire axle connected to the bottom of the tuning rod featured a ball-like shape, allowing it to make direct contact with the OFHC copper floor.% for heat dissipation while suppressing undesirable vibrations.

Notwithstanding the considerable size of our experimental cavity, we successfully crafted an ULC using OFHC copper. Our approach utilized various techniques, each tailored to address specific thermal interfaces. This ULC enabled us to attain an low cavity temperature of about 30\,mK, marking a significant milestone in achieving one of the lowest-temperatures and lowest-noise axion experiments to date.

\subsection{\label{subsec:readout}Readout electronics}
%\paragraph{Quantum noise limited amplifier} 
One critical element of CAPP-MAX's readout electronics is the low-noise amplifier, limited by quantum noise. The noise of these amplifiers is constrained by the uncertainty principle. Quantum squeezing technology~\cite{Zhong2013} offers a way to overcome this limitation. However, this approach doesn't readily apply to measuring signals with a stochastic phase component, as it generally necessitates prior knowledge of the anticipated signal's phase~\cite{Clerk2010}. Although the vacuum squeezing technique has been used in axion experiments to achieve a two-fold gain in scanning rate~\cite{Backes2021} for frequencies exceeding 4\,GHz, we have not yet implemented this technique in our work.
%\paragraph{JPA characteristics / HEMT}

\begin{figure}[ht]
\begin{center}
\includegraphics[width=.48\textwidth]{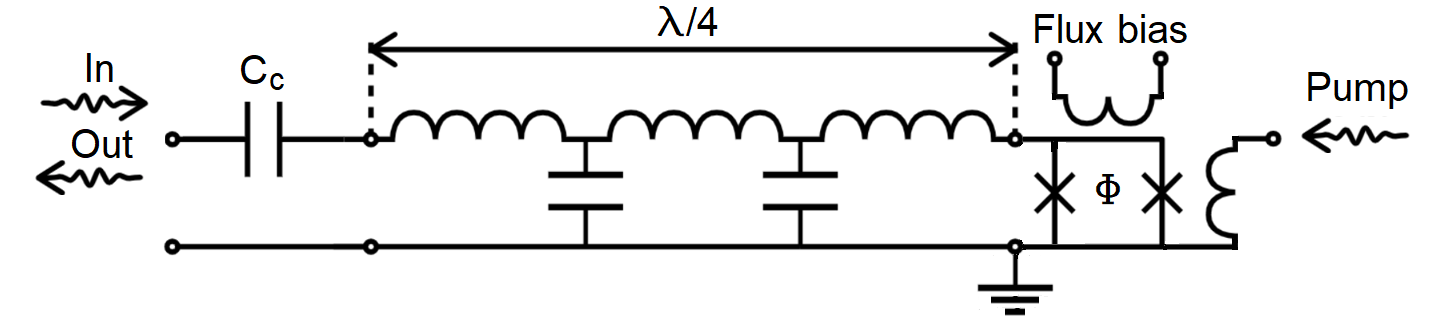}
\caption{\label{fig:FDJPA1}Equivalent circuit diagram of a JPA. The magnetic flux bias defines the average inductance of the SQUID loop (the Josephson junctions are denoted as $\times$ in the loop), while the pump is modulating this flux at twice the frequency of the signal to be amplified. Since the signal in and out ports are the same, a circulator is needed to operate the amplifier (see Fig.~\ref{fig:Parallel_serial_JPA_Leo1}(a)).}
\end{center}
\end{figure}

We utilize flux-driven Josephson parametric amplifiers~(JPAs)~\cite{Yamamoto08} as quantum noise-limited amplifiers for the faint signals from our haloscopes.
The equivalent circuit of our JPAs includes a coplanar-waveguide $\lambda$/4 resonator that is grounded through a nonhysteretic superconducting quantum interference device (SQUID) (Fig.~\ref{fig:FDJPA1}). This resonator is coupled to the external transmission line through a capacitor $C_c$. The signal input and output share the same port. The JPA operate by modulating the SQUID inductance through an external pump signal.
A SQUID acts as a non-linear inductance whose value can be modulated by sending a pump signal through a line inductively coupled to it.
The operating frequency can be tuned in a certain range  by adjusting the average inductance of the SQUID by applying DC flux using a superconducting coil. 

\begin{figure}[ht]
    \centering
    \subfigure[]{
        \includegraphics[width=.3\textwidth]{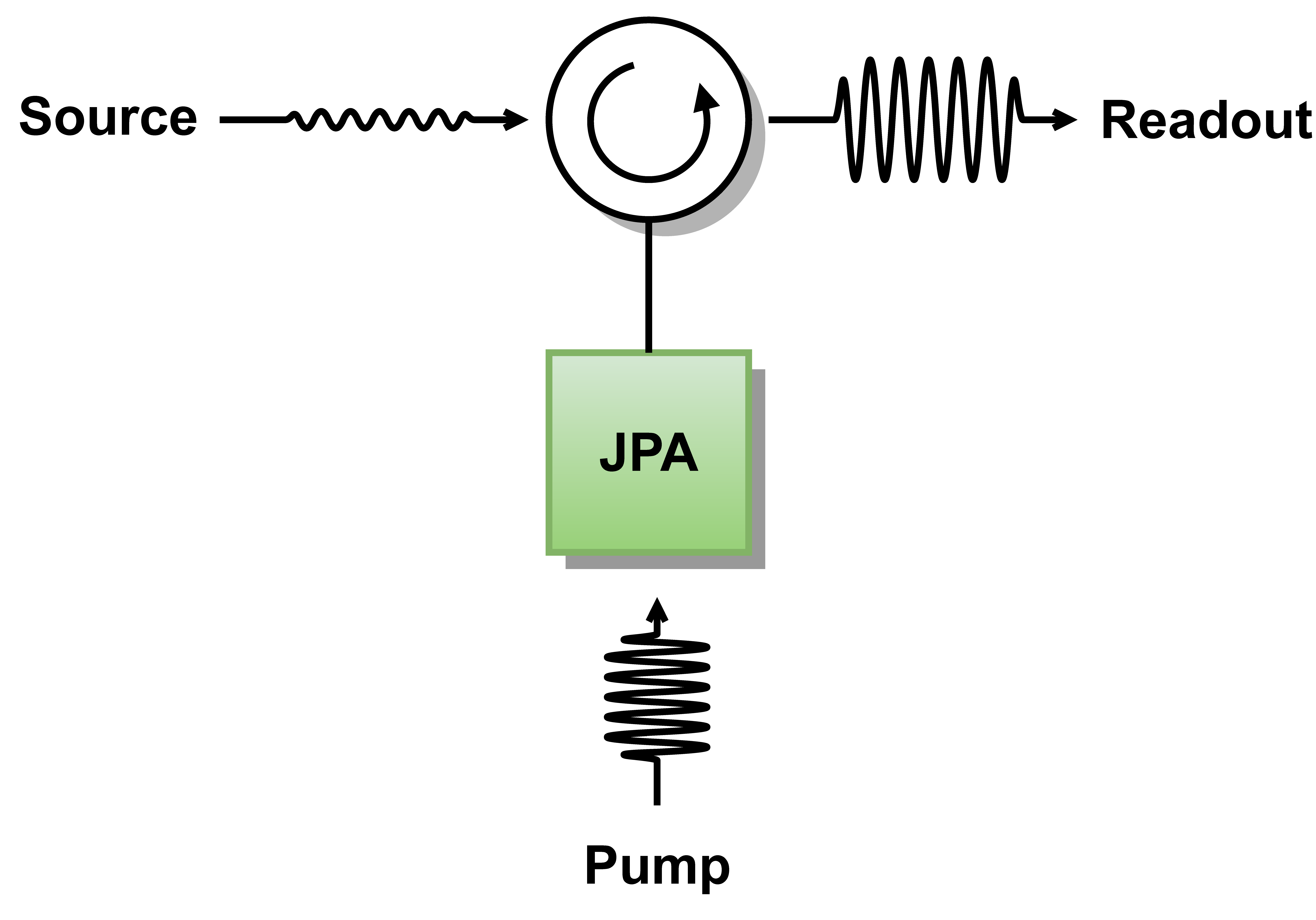}
     }
    \subfigure[]{
        \includegraphics[width=.48\textwidth]{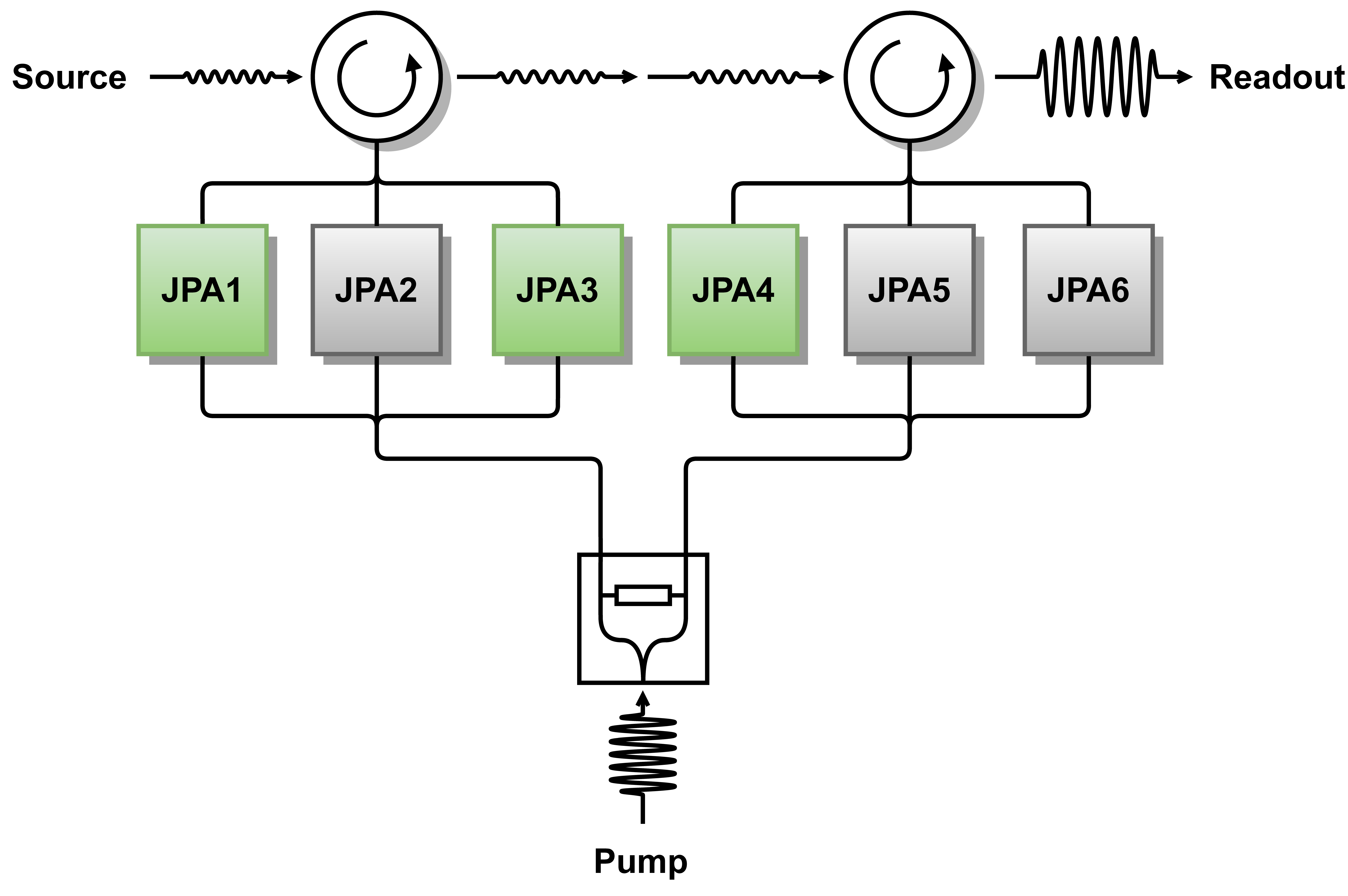}
    }
    \caption{\label{fig:Parallel_serial_JPA_Leo1} Simplified diagrams of the JPA connection in the readout chain. (a) Single-JPA configuration used in Runs 4 and 5. (b) JPA assembly used in Run 6 with two holders connected in series. Each holder has the capacity to accommodate up to 3 JPAs. To minimize any potential interference between them, two of them were positioned in Holder 1 and one in Holder 2. The green color highlights the actual placement of JPAs.}
 \end{figure}

Our JPAs operate in the three-wave mixing mode with the pump ($f_p$), signal ($f_s$), and idler ($f_i$) frequencies satisfying the relation~\cite{Roy2016} 

\begin{align}
    f_p = f_s+f_i.
    \label{eq:three_wave_mixing}
\end{align}

The JPA amplifies the signal injected from a source port and reflects it back.
The source can be either noise generated by a wideband 50\,$\Omega$ terminator or the cavity.
The signal paths for both cases are shown in Fig.~\ref{fig:Parallel_serial_JPA_Leo1}(a).
A circulator is employed to separate the incoming and outgoing signals.
The subsequent amplification is achieved by two HEMT amplifiers mounted at the 4\,K stage of the fridge.
Further signal processing is performed by a room-temperature amplifier and Spectrum Analyzer~(SA) or a custom DAQ system.

\begin{figure}[t]
    \begin{center}
    \includegraphics[width=.45\textwidth]{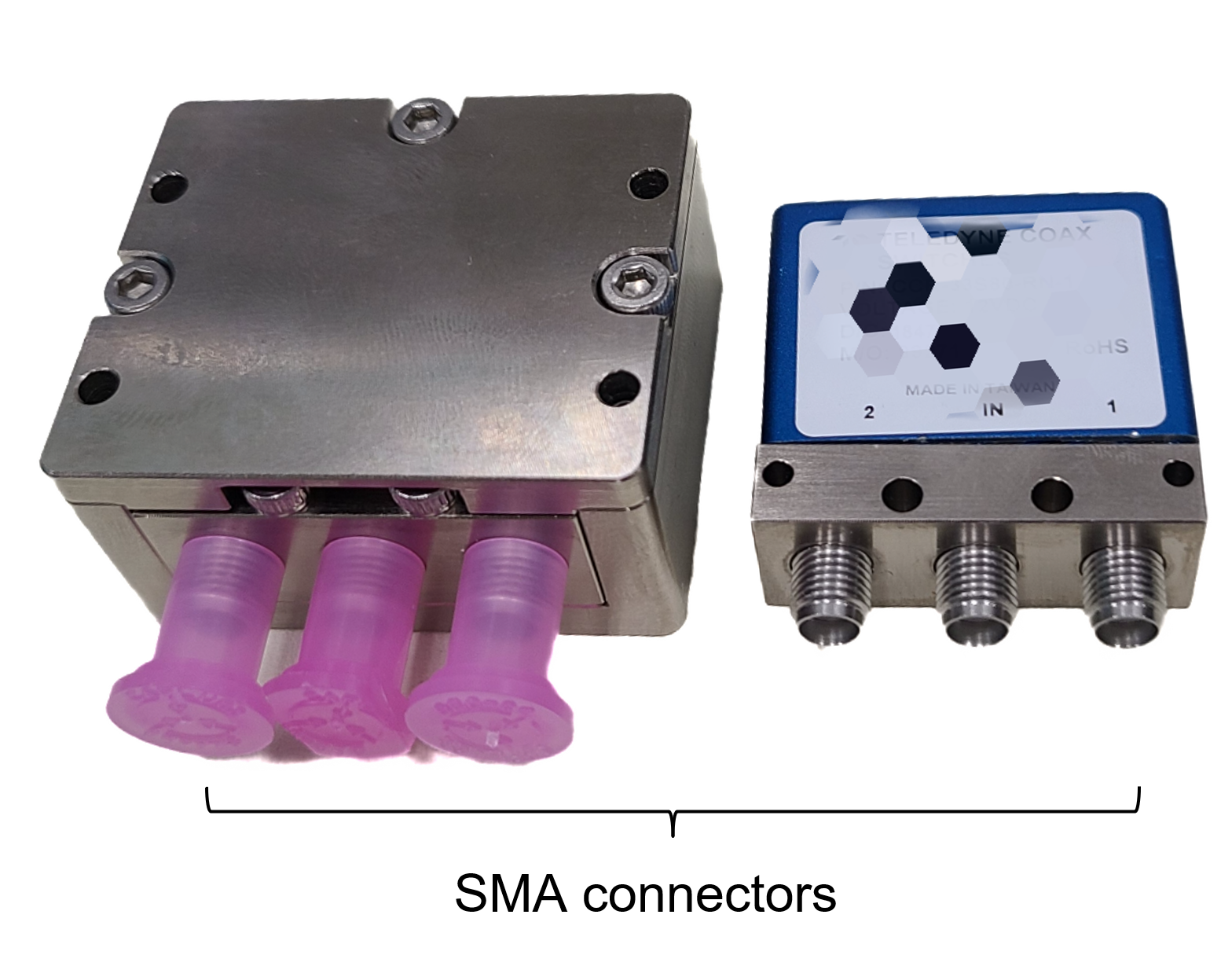}
    \caption{\label{fig:switch_shield} RF switches with (left) and without (right) a magnetic shield.}
    \end{center}
\end{figure}
%\paragraph{Operation of RF chain in the proximity of a strong magnet}

In our experimental setup, we encounter the challenge of potential magnetic-field effects on various components, including circulators, isolators, HEMT amplifiers, and most importantly, the JPA.
The magnet compensation region, which minimizes the impact of magnetic fields, is predetermined by the magnet manufacturer and may not accommodate all the components in our setup due to space limitations.

To overcome this challenge, we employ shielded components, such as circulators and isolators, and have also developed custom shields (see Fig.~\ref{fig:switch_shield}) dedicated to safeguarding sensitive components.

\begin{figure*}[ht]
    \begin{center}
    \includegraphics[angle=0, width=\textwidth]{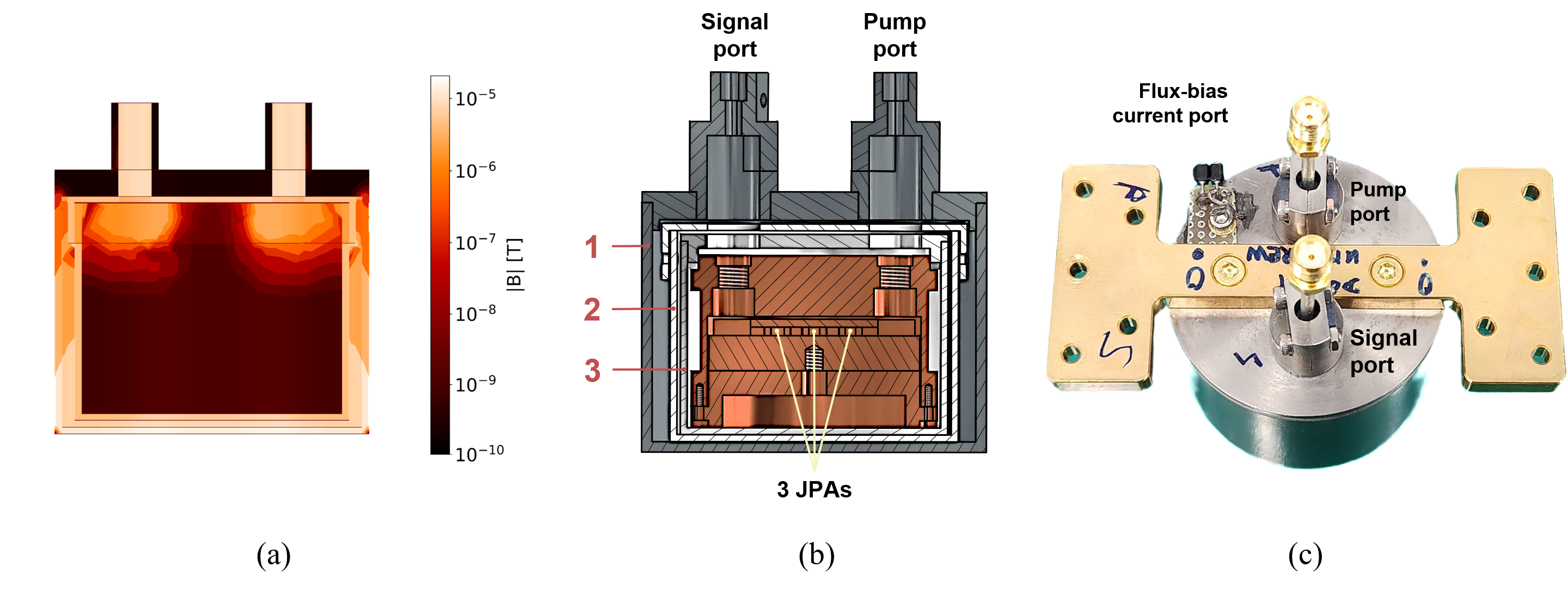}
    \caption{\label{fig:Gemini_holder} {Three-layer magnetic shield for a JPA.} (a)  Simulated magnetic field map inside the shield under an external background of 1\,T. (b) Drawing of the cross-sectional view of the shield:  Each layer, from the outside to the inside, is made of NbTi (1), cryoperm (2), and Al (3), respectively. (c) Photograph of the assembly with the SMA connectors  shown.}
    \end{center}
\end{figure*}

To protect the JPAs from strong magnetic fields, we developed a multilayer nested shield~\cite{Uchaikin23-LT29}.
The shield comprises three components, as shown in Fig.~\ref{fig:Gemini_holder}.
The outermost layer is crafted from a superconducting NbTi alloy with a high critical magnetic field.
The second layer is made of cryoperm alloy.
Its purpose is to effectively redirect and absorb external magnetic fields, significantly diminishing the overall magnetic field intensity.
Finally, the innermost layer is composed of a superconducting Al alloy, adding an extra layer of protection against magnetic field penetration.

In the initial conditions at non-cryogenic temperatures, the superconducting shields exhibit behavior akin to nonmagnetic materials. The reduction of the initial magnetic field is accomplished  by the second shield layer.
During the cooling process, the first layer, constructed from a superconducting material with a high critical magnetic field, undergoes a transition from its normal state to the superconducting state. This transformation enables it to effectively trap the initial magnetic field within the shield.
Meanwhile, the magnetic field within the second shield remains largely unaffected. As the temperature approaches the critical point for the innermost shield layer, it freezes the remaining magnetic field from the second shield.

In a detailed simulation using the ANSYS Electronics Desktop~\cite{ANSYS}, it was found that the magnetic field within the shield dropped to below 100\,nT when exposed to an external field ranging from 0 to 0.1\,T, as shown in Fig.~\ref{fig:Gemini_holder}.
The simulation is confirmed by a measurement at 4\,K~\cite{Uchaikin23-LT29}.

\begin{figure}[ht]
    \begin{center}
    \includegraphics[angle=0, width=.48\textwidth]{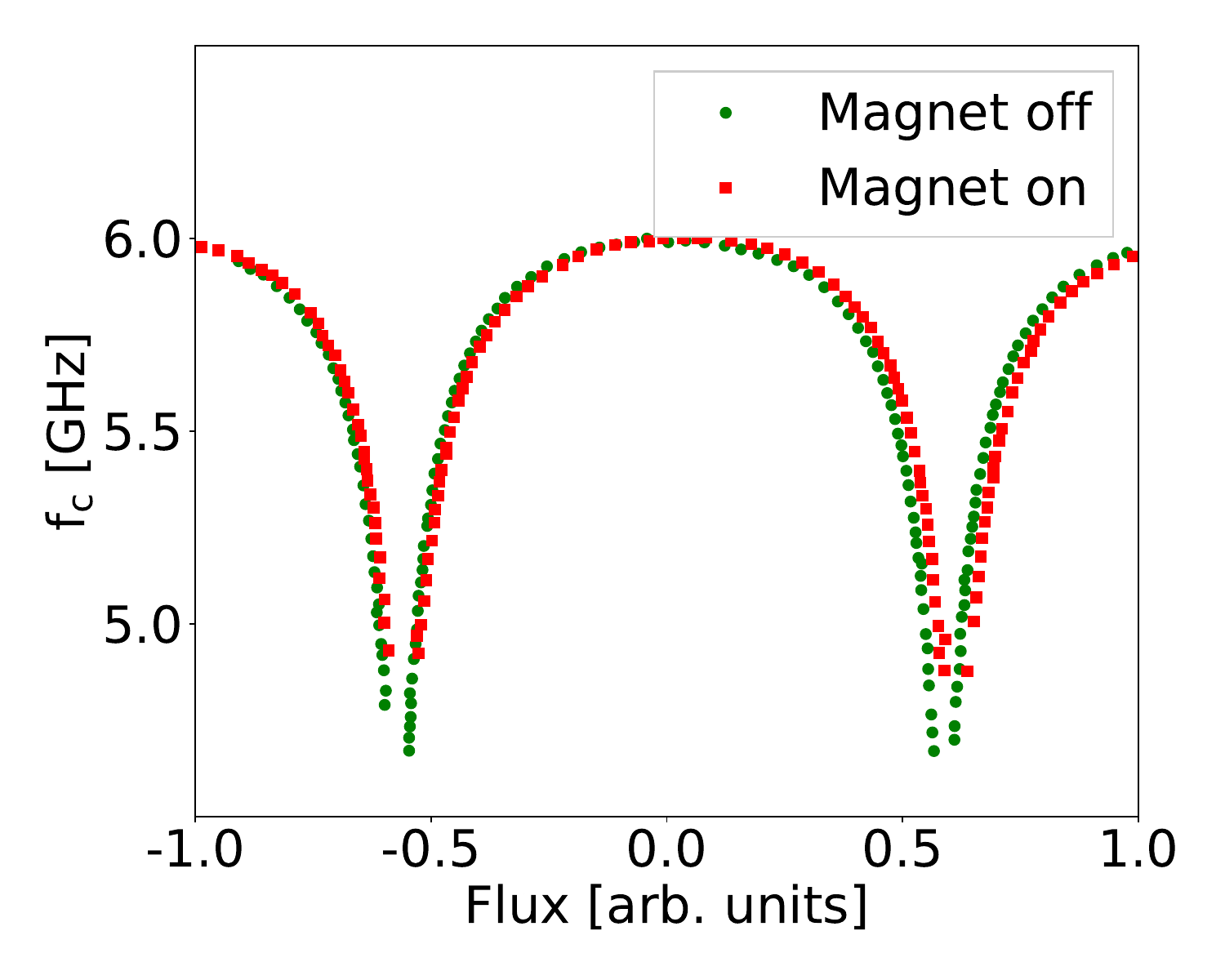}
    \caption{\label{fig:f_vs_bias_Magnet_off_and_on_3} Dependence of resonance frequency ($f_{c}$) in a passive mode of the JPA on applied DC flux bias, as measured with a separate device with the resonance frequency at 6\,GHz. The two colors correspond to the data taken in the presence and absence of the residual magnetic field of 0.1\,\si{\tesla}, which is ten times higher than expected during axion search experiment. 
    }
    \end{center}
\end{figure}

The magnet's residual field at the JPA position, without the shield, measures approximately 0.01\,T. With the implementation of this shield, the field strength at the JPA is effectively reduced, giving a small shift of the JPA DC bias  as shown in Fig.~\ref{fig:f_vs_bias_Magnet_off_and_on_3} for ten times as much (0.1\,T) external magnetic field.
The corresponding shielding factor is estimated between $10^{5}$ and $10^{6}$.

%\paragraph{Extension of JPA bandwidth}

While the flux-driven JPA offers extremely low noise close to the quantum noise limit, its original design had a limited tuning range of 45--60\,MHz~\cite{Uchaikin23-LT29}.
In Runs 4 and 5 we successfully used one amplifier to scan 20 and 58\,MHz, respectively~\cite{12TB-PRL}. 
This presented a challenge for CAPP-MAX as it required frequent warm-up and replacement, resulting in significant time loss and the consumption of large amounts of LHe. 
Our cavity possesses a tunable range of 200--300\,MHz, and the scanning range of each Run was determined by the tuning capabilities of the JPA. Therefore, our primary goal was to extend the amplifier's range without compromising the exceptional noise properties of our JPAs.
To achieve this, we employed two methods, which we recently reported: serial and parallel connections of the JPAs~\cite{Uchaikin-LTD20}.

\begin{figure}[ht]
    \begin{center}
    \includegraphics[width=\linewidth]{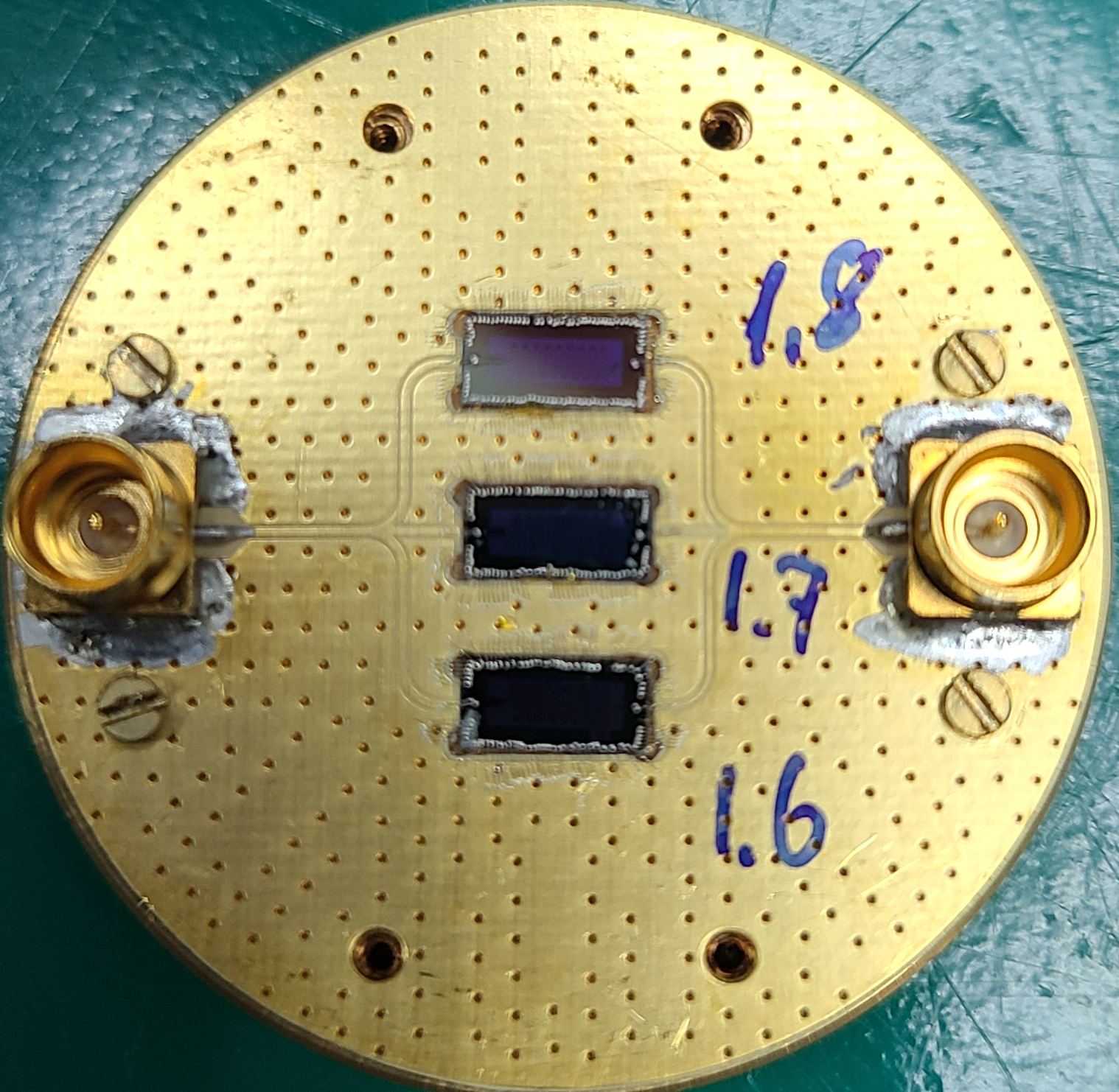}
    \caption{\label{fig:Gorynych3} Top view of the gold-plated PCB with three JPA chips bonded. The pump and signal lines are connected through  SMP connectors.}
    \end{center}
\end{figure}

During Run 6, we used a combination of parallel and serial connections, using Holder 1 with two JPAs and Holder 2 with one JPA; see Fig.~\ref{fig:Parallel_serial_JPA_Leo1}(b).
Holder 1 encompasses JPAs designed for frequency ranges of approximately 1.080--1.122\,GHz and 1.135--1.183\,GHz, while Holder 2 houses a JPA with a tunable range spanning 1.045--1.060\,GHz. 
To accommodate three JPAs within a single holder, we designed a specialized printed circuit board (PCB), as illustrated in Fig.~\ref{fig:Gorynych3}.
During operation, only a dedicated JPA is activated, with the other two being disabled by adjusting the resonance frequency.
Experimental results demonstrate that all three JPAs operate effectively without compromising their noise characteristics.
There was an overlap in frequencies between Runs~5 and~6.
To enhance sensitivity, we combined the shared data from these runs.

Our setup has the capacity to incorporate up to 6 JPAs through a combination of both parallel and serial connections.
For our upcoming Run 7, we are using 6\,JPAs covering a total range of approximately 300\,MHz from 1.2--1.5\,GHz~\cite{Uchaikin-LTD20}.
%\paragraph{Readout electronics diagram}

\begin{figure*}[ht]
    \begin{center}
    \includegraphics[width=1\textwidth]{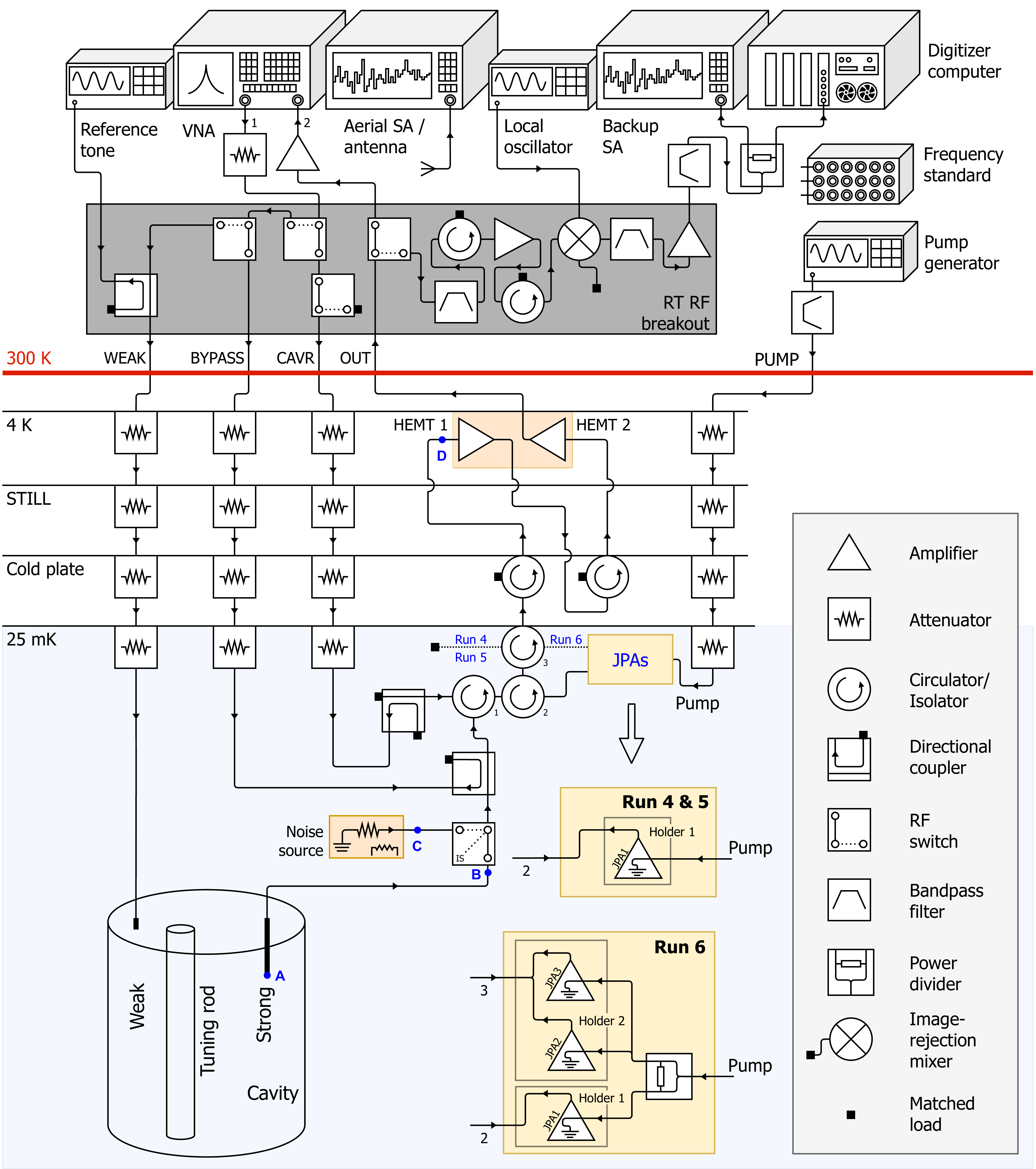}
    \caption{\label{fig:meas_diagram} Experimental setup including the cavity, a cryogenic receiver, measurement chains and the room temperature DAQ system. 
    The major components of the receiver chain are JPAs, a three-junction circulator/isolator, and a series of HEMT amplifiers (HEMT1 and HEMT2). The chain has two JPA configurations for two data taking runs, Runs~4 and~5, and Run~6. The cavity properties are measured through several lines: weak cavity coupler (WEAK), bypass cavity test (BYPASS), and cavity reflection (CAVR). The DAQ system consists of a radio-frequency signal generator (RF generator), a VNA, a SA, a radio-frequency local oscillator (LO), an image-rejection mixer (IRM), and a digitizer. The noise source is used in the equivalent-noise-temperature calibration of  elements in the RF-measurement chain. Transmission coefficients between points A--B and C--D are defined as $\eta_u$, $\eta_c$, respectively, see text.}
    \end{center}
\end{figure*}

The readout electronics diagram is depicted in Fig.~\ref{fig:meas_diagram}, providing an overview of the experimental setup used for all characterization measurements and the running of the experiment.
The diagram displays various temperature stages, beginning from room temperature (300\,K), 4\,K, 1\,K, cold plate, and mixing chamber (MXC) plate. Each temperature stage includes an attenuator for each RF line to prevent Johnson noise due to higher temperature stages from affecting the colder stages.

The setup is not only designed to conduct experiments but is also suitable for characterizing our JPA. In terms of noise characterization, our setup incorporates a noise source that enables the measurement of noise temperature for both the JPA and the HEMT.

Facilitating noise checks, cavity tests, and experiments, a cold RF switch is integrated within the setup. This RF switch enables seamless switching between the noise source and the cavity. Additionally, the RF switch features an intermediate state that allows for the measurement of line losses between the cavity and the switch, as detailed in reference~\cite{12TB-PRL}.

The JPA signal is amplified using low-noise cold HEMT amplifiers. Two HEMT amplifiers LNF-LNC0.6\textunderscore 2A~\cite{LNF-LNC0.6_2A} are connected in series and thermally stabilized at the 4\,K stage of the refrigerator.
To mitigate the adverse effects of signal reflections from the amplifier, an isolator is placed between the JPA assembly and the HEMT.
This isolator is thermally anchored to the fridge Cold plate.
Temperature sensors attached to the amplifiers are used to monitor the enclosure temperature, which typically reads 5\,K and 6\,K during operation at optimal voltage and current biases while the fridge is operational.

%\paragraph{Cryogenic noise source}

\begin{figure}[ht]
\begin{center}
\includegraphics[width=.48\textwidth]{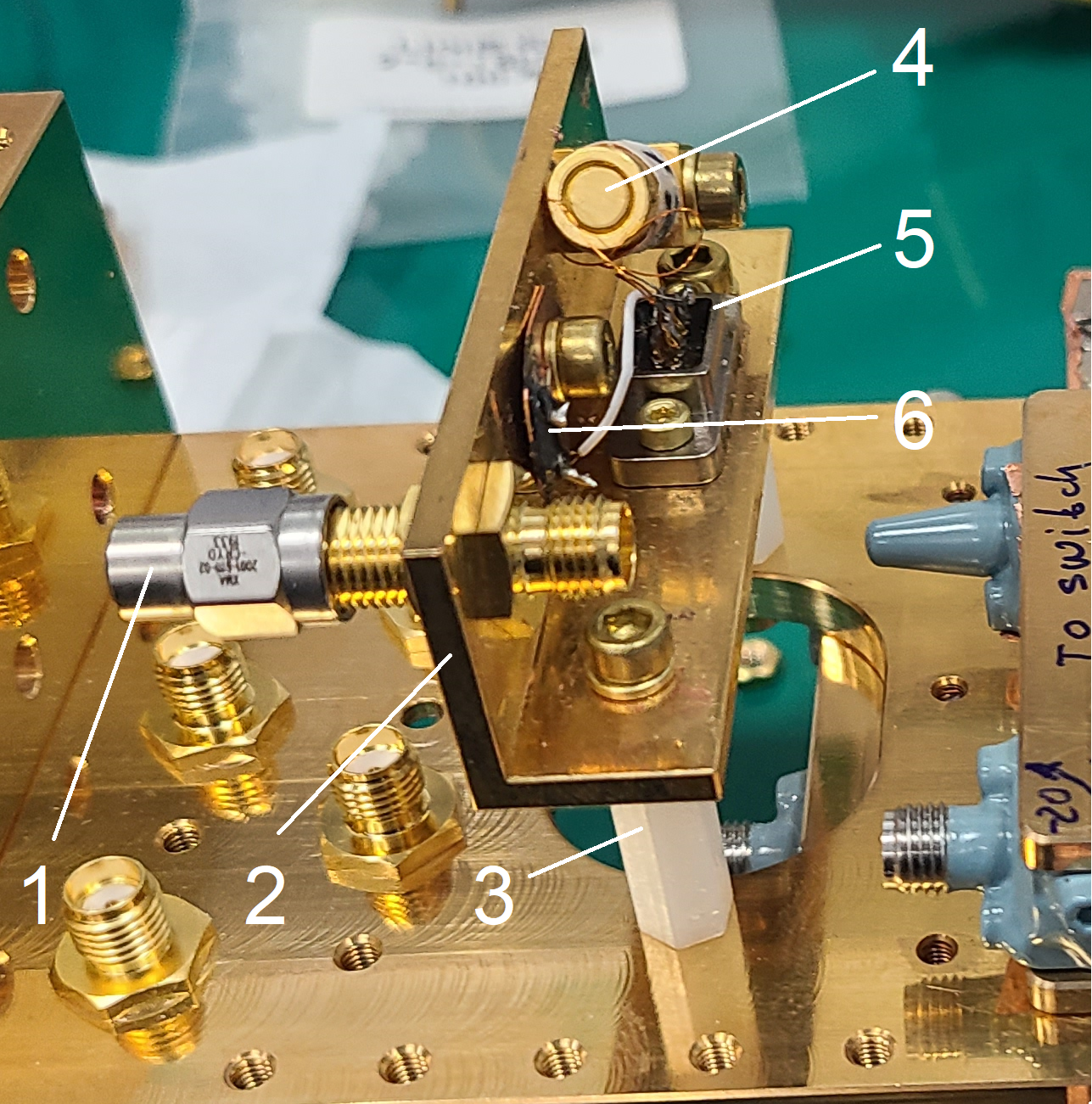}
\caption{\label{fig:NoiseSource}RF-noise-source assembly: (1) 50\,$\Omega$ terminator connected to SMA bulkhead, (2) enclosure, (3) thermally isolating plastic stand, (4) thermometer, (5) DC connector, and 6~-~heater.  The assembly is weakly thermally anchored to the MXC by a thin copper wire~(not shown).}
%Thermalization wire connected enclosure with the rest of the assembly is not shown.
\end{center}
\end{figure}

To measure the noise temperature of the HEMT amplifiers, we employed a methodology similar to the widely used Y-factor method~\cite{Engen70}.
We use a single-channel noise source (NS). The NS consists of a wideband matched 50~$\Omega$ terminator, a heater, a thermometer, and 9-pin Micro-D connectors for the thermometer and heater.
These components are assembled on a gold-plated copper holder (see Fig.~\ref{fig:NoiseSource}) and are specifically designed to be installed on the mixing-chamber RF assembly. 
In our test fridge, we utilize a four-channel version of the NS described in~\cite{Ivanov23-LT29}.

The NS is connected to the measurement chain using a 0.047-inch-diameter NbTi microwave coaxial cable, which is superconducting below 11\,K and provides excellent thermal isolation between the NS and other circuits.
Superconducting NbTi twisted pair cables with negligible thermal conductivity are used to connect the temperature sensor and heater. The temperature sensor is a calibrated RuO$_2$ sensor produced by Lake Shore Cryotronics, and the heater is a metal-film 100~$\Omega$ resistance that maintains stability within 1\% over a wide temperature range, from room temperature to the base temperature of the MXC.

For precise temperature measurement and control, we utilize the 372~AC resistance bridge and temperature controller provided by Lake Shore Cryotronics.
The NS is securely fixed to the MXC assembly using two 20-mm-long plastic spacers, and thermal anchoring to the MXC plate is achieved through a 15-cm-long copper wire (see Fig.~\ref{fig:NoiseSource}).

The \text{Bypass} and \text{Output} ports (see Fig.~\ref{fig:meas_diagram}) were used for direct measurements of JPA characteristics that effectively bypass the cavity.
To assess the noise level of the HEMT amplifier, we deactivate the JPA pump signal and adjust the DC flux bias current to shift the passive resonance of the JPA away from the measurement frequency.
In doing so, the JPA behaves as an almost ideal reflector, minimizing its influence on other circuits within the setup.

Subsequently, we proceed to establish and stabilize the temperature of the noise source at a specific value. We then measured the power spectral density~(PSD) within the desired frequency range. This process is repeated for 3--4 different temperature settings, allowing us to acquire a range of PSD values corresponding to various noise temperatures.

The measurement of JPA noise can be accomplished using the modified Y-method, as described earlier in this paragraph and in articles~\cite{article:Kutlu21, article:CAPP-PACE-JPA}.
However, this method has some drawbacks, including the lengthy setup and cooling time required for the noise source and the potential for saturation due to the limited dynamic range of the JPA.
To address these challenges, we employ a spectrum comparison method.
In this approach, we compare the SNR with the JPA turned off to the SNR with the JPA turned on.
This method provides a more time-efficient measurement technique to assess the noise characteristics of the JPA~\cite{article:ADMX-2}.
The result of the direct Y-factor method on the JPA for a limited temperature range is consistent with the SNR-comparison method. Our flux-driven JPAs demonstrate noise levels that are close to the quantum noise limit (QNL), as reported in~\cite{Uchaikin23-LT29}.

In our frequency range of 1--8\,GHz, the quantum noise limit ranges from 48 to 384\,mK, while the minimum added noise due to the amplifier increases from 24 to 192\,mK. This highlights the significance of cooling all the readout components of the initial stages to temperatures below these limits in order to achieve the lowest possible system noise temperature.

At temperatures below 100\,mK, certain effects and phenomena become more pronounced, which are usually hidden by the thermal motion of the particles at higher temperatures. Among these effects, acoustic mismatch~\cite{Kapitza41, Little59} and electron--phonon decoupling play a significant role. Also, phonon thermal conductivity decreases as a result of the reduction in the number and energy of phonons. Thus, it is advisable to avoid dielectrics during thermalization, even layers of oxides that form on metals. 

By employing metal interfaces and applying a gold coating between the mixing chamber (MXC) and the various parts and components installed on it, we successfully achieved notably low temperatures for the cavity, JPA, and cold RF components in our axion experiment, reaching 22--25\,mK. This approach led to the reduction of the total system noise, bringing it as low as 150\,mK.

\section{Data Acquisition and Monitoring}

\begin{figure*}[ht]
\begin{center}
\includegraphics[width=.95\textwidth]{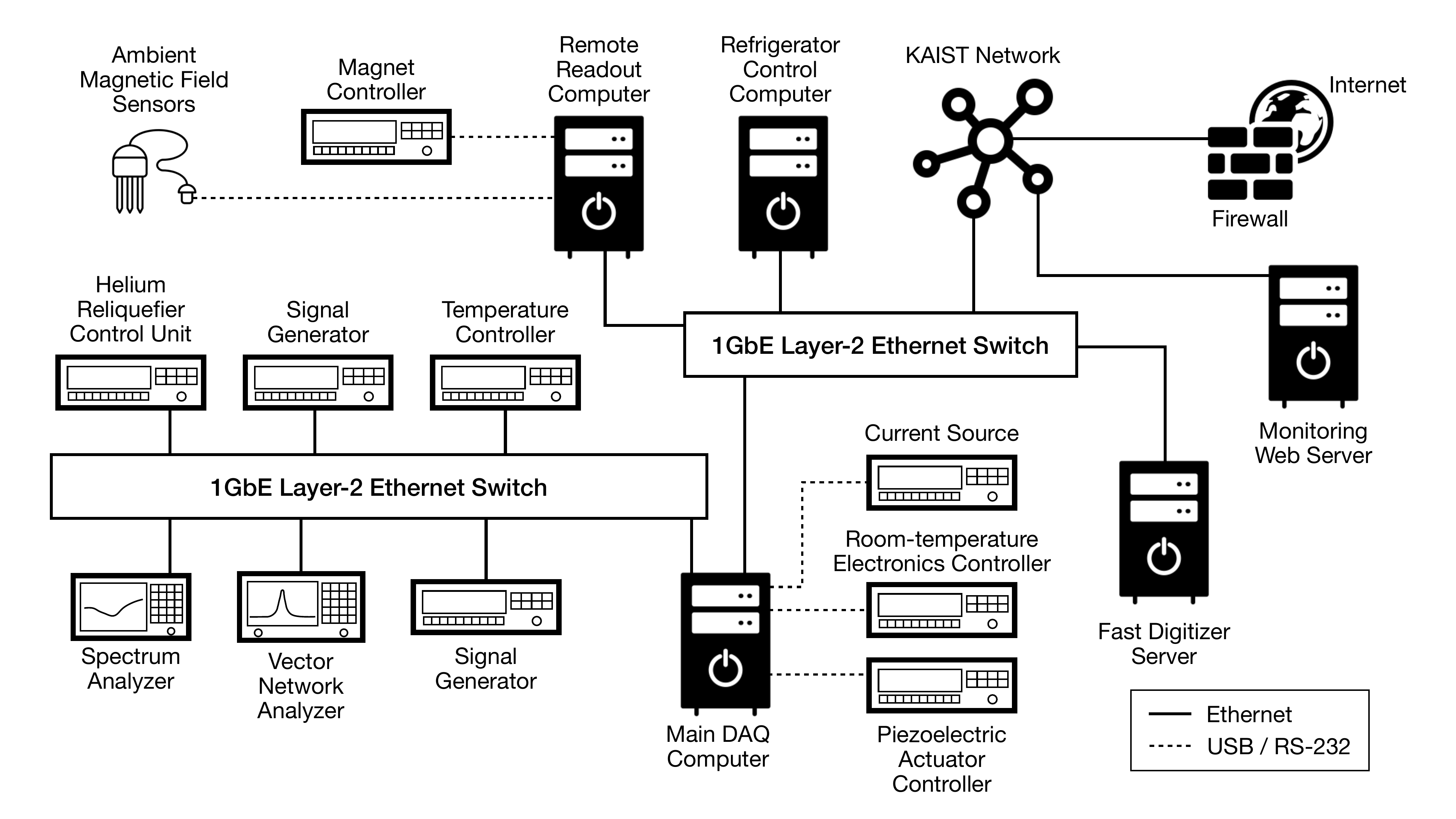}
\caption{\label{fig:daq_overview}Instrument connections for  data acquisition. Most  instruments are connected to the main DAQ computer via either Ethernet (solid line)  distributed by Ethernet switches  or  USB/RS-232 (dashed line) directly.}
\end{center}
\end{figure*}

\begin{figure*}[t]
\begin{center}
\includegraphics[width=.98\textwidth]{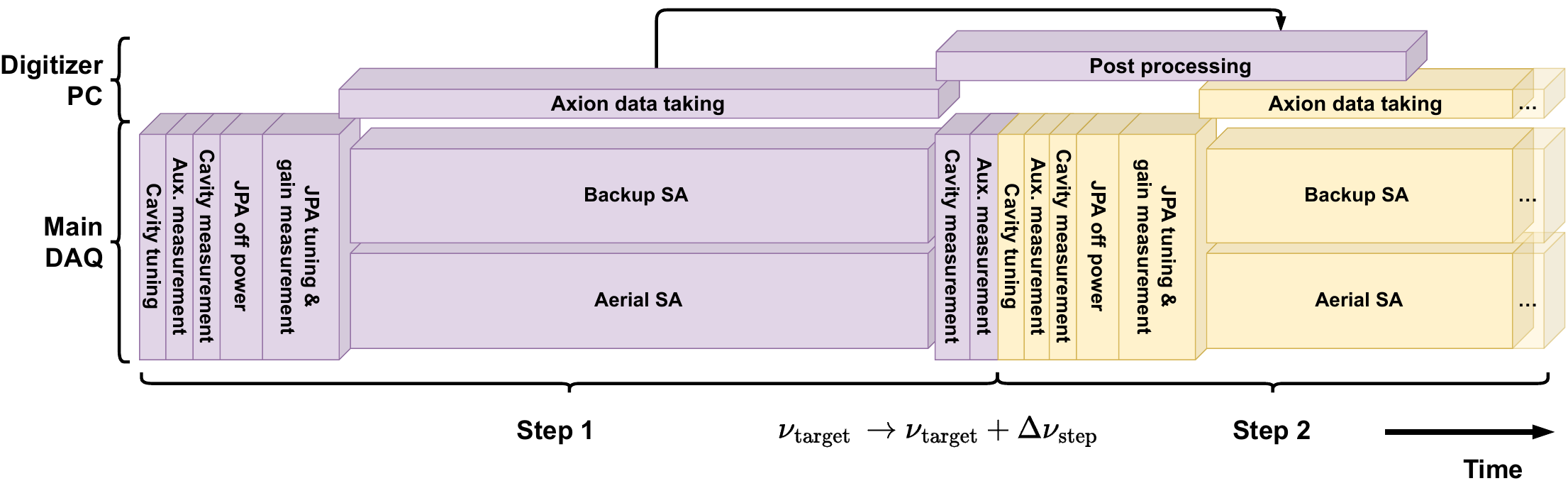}
\caption{\label{fig:measurement_sequence}  Measurement sequence during data taking. It includes physics data taking, cavity measurement, and system noise calibration. $\nu_{\textrm{target}}$ is the target frequency that the resonance frequency of the cavity and the JPA (with -100\,kHz offset) is tuned. $\Delta\nu_\textrm{target}$ is the frequency tuning step, which is kept 10\,kHz throughout the data collection. 
}
\end{center}
\end{figure*}

As shown in Fig.~\ref{fig:daq_overview}, most of the instruments are connected to the main DAQ computer over Ethernet within a local network. Others are directly connected to the main DAQ computer via the universal serial bus (USB) and the recommended standard 232 (RS-232).
An additional readout for the magnet information and measurements of the ambient magnetic field around the system is also connected over Ethernet. Finally, a dedicated computer for controlling a fast digitizer is connected over Ethernet as well. 

The DAQ is governed by a home-developed data acquisition software, CULDAQ~\cite{jphysconfser_898_032035_2017}.
The software provides features of controls over devices in the experiment via various communication protocols, data readout and packaging, storing monitoring data, and experiment run sequence. In the DAQ software, there are three layers of classes for the device controls: abstract classes to support communication protocols such as Ethernet, USB, RS-232, and so on; wrapper classes that translate the standard commands for programmable instruments or instructions provided by the manufacturers into user-friendly functions; classes providing higher level functions that utilize and combine a set of functions implemented in the wrapper classes to perform practical tasks conveniently. This hierarchy of classes keeps the software in a modular structure and provides a good re-usability and readability of the code.
The software also provides utility classes for the convenience that support database access, logging, and data packaging. 
The data acquired in the experiment are packaged in a ROOT format~\cite{ROOT}. 
The data read by a fast digitizer are also packaged as separate ROOT files, and the former ROOT file contains a map to match the experiment runs to the fast digitizer data. 
In addition, critical information for the experiment operation such as the total system physical temperatures, pressures, and magnetic fields are recorded in a database so that an online monitoring is available during the experiment.

When the components implemented in the software are combined, a logical sequence of the experiment can be formed. An experiment sequence requires a set of configuration parameters such as a frequency tuning range and step, a number of spectra to be averaged, parameters of instruments (center and span frequencies, resolution bandwidth, and so on), and they are given in the beginning of a measurement. A typical experiment sequence is as follows:

\begin{enumerate}
    \item Initializing the instruments by the given parameters.
    \item Tuning of the cavity resonant frequency to the initial target frequency.
    \item Measuring the auxiliary data such as the system physical temperatures, pressures, magnetic field, and so on.
    \item Tuning of the JPA at the current resonant frequency and a target gain.
    \item Characterizing the cavity by measuring the cavity reflection and transmission.
    \item Characterizing the receiver chain with turning on and off the JPA.
    \item Taking power spectrum data for about 192\,s with a fast digitizer (high efficiency) and a spectrum analyzer (for consistency checks).
    \item Characterizing the cavity again.
    \item Measuring the auxiliary data again.
    \item Tuning the cavity resonant frequency to the next target frequency, with a step of about 10\,kHz, and repeating the sequence from 3.
\end{enumerate}

The sequence diagram is depicted in Fig.~\ref{fig:measurement_sequence}.
Power spectrum data is taken using a fast digitizer. In addition, we acquire a small data sample simultaneously with a spectrum analyzer for consistency checking. We employ an M4i.4480-x8 manufactured by Spectrum Instrumentation GmbH~\cite{spectrum_instrumentation} as the fast digitizer. For the experiment, we implemented the fast digitizer with a sampling rate of 45\,MSamples/s and a resolution bandwidth of 10\,Hz, therefore, the acquired data is capable of searching not only for virialized axions, but also for non-virialized axions~\cite{Sagitarius}. The practical DAQ efficiency of the fast digitizer approaches 100\%~\cite{jinst_17_p05025_2022}.

\begin{figure*}[t]
\begin{center}
\includegraphics[width=.98\textwidth]{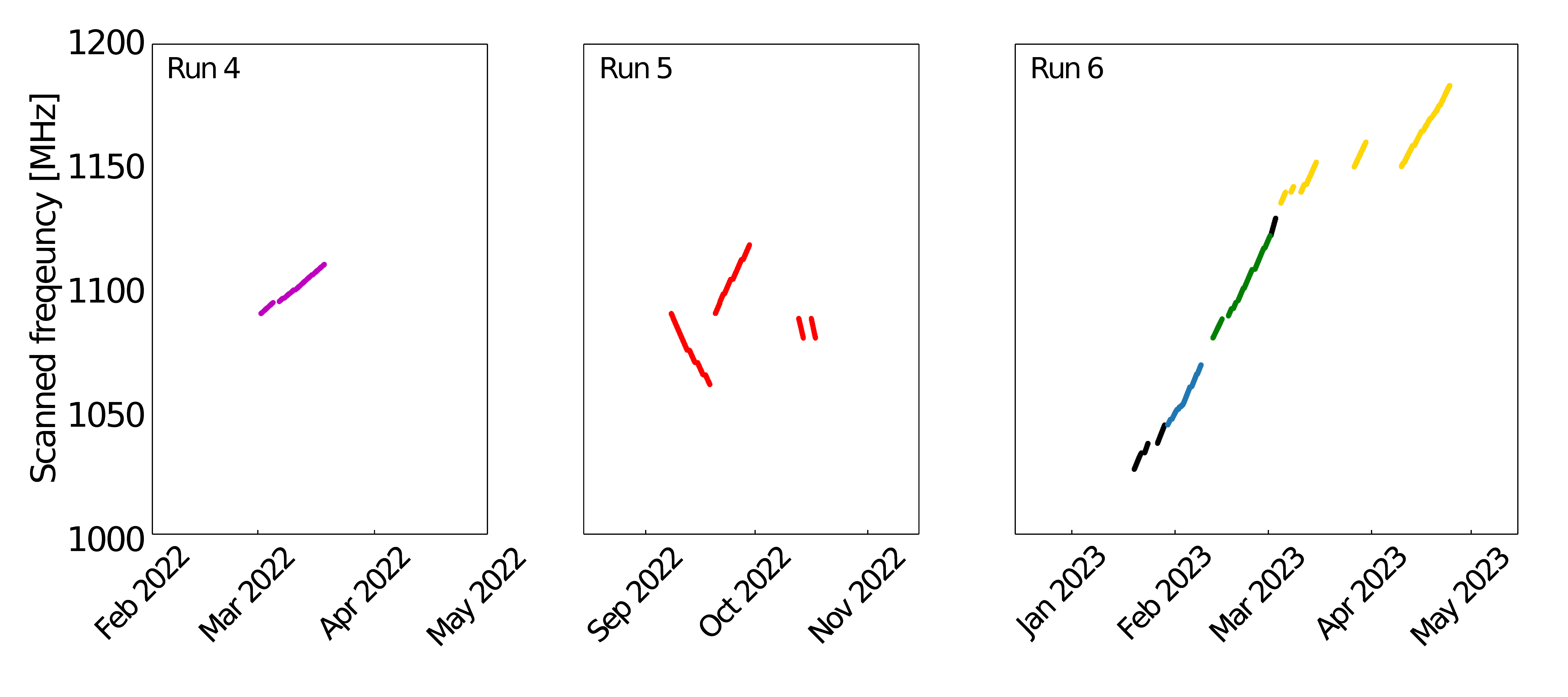}
\caption{\label{fig:acqusition_history} Scanned frequency as a function of calendar time for Runs~4,~5, and~6.  Different colors represent different JPAs, except black, which corresponds to the scan with no JPA operating in the RF-chain.
}
\end{center}
\end{figure*}

The total scanned frequency range within Runs~4,~5, and~6 with  different RF configurations in Fig.~\ref{fig:meas_diagram}, is shown in Fig.~\ref{fig:acqusition_history}. 
Runs 4 and 5 were taken with a single JPA while Run 6 utilizes 2 JPA holders with a total of 3 JPA chips. 

\section{Analysis and results \label{sec:analysls}}

\begin{figure*}[t]
    \begin{center}
    \includegraphics[width=\textwidth]{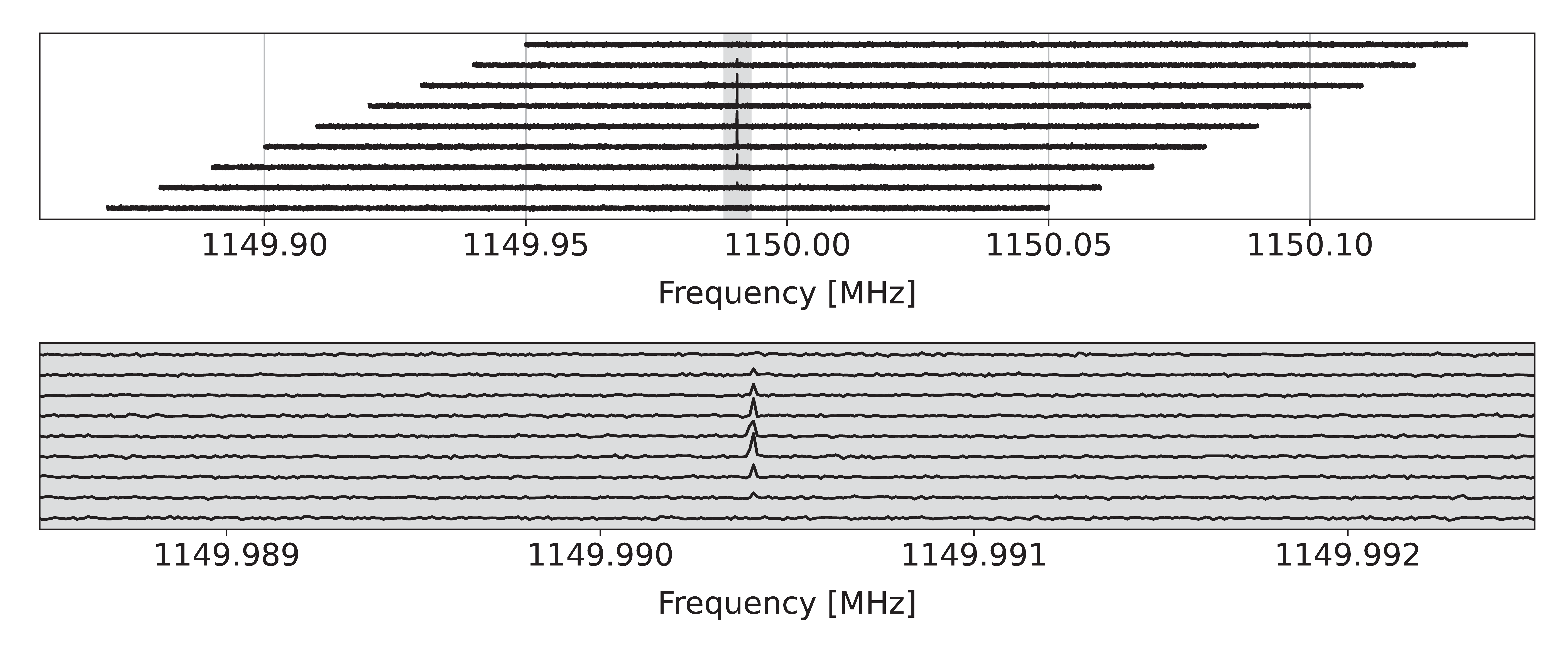}
    \caption{\label{fig:candidate_run6} Candidate signal with a large SNR. The signal strength follows the cavity response, indicating the signal goes through the cavity. The signal was finally rejected as it  was also picked up by the aerial antenna plugged into a spectrum analyzer  placed near the experiment to monitor ambient interferences. The frequency and width of the interference was at approximately 1149.9905\,MHz and 20\,Hz, respectively.
}
    \end{center}
\end{figure*}

To confirm the integrity of this data, a parallel system using a commercial spectrum analyzer, is also used in the data stream.
In addition, a second spectrum analyzer is used to collect data using an aerial antenna tuned to the frequency of interest in order to detect unwanted interference from outside~\cite{Fritz}.
Those interference signals are usually narrower than those expected from standard halo axion dynamics, but perhaps compatible with narrow axion lines that appear as transients due to the focusing effects of the sun, moon, planets, and the earth itself~\cite{article:CAST-CAPP,article:stream1,article:KZ1,article:KZ2,article:zioutas2017,Fischer2017}.
One such signal is shown in Fig.~\ref{fig:candidate_run6}.
The signal was intermittent and appeared simultaneously at the SA connected to the aerial antenna, indicating external interference.
It was most likely due to a radio communication with approaching aircraft into a nearby airport using the same frequency~\cite{FrequencyDistribution}.

Fig.~\ref{fig:candidate_run6} shows the main analysis strategy for signals with a large SNR ratio. The criteria for considering those signals as candidates are as follows:

\begin{enumerate}
    \item The signal amplitude needs to follow the cavity response, and
    \item The signal needs to be absent from the spectrum analyzer connected to the aerial antenna.
\end{enumerate}

If those conditions are met, we acquire more data for a long time, enough to establish the stability of the signal as a function of time.
If other modes than the $\rm TM_{010}$ are close to the candidate frequency, the cavity is tuned to that mode as well~\cite{article:ADMX-3}.
If all is consistent with the signal being an axion, the final check is testing its magnetic field dependence.
\begin{figure*}[t]
    \begin{center}
    \includegraphics[width=.98\textwidth]{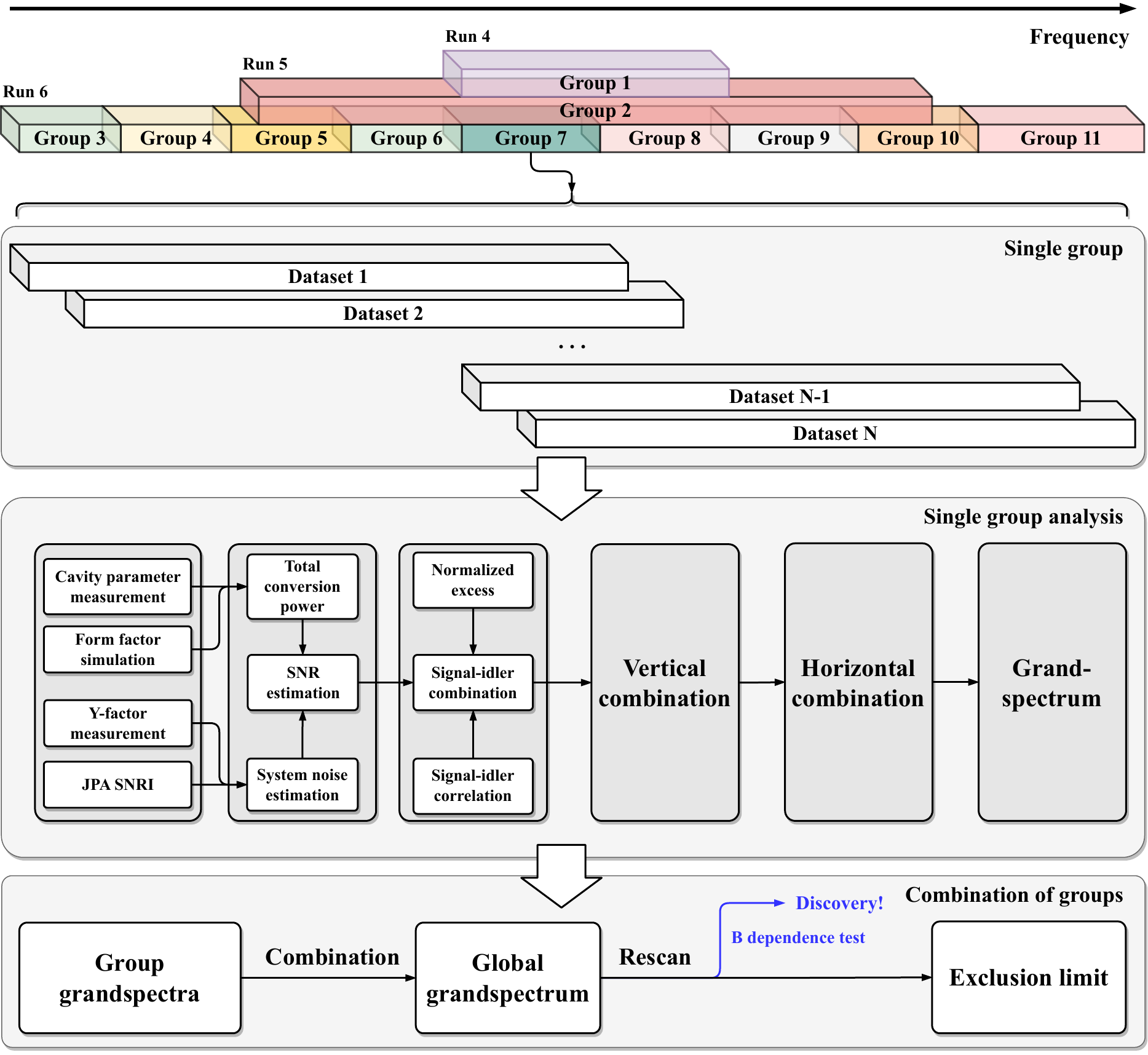}
    \caption{\label{fig:analysis_overview} Overview of the analysis procedures in three different runs (Runs 4, 5, and 6). The group refers to a bundle of the dataset within approximately the same statistical characteristics. Each group of data is processed according to the steps  outlined in the single-group analysis part. Subsequently, the resulting grandspectra are optimally combined in order to  construct the global grandspectrum.  Re-scan candidates are identified based on this grandspectrum for additional statistical testing.
    }
    \end{center}
\end{figure*}
The rest of the signals with small SNR require more steps to optimize signal detection. The overall data-analysis steps are shown in Fig.~\ref{fig:analysis_overview} and are described in detail below.
The flow of the analysis is similar to the procedures described in~\cite{12TB-PRL, article:Ahn-anal, article:ADMX-anal, article:HAYSTAC-anal, article:TASEH-anal}.
The goal is to construct grandspectra where the candidates for rescan can be classified or the exclusion limit can be set. 
The underlying probability distribution of the noise fluctuation is Gaussian, as a large number of spectra are averaged together. 
Unbiased combination procedures throughout the analysis guarantee that the final product of the analysis also yields a standard normal distribution, by properly normalizing the uncertainty and compensating for the mean offset.
The grandspectra from predefined subsets and their rescan datasets are aligned and combined using the Bayesian power measured (BPM) analysis framework~\cite{article:bayesian_method}.

The first step of the analysis involves the estimation of the SNR of axions with an arbitrary coupling strength $g_\gamma$ at each data point. 
This step plays two important roles: first, the SNR is used for estimating the experimental sensitivity on the axion--photon coupling $g_\gamma$. Next, it is used as a weighting factor for optimal combination of the data where axions possibly exist. 
As defined in Eq.~(\ref{eq:snr}), it requires the axion-to-photon conversion power and the total system noise temperature at each frequency.

The conversion power depends on experimental parameters such as cavity bandwidth ($\Delta \omega_{c}$), resonance frequency ($\omega_c$), loaded quality factor ($Q = \omega_c/\Delta \omega_c$) of the cavity, coupling coefficient of the strong antenna ($\beta$), as well as the root mean square of the external magnetic flux ($B_{\textrm{rms}}$) inside the cavity volume ($V$) and form factor of the TM$_{010}$ mode ($C$). The loaded quality factor, the resonance frequency of the cavity, and the antenna coupling are obtained using the real and imaginary part of the complex S-parameter through the reflection line. $B_{\textrm{rms}}$, $V$ and $C$ are obtained from simulation using COMSOL Multiphysics~\cite{COMSOL}, together with the magnetic field profile, which is a simulation result provided by the manufacturer. 

The system noise temperature when the JPA is off (the pump is off and the resonant frequency of the JPA is tuned far away from the cavity frequency) is estimated using the Y-factor method.
The noise power spectrum is modeled as
\begin{equation}
    P_{\textrm{off}} = k_B \Delta \nu_b G_\textrm{off}(T_\textrm{NS}^{\textrm{eff}} + T_{\textrm{off}}). 
    \label{eq:jpaoff_power}
\end{equation}
$P_{\textrm{off}}$ is the noise power when the JPA is off, $\Delta \nu_b$ is the resolution bandwidth and $T_{\textrm{NS}}^{\textrm{eff}}$ is the thermal noise from the matched load of the noise source.
$T_{\textrm{off}} = T_{\textrm{down}}/\eta_{u}$ and $G_{\textrm{off}} = G_{\textrm{down}}\eta_{u}$ are the system noise and gain correspondingly when the JPA is off. $\eta_{u}$ is defined as the transmission coefficient between the noise source and the first HEMT amplifier (between C and D in Fig.~\ref{fig:meas_diagram}).
$T_{\textrm{down}}$ and $G_{\textrm{down}}$ are the noise and gain, respectively, contributed from the downstream part of the chain, corresponding to the first HEMT amplifier and then all the subsequent components until the signal is digitized. 
$P_{\textrm{off}}$ is measured at more than three different temperatures of the noise source, and then the $P_{\textrm{off}}$ versus $T_{\textrm{NS}}^{\textrm{eff}}$ graph is fitted using Eq.~(\ref{eq:jpaoff_power}) to estimate $T_{\textrm{off}}$ and $G_{\textrm{off}}$.
For a given noise-source temperature, $P_{\textrm{off}}$ spectra are measured with sweeping LO frequencies in order to keep an intermediate frequency (IF) of 10.7\,MHz for the digitizer.
In practice, it is done with 1\,MHz frequency step for a given scanning measurement range.
After the data acquisition is finished and a scanning of $\sim$10\,MHz is accomplished, the whole process is repeated in order to update the $T_{\textrm{off}}$ slowly drifting due to HEMT amplifiers' gain stability limitations. 

The system noise temperature when the JPA is on (the pump power is on and the resonant frequency of the JPA is near the cavity frequency) can be estimated using the SNR improvement method (SNRI)~\cite{article:ADMX-2}.
SNRI is defined as,
\begin{equation}
    \textrm{SNRI}
    = \bigg( \frac{G_{\textrm{on}}}{P_{\textrm{on}}} \bigg) \bigg(\frac{G_{\textrm{off}}}{P_{\textrm{off}}} \bigg)^{-1} = G_{J}/r.
    \label{eq:snri}
\end{equation}
$P_{\textrm{on,off}}$ and $G_{\textrm{on,off}}$ are noise-power and gain when the JPA is on and off, respectively. After reorganizing the equation, $r=P_{\textrm{on}}/P_{\textrm{off}}$ is defined as the ratio spectrum, and $G_J = G_{\textrm{on}}/G_{\textrm{off}}$ as the JPA gain, typically around 50. The JPA gain can be further expressed as $|S_{\textrm{21,on}}|^2/|S_{\textrm{21,off}}|^2$ where $S_{21}$ is the S-parameter measured through the bypass line. 

Using the SNRI and the reference noise temperatures, the total system noise temperature of the JPA can be expressed as,
\begin{align}
    T_\textrm{sys} 
    &= (T_\textrm{off} + T_\textrm{MXC}^\textrm{eff})\cdot \frac{r}{\eta_cG_J}.
    \label{eq:tsys_estimation}
\end{align}
$T_\textrm{MXC}^\textrm{eff}$ is the effective noise temperature of the attenuator  located at the JPA assembly thermally anchored to the mixing chamber. $\eta_c$ is the transmission coefficient of the line between the cavity and the cryogenic RF switch (between A and B in Fig.~\ref{fig:meas_diagram}). It is estimated to be equal to 0.932 $\pm$ 0.003 using the intermediate state (IS) of the RF switch shown in Fig.~\ref{fig:meas_diagram}, under the assumption that it is frequency-independent within the frequency range of the data taking~\cite{12TB-PRL}. The measurement results are found to be consistent throughout all the runs, consistent with the fact that the same RF cables and the strongly coupled antenna are used. As a result, the total system noise temperature around the resonance frequency of the cavity is found to be 170\,mK and 250\,mK on average in Runs 4 and 5, respectively. In Run 6, it varies from 200\,mK to 400\,mK with JPA, and from 1.5\,K to 3\,K without JPA. 

\begin{figure*}[ht]
\begin{center}
    \includegraphics[width=0.98\linewidth]{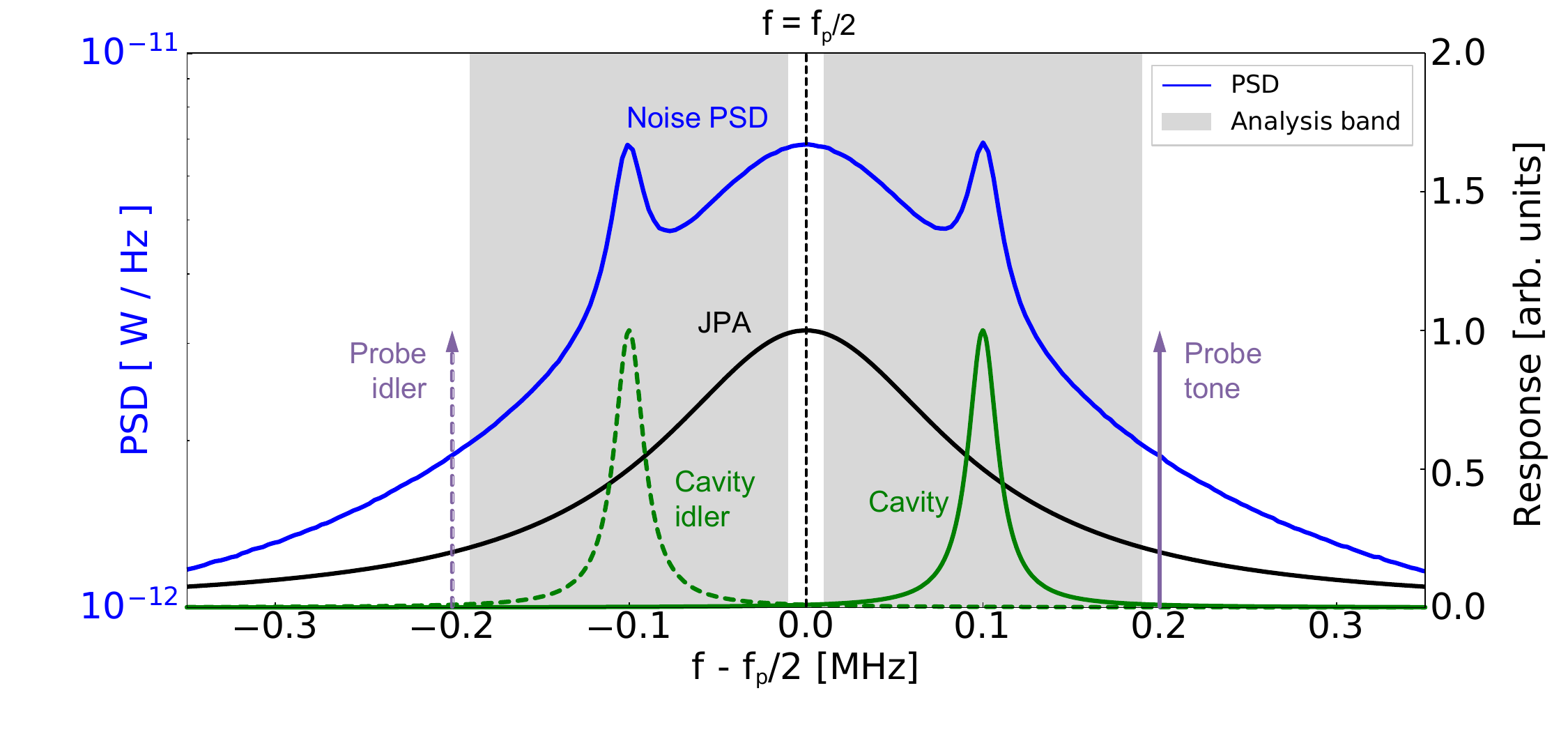}
    \caption{Example of the noise PSD (blue) and frequency responses of the JPA (black) and the cavity (green) as a function of the offset from the half pump frequency ($f_p/2)$. The total gain is $\sim$110\,dB and the frequency responses are normalized to their maximum value. The probe tone (purple) injected through the weak port of the cavity is used for cross-checking and systematic error estimation. The dashed lines represent the idler modes. Since the axion signal also appears on the idler side, the frequency bands (grey) on both sides are included in its analysis. 
    }
    \label{fig:data_example}
\end{center}
\end{figure*}

The baselines of the ratio spectra and the JPA gain are estimated to reduce the statistical uncertainty and define the normalized excess. 
For the ratio spectra, 5 neighboring data points are merged together. 
Of the 1\,MHz range, we utilize 360\,kHz of the data;  180\,kHz range of the data around the resonance frequency of the cavity and the same for the range around its idler frequency. The JPA essentially yields a highly-symmetric output around the half pump frequency, owing to its three-wave-mixing property as described in Eq.~(\ref{eq:three_wave_mixing})~\cite{article:Kutlu21}. 
Using this fact, both spectra are decomposed into their symmetrical and asymmetrical parts. Each part in each spectrum is fitted individually, then combined as the final baseline.

The baseline of $P_\textrm{off}$ is modeled and fitted with a fifth-order polynomial accommodating the frequency response of the circuit components after the frequency mixer. In doing so, the data close to the resonance frequency of the cavity are not included in the fit. It is assumed that the major contributions to the polynomial background are the thermal photons from the attenuators installed in the measurement lines as shown in Fig.~\ref{fig:meas_diagram}, which are responsible for the terms in the parenthesis of Eq.~(\ref{eq:tsys_estimation}).

The normalized power excess $\delta$ is defined using the ratio spectrum and its baseline.
\begin{gather}
    \delta  
    = \frac{r - b_{r}}{\delta r}
    = \bigg( \frac{r}{b_{r}}-1 \bigg)\cdot \sqrt{N_{\textrm{avg}}}
\end{gather}
$\delta {r} = b_{r}/\sqrt{N_\textrm{avg}}$ is the expected standard deviation of $r$ by the central limit theorem, where $N_\textrm{avg}$ is the number of averages, and $b_{r}$ is the baseline of the ratio spectrum. Thus $\delta$ is expected to follow the standard normal distribution, which leads the reduced $\chi^{2}$ goodness of the fit to have a mean of 1. Throughout the data, the normalized excesses fulfill these expectations within the margins of error.
\begin{figure*}[ht]
\begin{center}
\includegraphics[width=.98\textwidth]{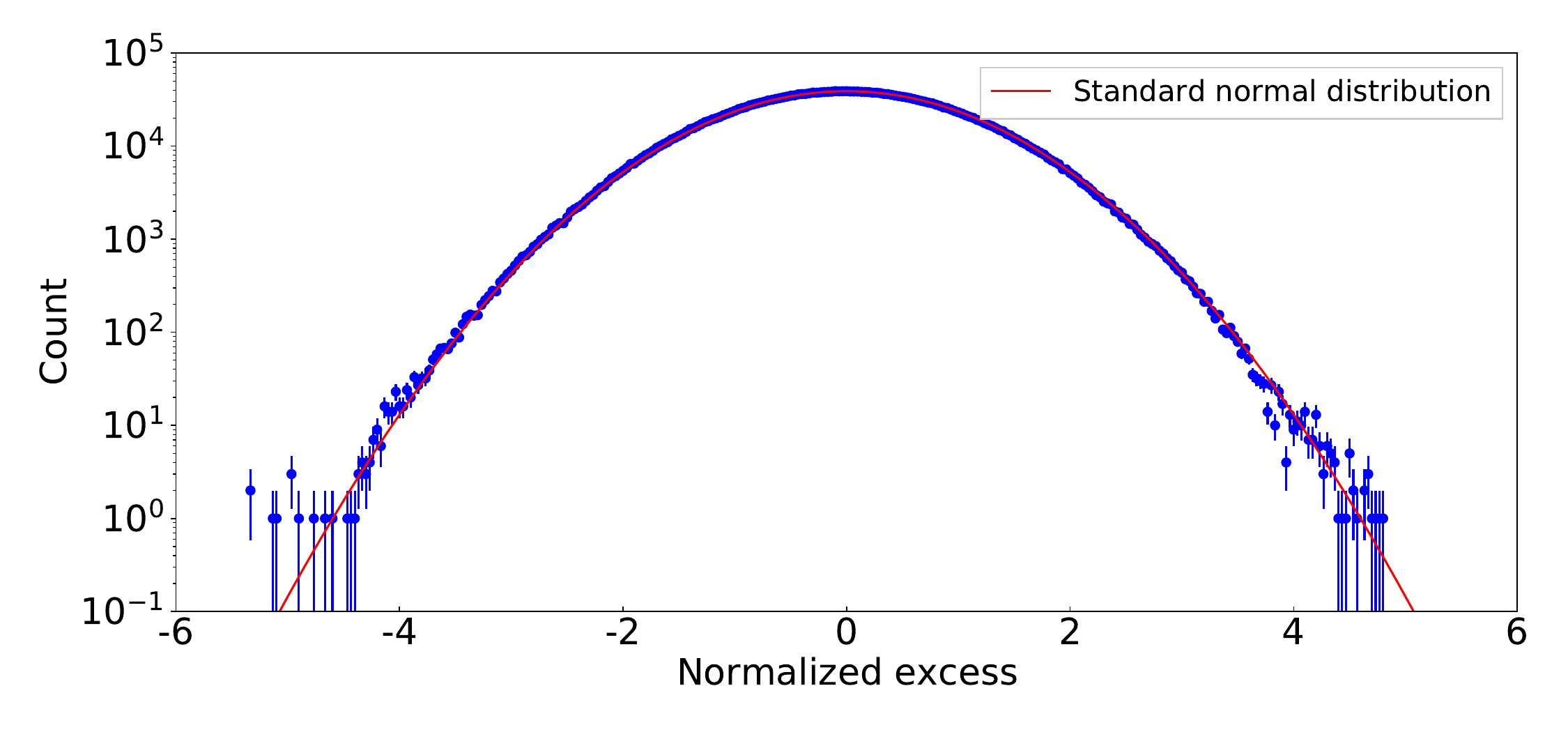}
\caption{\label{fig:Normal_Be} Statistical distribution of grandspectra excesses for Runs~5 and~6. The total number of entries is approximately 3.7 million. The negative excesses on the left are due to the distortion of the baseline  by minor mode crossings.
}
\end{center}
\end{figure*}

The combination process starts with the signal and idler sides of the data, shown as gray area in Fig.~\ref{fig:data_example}. The SNR in the signal ($R_{S}$) and the idler ($R_{I}$) sides can be defined using the system noise temperature on each side of the data. Equation~(\ref{eq:tsys_estimation}) is also used for the idler side, but with the JPA gain for the idler mode $G_{I} = G_{J} - 1$. The JPA-induced correlation between the two analysis bands is considered in optimizing the weights and the error combination. Then, the spectra are optimally weighted by their expected SNR to be summed vertically.

After the vertical combination, the SNR of the virialized axion, dispersed in a bandwidth of around 5\,kHz, is combined horizontally according to~\cite{article:Turner, article:HAYSTAC-anal}. Since the axion mass is unknown, the combination is repeated for each axion rest-mass frequency. The effect of a mismatch between the axion frequency and the closest frequency point is found to be negligible because the resolution bandwidth~(RBW) used is much smaller than the axion line shape. To account for the variation at different axion masses, the line shape is evaluated at every 1\,MHz.

The combined spectrum is normalized once more by dividing its standard deviation and subtracting its mean for the purpose of constructing a grandspectrum. 
In Run 6 the standard deviation is found to be frequency-dependent and is mostly larger than expected from simulation. For this reason, the data taken in Run 6 are divided into 9 different groups, where the standard deviation is approximately uniform within each group. 
The group grandspectra are obtained from their own means and standard deviations, then they are vertically combined once more to construct a single grandspectrum.
The rescan candidates are defined based on the 3.718$\sigma$ threshold given by 90\% confidence level for a target SNR of 5$\sigma$.
For the large narrow peaks, we follow the procedure outlined at the beginning of this section. 
Those peaks confirmed not to be due to axion conversion are removed together with their left and right single bins in the vertically combined spectra. 
For the rest of the candidates, data is taken for at least three times longer than the initial acquisition.  

\begin{figure*}[t]
    \begin{center}
    \includegraphics[width=\textwidth]{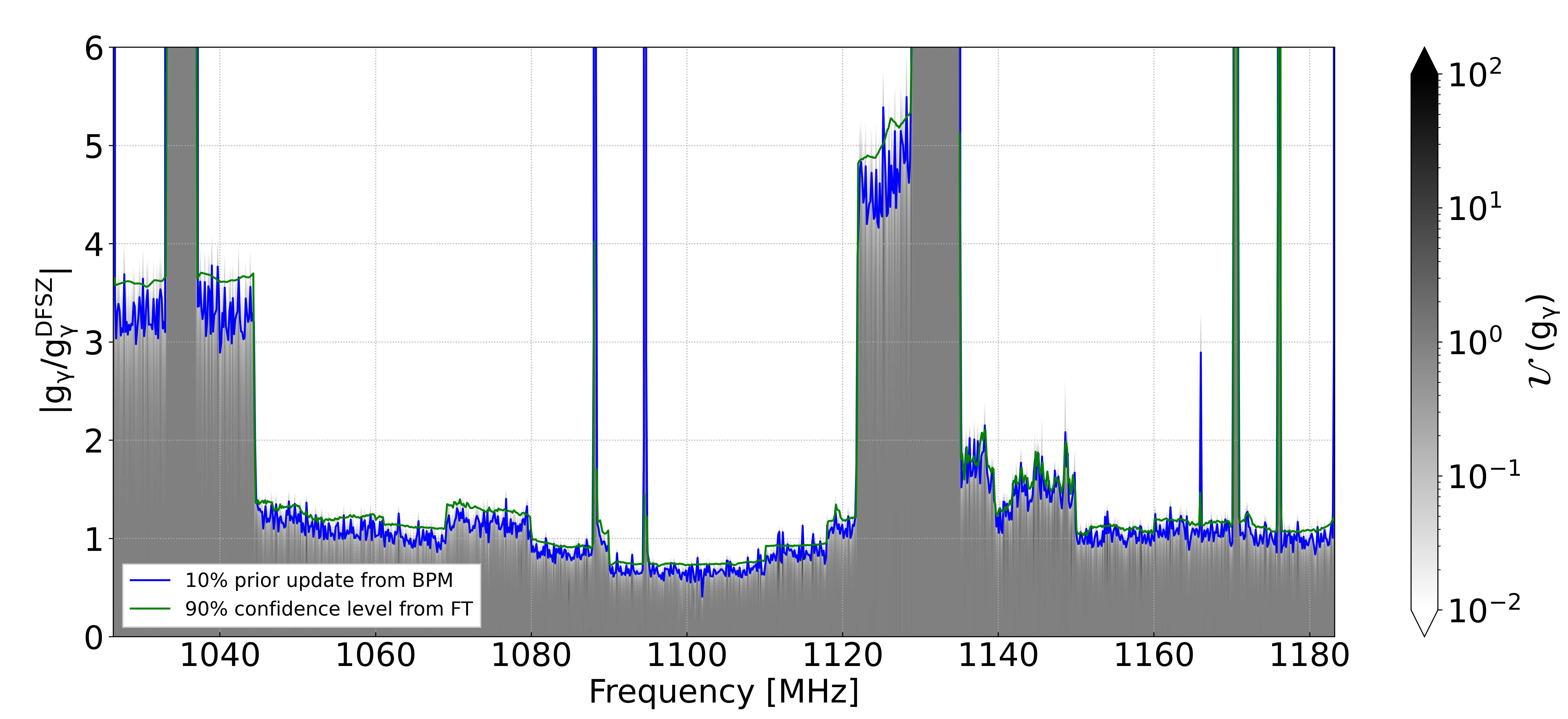}
    \caption{\label{fig:exclusion} 
    Exclusion limits for axion--photon coupling obtained through the Bayesian power measured (BPM) analysis (blue). The 10\% prior update contour of the BPM framework is equivalent to the standard 90\% exclusion line (green) achieved with the FT framework. The subaggregated prior update are scaled in grey.
    }
    \end{center}
\end{figure*}

\begin{figure*}
    \begin{center}
    \includegraphics[width=\textwidth]{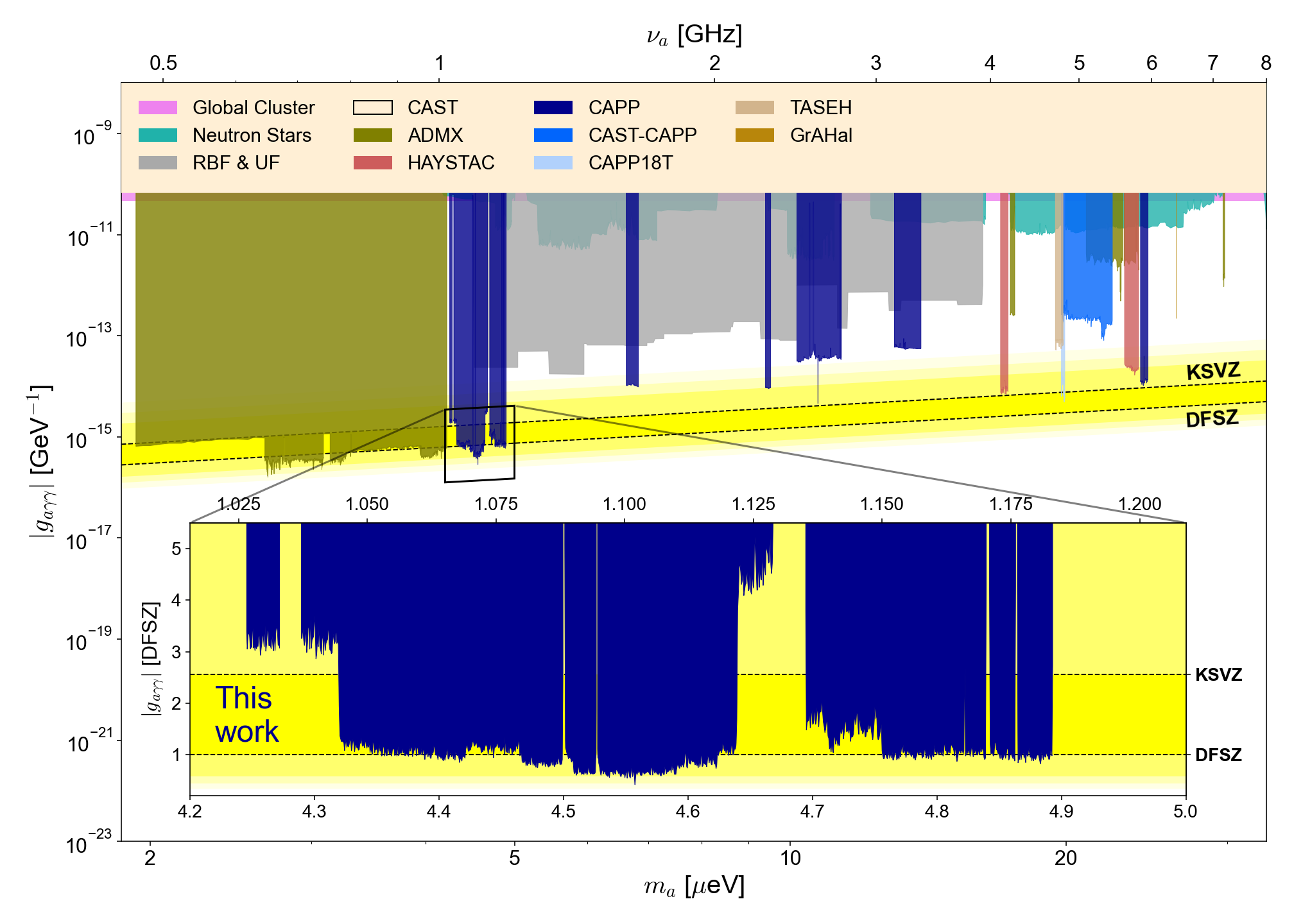}
    \caption{\label{fig:exclusion1} 
    Exclusion limits set by this work for axion--photon coupling at approximately 90\% confidence level assuming that axions are 100\% of the local dark matter density with a value of 0.45\,GeV/cm$^3$. The yellow band represents the areas predicted by the theoretical models of KSVZ and DFSZ. The magnified version is shown in the inset over the mass range between 4.2\,$\mu$eV and 5.0\,$\mu$eV. The gaps around 1.0881\,GHz, 1.0947\,GHz, 1.1290--1.1350\,GHz, 1.1705\,GHz and 1.1762\,GHz correspond to mode crossings. The frequency ranges of 1.0267--1.0443\,GHz and 1.1220--1.1229\,GHz were scanned with HEMT as the first-stage amplifier. The remaining gaps correspond to data whose analysis was not finalized by the publication time for a variety of reasons, including incomplete stored information.
    The exclusion  limits from other experiments (Global Cluster~\cite{Dolan_2022}, Neutron Stars~\cite{PhysRevLett.125.171301,Darling_2020,PhysRevLett.129.251102,battye2023searching}, RBF \& UF~\cite{article:Panfilis1987, article:Hagmann1990}, CAST~\cite{Anastassopoulos2017}, ADMX~\cite{article:ADMX-1, article:ADMX-2, article:ADMX-3}, HAYSTACK~\cite{article:HAYSTAC-1, article:HAYSTAC-2, article:HAYSTAC-3}, CAPP~\cite{article:CAPP-8TB, article:CAPP-PACE-JPA, article:CAPP-9T, article:CAPP-PACE, 12TB-PRL}, CAST-CAPP~\cite{article:CAST-CAPP}, CAPP18T~\cite{article:CAPP18T2023, CAPP-18T}, TASEH~\cite{TASEH-1}, GrAHal~\cite{grenet2021grenoble})  are taken from Ref.~\cite{AxionLimits}.
    }
    \end{center}
\end{figure*}

The grandspectra from all runs and their rescans are combined once more to report the limit on $g_{\gamma}$. The systematic effect of the baseline estimation on the final SNR is calibrated using a software-synthesized axion injected at every 200\,kHz in each Run. 
The overall signal detection efficiency is found to fluctuate slightly around 90\% in Runs 4, 5 and 6. The effect of the frequency-dependent Gaussian width in Run 6 is considered here during the normalization procedure. 
The combined sensitivity is calculated on the basis of the Bayesian power measurement framework~\cite{article:bayesian_method}. 
Figure~\ref{fig:Normal_Be} shows the statistical distribution of the formed excess from the combined Runs 5 and 6. About 10\% of the candidates above the given threshold are newly found after further improvement of the analysis after rescan, as well as combining Runs 4, 5, and 6. The effect of such candidates on exclusion of $g_{a\gamma\gamma}$ is naturally reflected by accomodating the Baysian approach.
The 10\% prior update contour with 400\,kHz of sub-aggregate window is set for the frequency-dependent limit corresponding to approximately 90\% confidence level. 
The limit from the frequentist threshold (FT) approach is reported as a reference, with the same sub-aggregate window, see Fig.~\ref{fig:exclusion}.
The exclusion plot shown in Fig.~\ref{fig:exclusion1} includes all major axion dark matter experiments and astrophysical limits between approximately 0.5\,GHz and 8\,GHz. The current work is indicated in the inset.
The exclusion plot here corresponds to the blue line of Fig.~\ref{fig:exclusion}.


\section{Discussion}
% SC cavity, variance method
Over the past few decades, intense experimental effort has been made internationally to probe the vast parameter space for axions and axion-like particles.
Cavity-based haloscopes have provided the most sensitive searches in the $\mathcal{O}(10^0)$\,GHz range.
However, there are only a few attempts to go beyond that, and even then, the sensitivities are still far away from the hadronic axion models.
This is mainly due to the loss of cavity detection volume, the decrease in the cavity quality factor at higher frequencies, and the linear increase of quantum noise with frequency.
Therefore, different approaches are needed to achieve high sensitivity to frequencies higher than 1\,GHz. 

%\paragraph{SC cavities in strong magnetic fields.} 

It was thought for a long time that superconducting (SC) cavities would perform worse than copper cavities in a strong magnetic field. Certainly, there have been previous attempts to manufacture SC cavities but they all failed to show any improvement until CAPP's breakthrough achievement using high-temperature-superconducting (HTS) tapes~\cite{Danho19_1, Danho19_2, Danho22}. CAPP demonstrated a steady improvement in this field over the years, currently reaching a quality factor of $13\times10^6$ in  8\,T magnetic field.
When data acquisition is optimized, the axion scanning rate can be proportional to the cavity quality factor even when it exceeds the axion quality factor~\cite{Jinsu_Kim2023,article:Cervantes2022}.

%\paragraph{High frequency}
CAPP has developed several novel cavity designs that efficiently increase the search frequency, along with practical tuning mechanisms.
They include (i)~a pizza-cavity design consisting of multiple cells divided by equidistant partitions~\cite{article:pizza_cavity}, tuned by a single carousel-like structure and read-out by a single antenna (ii)~a wheel-tuning mechanism suitable for exploiting higher-order resonant modes such as TM$_{030}$~\cite{article:wheel_mechanism}, and (iii)~a tunable photonic crystal design featuring an array of dielectric rods whose position can be adjusted using an auxetic structure of rotating squares~\cite{article:photonic_crystal}.
In addition, CAPP recently proposed a new search method based on heterodyne interferometry that mixes weak axion-induced photons with strong reference photons~\cite{article:variance}.
The variance of the interference, whose intensity is proportional to the amplitude of the reference probe, can be measured using a power detector even at high dark count rates, enabling sensitive detection nearly quantum-noise limited.

%\paragraph{Single photon detector}
The long-term direction will go towards the development of single-photon detectors, which are overwhelmingly advantageous, particularly at high frequencies and low temperatures. 
%Compared to quantum-limited linear amplifiers that are subject to fundamental quantum noise, which is linear in frequency, single-photon detectors that are in principle noiseless improve the scanning speed dramatically with frequency at the low temperature limits, becoming competitive and ultimately favored for searches above 10\,GHz~\cite{article:SPC_QLA}.
%CAPP plans to develop a microwave photon counter based on Rydberg atom quantum technologies.
%The general features of Rydberg atoms, such as large transition dipole moments, energy transition levels in the GHz range, tunable transition frequencies via the Stark effect, and long lifetimes on the order of milliseconds, make them suitable for efficient detection of microwave photons.
Quantum-limited linear amplifiers are subject to fundamental quantum noise, which is linear in frequency. Single-photon detectors at low temperatures could improve the scanning speed dramatically at high frequencies when their dark count is kept low~\cite{article:SPC_QLA, article:SPD, article:Kuzmin2020, article:Gatti2023} and above 10\,GHz~\cite{article:SPC_QLA}.
For frequencies below 10\,GHz, we are developing axion detection chains based on single-photon counters. 
The main component of one approach is a chip containing a microwave coplanar waveguide as an input line, terminated by the current-biased Josephson junction (CBJJ) used as a photon switching detector~\cite{article:Kuzmin2020}.
The second approach involves nanobolometers with a resolution of several zeptojoules~\cite{article:Mottonen2016}.  
Both approaches, together with the heterodyne interferometry method~\cite{article:variance}, will significantly increase the scanning speed of the experiment.

In addition, CAPP plans to develop a microwave photon counter based on Rydberg atom quantum technologies~\cite{article:CARRACK, article:Ahn23}.
The general features of Rydberg atoms, such as large transition dipole moments, energy transition levels in the GHz range, tunable transition frequencies via the Stark effect, and long lifetimes on the order of milliseconds, make them suitable for efficient detection of microwave photons.

\section{Summary}
We reported on the first high-sensitivity axion haloscope search for frequencies above 1\,GHz and a scanned range of more than 150\,MHz.
The achievement described above was made possible by combining a large aperture 12-T magnet with a 37-liter ultralight microwave resonant cavity whose physical temperature was kept between 40\,mK and 22\,mK, and JPAs with total system noise temperature around 200\,mK.

The 12\,T magnet consists of two coils, an inner coil based on Nb$_3$Sn low temperature superconducting (LTS) cable and an outer coil based on the more conventional NbTi LTS cable, made by Oxford Instruments. The total energy content of the magnet, when fully energized, is 5.562\,MJ, comes with a persistent mode switch, and operates reliably within its specifications. We have successfully operated for the first time in the world a large volume, 12\,T magnet specifically designed to host cavities in the 1-8\,GHz frequency range.

The sensitivity of the experiment relies equally well on the ability to thermalize and cool the 37-liter microwave cavity and the quantum noise-limited amplifiers to the lowest possible temperature, preferably below 40\,mK. Our ultra-light-cavity is also implemented for the first time in such large volume successfully without jeopardizing the reliability of its critical parameters in axion dark matter research. Combined with the state of the art thermalization techniques applied in all the important interfaces we managed to keep the cavity for most of the time well below 40\,mK. We gold-plated the interfaces when good thermalization was needed and partially or totally isolated the elements when it was appropriate.

The maximum temperature difference between the top and bottom of the cavity was 15\,mk, which was slowly reduced as a function of time. Eventually, since this run, we have also improved their thermalization and achieved less than 5\,mk temperature difference by using annealed copper tapes between the top and bottom plates of the cavity.

Finally, our JPAs at around 1\,GHz are performing near quantum noise limit but with limited bandwidth between 50 to 100\,MHz. In order to reduce the need for warmup and cooldown in the experiment, we have bundled up several JPAs together in an optimum manner in order to maintain most of their performance characteristics without compromising significantly the total system noise. 

For all of the above projects we have had a very steep learning curve and by the time we finished the 150\,MHz scan, we have learned how to effectively combine all of them with high efficiency. The next project is ready to scan between 1.2 and 1.5\,GHz at DFSZ sensitivity at a scan rate of more than 3\,MHz per day, with the added benefit of using a tuning rod covered by high-temperature superconducting tape. Superconducting cavities combined with the variance method promise an additional up to two orders of magnitude in real calendar scanning efficiency, and it is the main current activity at CAPP.

\begin{acknowledgments}
This work was supported by the Institute for Basic Science (IBS-R017-D1) and  
JSPS KAKENHI (Grant No.~JP22H04937).
Arjan F. van Loo was supported by a JSPS postdoctoral fellowship. 
\end{acknowledgments}

\appendix
\section{Bayesian power measured analysis}

The exclusion limit on $g_{\gamma}$ from the data acquired in all runs and their rescans is evaluated using a Bayesian power measured framework. The posterior probability $P_{\textrm{posterior}}$ of axion presented at a certain frequency bin is updated by number of observation made by the initial scans and the rescans. For a given observation, the ratio $P_{\textrm{prior}}/P_{\textrm{posterior}}$ can be approximated by the Bayes factor of two competing hypotheses (whether or not the signal exists). Then the total update is given by the product of the Bayes factors from all observations at the given frequency. Since the coupling strength $g_{\gamma}$ is also unknown, the total update is evaluated until it is meaningful ($10^{-2}-10^{2}$). The color map in Fig.~\ref{fig:exclusion} shows the total update in each frequency bin, aggregated to 300\,kHz. The aggregate update can be obtained by the average of the total update in individual bins, by neglecting the frequency dependence of the posterior probability in each aggregate window. 
The $g_{\gamma}$ which makes the aggregate update to be 0.1 is drawn as blue line in Fig.~\ref{fig:exclusion}, corresponding to the 90\% confidence level on $g_\gamma$ to be excluded. For comparison, exclusion limit from the threshold based frequentist approach is also drawn as the green line, aggregated with the same window size. 

\section{Systematic-error calibration}
As previously reported in~\cite{12TB-PRL}, the systematic error on $g_{a\gamma\gamma}$ in Run 4 is estimated to be about 3\% and is similar in Run 5 as the system configuration is essentially the same in those Runs. It is estimated to be about 8\% in Run 6, and the main sources of error in comparison to other runs are the calibration of the SNR for the idler side signal and the drift of $T_\textrm{off}$ due to active devices other than JPAs. The SNR error from the idler side is estimated using the single-tone calibration as described in Sec.~\ref{sec:analysls} of the main text. This error is frequency-dependent and tends to be higher in the region where the JPA performance is not optimal. The drift of $T_\textrm{off}$ is the dominant source of systematic error for the data taken without the JPAs in Run 6, and its effect on $g_{a\gamma\gamma}$ is around 2\%. Consequently, the average systematic error on the combined exclusion limit is approximately 4\%, assuming that there is no correlation between the runs.

\bibliography{main}

\begin{thebibliography}{122}%
\makeatletter
\providecommand \@ifxundefined [1]{%
 \@ifx{#1\undefined}
}%
\providecommand \@ifnum [1]{%
 \ifnum #1\expandafter \@firstoftwo
 \else \expandafter \@secondoftwo
 \fi
}%
\providecommand \@ifx [1]{%
 \ifx #1\expandafter \@firstoftwo
 \else \expandafter \@secondoftwo
 \fi
}%
\providecommand \natexlab [1]{#1}%
\providecommand \enquote  [1]{``#1''}%
\providecommand \bibnamefont  [1]{#1}%
\providecommand \bibfnamefont [1]{#1}%
\providecommand \citenamefont [1]{#1}%
\providecommand \href@noop [0]{\@secondoftwo}%
\providecommand \href [0]{\begingroup \@sanitize@url \@href}%
\providecommand \@href[1]{\@@startlink{#1}\@@href}%
\providecommand \@@href[1]{\endgroup#1\@@endlink}%
\providecommand \@sanitize@url [0]{\catcode `\\12\catcode `\$12\catcode
  `\&12\catcode `\#12\catcode `\^12\catcode `\_12\catcode `\%12\relax}%
\providecommand \@@startlink[1]{}%
\providecommand \@@endlink[0]{}%
\providecommand \url  [0]{\begingroup\@sanitize@url \@url }%
\providecommand \@url [1]{\endgroup\@href {#1}{\urlprefix }}%
\providecommand \urlprefix  [0]{URL }%
\providecommand \Eprint [0]{\href }%
\providecommand \doibase [0]{https://doi.org/}%
\providecommand \selectlanguage [0]{\@gobble}%
\providecommand \bibinfo  [0]{\@secondoftwo}%
\providecommand \bibfield  [0]{\@secondoftwo}%
\providecommand \translation [1]{[#1]}%
\providecommand \BibitemOpen [0]{}%
\providecommand \bibitemStop [0]{}%
\providecommand \bibitemNoStop [0]{.\EOS\space}%
\providecommand \EOS [0]{\spacefactor3000\relax}%
\providecommand \BibitemShut  [1]{\csname bibitem#1\endcsname}%
\let\auto@bib@innerbib\@empty
%</preamble>
\bibitem [{\citenamefont {Copi}\ \emph {et~al.}(1995)\citenamefont {Copi},
  \citenamefont {Schramm},\ and\ \citenamefont {Turner}}]{article:dark_matter}%
  \BibitemOpen
  \bibfield  {author} {\bibinfo {author} {\bibfnamefont {C.~J.}\ \bibnamefont
  {Copi}}, \bibinfo {author} {\bibfnamefont {D.~N.}\ \bibnamefont {Schramm}},\
  and\ \bibinfo {author} {\bibfnamefont {M.~S.}\ \bibnamefont {Turner}},\
  }\bibfield  {title} {\bibinfo {title} {Big-bang nucleosynthesis and the
  baryon density of the universe},\ }\href
  {https://doi.org/10.1126/science.7809624} {\bibfield  {journal} {\bibinfo
  {journal} {Science}\ }\textbf {\bibinfo {volume} {267}},\ \bibinfo {pages}
  {192} (\bibinfo {year} {1995})}\BibitemShut {NoStop}%
\bibitem [{\citenamefont {Aghanim}\ \emph {et~al.}(2020)\citenamefont {Aghanim}
  \emph {et~al.}}]{article:Planck}%
  \BibitemOpen
  \bibfield  {author} {\bibinfo {author} {\bibfnamefont {N.}~\bibnamefont
  {Aghanim}} \emph {et~al.},\ }\bibfield  {title} {\bibinfo {title} {Planck
  2018 results - vi. cosmological parameters},\ }\href
  {https://doi.org/10.1051/0004-6361/201833910} {\bibfield  {journal} {\bibinfo
   {journal} {Astron. Astrophys.}\ }\textbf {\bibinfo {volume} {641}},\
  \bibinfo {pages} {A6} (\bibinfo {year} {2020})}\BibitemShut {NoStop}%
\bibitem [{\citenamefont {Roszkowski}\ \emph {et~al.}(2018)\citenamefont
  {Roszkowski}, \citenamefont {Sessolo},\ and\ \citenamefont
  {Trojanowski}}]{article:WIMP1}%
  \BibitemOpen
  \bibfield  {author} {\bibinfo {author} {\bibfnamefont {L.}~\bibnamefont
  {Roszkowski}}, \bibinfo {author} {\bibfnamefont {E.~M.}\ \bibnamefont
  {Sessolo}},\ and\ \bibinfo {author} {\bibfnamefont {S.}~\bibnamefont
  {Trojanowski}},\ }\bibfield  {title} {\bibinfo {title} {{WIMP} dark matter
  candidates and searches—current status and future prospects},\ }\href
  {https://doi.org/10.1088/1361-6633/aab913} {\bibfield  {journal} {\bibinfo
  {journal} {Reports on Progress in Physics}\ }\textbf {\bibinfo {volume}
  {81}},\ \bibinfo {pages} {066201} (\bibinfo {year} {2018})}\BibitemShut
  {NoStop}%
\bibitem [{\citenamefont {Schumann}(2019)}]{article:WIMP2}%
  \BibitemOpen
  \bibfield  {author} {\bibinfo {author} {\bibfnamefont {M.}~\bibnamefont
  {Schumann}},\ }\bibfield  {title} {\bibinfo {title} {Direct detection of
  {WIMP} dark matter: concepts and status},\ }\href
  {https://doi.org/10.1088/1361-6471/ab2ea5} {\bibfield  {journal} {\bibinfo
  {journal} {Journal of Physics G: Nuclear and Particle Physics}\ }\textbf
  {\bibinfo {volume} {46}},\ \bibinfo {pages} {103003} (\bibinfo {year}
  {2019})}\BibitemShut {NoStop}%
\bibitem [{\citenamefont {Chadha-Day}\ \emph {et~al.}(2022)\citenamefont
  {Chadha-Day}, \citenamefont {Ellis},\ and\ \citenamefont
  {Marsh}}]{article:axion1}%
  \BibitemOpen
  \bibfield  {author} {\bibinfo {author} {\bibfnamefont {F.}~\bibnamefont
  {Chadha-Day}}, \bibinfo {author} {\bibfnamefont {J.}~\bibnamefont {Ellis}},\
  and\ \bibinfo {author} {\bibfnamefont {D.~J.~E.}\ \bibnamefont {Marsh}},\
  }\bibfield  {title} {\bibinfo {title} {Axion dark matter: What is it and why
  now?},\ }\href {https://doi.org/10.1126/sciadv.abj3618} {\bibfield  {journal}
  {\bibinfo  {journal} {Science Advances}\ }\textbf {\bibinfo {volume} {8}},\
  \bibinfo {pages} {eabj3618} (\bibinfo {year} {2022})}\BibitemShut {NoStop}%
\bibitem [{\citenamefont {Semertzidis}\ and\ \citenamefont
  {Youn}(2022)}]{article:axion2}%
  \BibitemOpen
  \bibfield  {author} {\bibinfo {author} {\bibfnamefont {Y.~K.}\ \bibnamefont
  {Semertzidis}}\ and\ \bibinfo {author} {\bibfnamefont {S.}~\bibnamefont
  {Youn}},\ }\bibfield  {title} {\bibinfo {title} {Axion dark matter: How to
  see it?},\ }\href {https://doi.org/10.1126/sciadv.abm9928} {\bibfield
  {journal} {\bibinfo  {journal} {Science Advances}\ }\textbf {\bibinfo
  {volume} {8}},\ \bibinfo {pages} {eabm9928} (\bibinfo {year}
  {2022})}\BibitemShut {NoStop}%
\bibitem [{\citenamefont {Green}\ and\ \citenamefont
  {Kavanagh}(2021)}]{article:PBH1}%
  \BibitemOpen
  \bibfield  {author} {\bibinfo {author} {\bibfnamefont {A.~M.}\ \bibnamefont
  {Green}}\ and\ \bibinfo {author} {\bibfnamefont {B.~J.}\ \bibnamefont
  {Kavanagh}},\ }\bibfield  {title} {\bibinfo {title} {Primordial black holes
  as a dark matter candidate},\ }\href
  {https://doi.org/10.1088/1361-6471/abc534} {\bibfield  {journal} {\bibinfo
  {journal} {Journal of Physics G: Nuclear and Particle Physics}\ }\textbf
  {\bibinfo {volume} {48}},\ \bibinfo {pages} {043001} (\bibinfo {year}
  {2021})}\BibitemShut {NoStop}%
\bibitem [{\citenamefont {Villanueva-Domingo}\ \emph
  {et~al.}(2021)\citenamefont {Villanueva-Domingo}, \citenamefont {Mena},\ and\
  \citenamefont {Palomares-Ruiz}}]{article:PBH2}%
  \BibitemOpen
  \bibfield  {author} {\bibinfo {author} {\bibfnamefont {P.}~\bibnamefont
  {Villanueva-Domingo}}, \bibinfo {author} {\bibfnamefont {O.}~\bibnamefont
  {Mena}},\ and\ \bibinfo {author} {\bibfnamefont {S.}~\bibnamefont
  {Palomares-Ruiz}},\ }\bibfield  {title} {\bibinfo {title} {A brief review on
  primordial black holes as dark matter},\ }\href
  {https://doi.org/10.3389/fspas.2021.681084} {\bibfield  {journal} {\bibinfo
  {journal} {Front. Astron. Space Sci.}\ }\textbf {\bibinfo {volume} {8}}
  (\bibinfo {year} {2021})}\BibitemShut {NoStop}%
\bibitem [{\citenamefont {Abel}\ \emph {et~al.}(2020)\citenamefont {Abel} \emph
  {et~al.}}]{article:nEDM}%
  \BibitemOpen
  \bibfield  {author} {\bibinfo {author} {\bibfnamefont {C.}~\bibnamefont
  {Abel}} \emph {et~al.},\ }\bibfield  {title} {\bibinfo {title} {Measurement
  of the permanent electric dipole moment of the neutron},\ }\href
  {https://doi.org/10.1103/PhysRevLett.124.081803} {\bibfield  {journal}
  {\bibinfo  {journal} {Phys. Rev. Lett.}\ }\textbf {\bibinfo {volume} {124}},\
  \bibinfo {pages} {081803} (\bibinfo {year} {2020})}\BibitemShut {NoStop}%
\bibitem [{\citenamefont {Alexander}\ \emph {et~al.}(2022)\citenamefont
  {Alexander} \emph {et~al.}}]{article:pEDM}%
  \BibitemOpen
  \bibfield  {author} {\bibinfo {author} {\bibfnamefont {J.}~\bibnamefont
  {Alexander}} \emph {et~al.},\ }\bibfield  {title} {\bibinfo {title} {The
  storage ring proton {EDM} experiment},\ }\bibfield  {journal} {\bibinfo
  {journal} {ArXiv}\ }\href {https://doi.org/10.48550/arXiv.2205.00830}
  {10.48550/arXiv.2205.00830} (\bibinfo {year} {2022}),\ \Eprint
  {https://arxiv.org/abs/2205.00830} {2205.00830} \BibitemShut {NoStop}%
\bibitem [{\citenamefont {Peccei}\ and\ \citenamefont
  {Quinn}(1977)}]{article:PQ77}%
  \BibitemOpen
  \bibfield  {author} {\bibinfo {author} {\bibfnamefont {R.~D.}\ \bibnamefont
  {Peccei}}\ and\ \bibinfo {author} {\bibfnamefont {H.~R.}\ \bibnamefont
  {Quinn}},\ }\bibfield  {title} {\bibinfo {title} {$\mathrm{CP}$ conservation
  in the presence of pseudoparticles},\ }\href
  {https://doi.org/10.1103/PhysRevLett.38.1440} {\bibfield  {journal} {\bibinfo
   {journal} {Phys. Rev. Lett.}\ }\textbf {\bibinfo {volume} {38}},\ \bibinfo
  {pages} {1440} (\bibinfo {year} {1977})}\BibitemShut {NoStop}%
\bibitem [{\citenamefont {Weinberg}(1978)}]{article:Weinberg78}%
  \BibitemOpen
  \bibfield  {author} {\bibinfo {author} {\bibfnamefont {S.}~\bibnamefont
  {Weinberg}},\ }\bibfield  {title} {\bibinfo {title} {A new light boson?},\
  }\href {https://doi.org/10.1103/PhysRevLett.40.223} {\bibfield  {journal}
  {\bibinfo  {journal} {Phys. Rev. Lett.}\ }\textbf {\bibinfo {volume} {40}},\
  \bibinfo {pages} {223} (\bibinfo {year} {1978})}\BibitemShut {NoStop}%
\bibitem [{\citenamefont {Wilczek}(1978)}]{article:Wilczek78}%
  \BibitemOpen
  \bibfield  {author} {\bibinfo {author} {\bibfnamefont {F.}~\bibnamefont
  {Wilczek}},\ }\bibfield  {title} {\bibinfo {title} {Problem of strong
  $\mathrm{P}$ and $\mathrm{T}$ invariance in the presence of instantons},\
  }\href {https://doi.org/10.1103/PhysRevLett.40.279} {\bibfield  {journal}
  {\bibinfo  {journal} {Phys. Rev. Lett.}\ }\textbf {\bibinfo {volume} {40}},\
  \bibinfo {pages} {229} (\bibinfo {year} {1978})}\BibitemShut {NoStop}%
\bibitem [{\citenamefont {Preskill}\ \emph {et~al.}(1983)\citenamefont
  {Preskill}, \citenamefont {Wise},\ and\ \citenamefont
  {Wilczek}}]{article:axion_cosmology1}%
  \BibitemOpen
  \bibfield  {author} {\bibinfo {author} {\bibfnamefont {J.}~\bibnamefont
  {Preskill}}, \bibinfo {author} {\bibfnamefont {M.~B.}\ \bibnamefont {Wise}},\
  and\ \bibinfo {author} {\bibfnamefont {F.}~\bibnamefont {Wilczek}},\
  }\bibfield  {title} {\bibinfo {title} {Cosmology of the invisible axion},\
  }\href {https://doi.org/https://doi.org/10.1016/0370-2693(83)90637-8}
  {\bibfield  {journal} {\bibinfo  {journal} {Physics Letters B}\ }\textbf
  {\bibinfo {volume} {120}},\ \bibinfo {pages} {127} (\bibinfo {year}
  {1983})}\BibitemShut {NoStop}%
\bibitem [{\citenamefont {Abbott}\ and\ \citenamefont
  {Sikivie}(1983)}]{article:axion_cosmology2}%
  \BibitemOpen
  \bibfield  {author} {\bibinfo {author} {\bibfnamefont {L.~F.}\ \bibnamefont
  {Abbott}}\ and\ \bibinfo {author} {\bibfnamefont {P.}~\bibnamefont
  {Sikivie}},\ }\bibfield  {title} {\bibinfo {title} {A cosmological bound on
  the invisible axion},\ }\href
  {https://doi.org/https://doi.org/10.1016/0370-2693(83)90638-X} {\bibfield
  {journal} {\bibinfo  {journal} {Phys. Lett. B}\ }\textbf {\bibinfo {volume}
  {120}},\ \bibinfo {pages} {133} (\bibinfo {year} {1983})}\BibitemShut
  {NoStop}%
\bibitem [{\citenamefont {Dine}\ and\ \citenamefont
  {Fischler}(1983)}]{article:axion_cosmology3}%
  \BibitemOpen
  \bibfield  {author} {\bibinfo {author} {\bibfnamefont {M.}~\bibnamefont
  {Dine}}\ and\ \bibinfo {author} {\bibfnamefont {W.}~\bibnamefont
  {Fischler}},\ }\bibfield  {title} {\bibinfo {title} {The not-so-harmless
  axion},\ }\href
  {https://doi.org/https://doi.org/10.1016/0370-2693(83)90639-1} {\bibfield
  {journal} {\bibinfo  {journal} {Phys. Lett. B}\ }\textbf {\bibinfo {volume}
  {120}},\ \bibinfo {pages} {137} (\bibinfo {year} {1983})}\BibitemShut
  {NoStop}%
\bibitem [{\citenamefont {Takahashi}\ \emph {et~al.}(2018)\citenamefont
  {Takahashi}, \citenamefont {Yin},\ and\ \citenamefont
  {Guth}}]{article:pre_infl1}%
  \BibitemOpen
  \bibfield  {author} {\bibinfo {author} {\bibfnamefont {F.}~\bibnamefont
  {Takahashi}}, \bibinfo {author} {\bibfnamefont {W.}~\bibnamefont {Yin}},\
  and\ \bibinfo {author} {\bibfnamefont {A.~H.}\ \bibnamefont {Guth}},\
  }\bibfield  {title} {\bibinfo {title} {{QCD} axion window and low-scale
  inflation},\ }\href {https://doi.org/10.1103/PhysRevD.98.015042} {\bibfield
  {journal} {\bibinfo  {journal} {Phys. Rev. D}\ }\textbf {\bibinfo {volume}
  {98}},\ \bibinfo {pages} {015042} (\bibinfo {year} {2018})}\BibitemShut
  {NoStop}%
\bibitem [{\citenamefont {Graham}\ and\ \citenamefont
  {Scherlis}(2018)}]{article:pre_infl2}%
  \BibitemOpen
  \bibfield  {author} {\bibinfo {author} {\bibfnamefont {P.~W.}\ \bibnamefont
  {Graham}}\ and\ \bibinfo {author} {\bibfnamefont {A.}~\bibnamefont
  {Scherlis}},\ }\bibfield  {title} {\bibinfo {title} {Stochastic axion
  scenario},\ }\href {https://doi.org/10.1103/PhysRevD.98.035017} {\bibfield
  {journal} {\bibinfo  {journal} {Phys. Rev. D}\ }\textbf {\bibinfo {volume}
  {98}},\ \bibinfo {pages} {035017} (\bibinfo {year} {2018})}\BibitemShut
  {NoStop}%
\bibitem [{\citenamefont {Borsanyi}\ \emph {et~al.}(2016)\citenamefont
  {Borsanyi} \emph {et~al.}}]{article:post_infl1}%
  \BibitemOpen
  \bibfield  {author} {\bibinfo {author} {\bibfnamefont {S.}~\bibnamefont
  {Borsanyi}} \emph {et~al.},\ }\bibfield  {title} {\bibinfo {title}
  {Calculation of the axion mass based on high-temperature lattice quantum
  chromodynamics},\ }\href {https://doi.org/10.1038/nature20115} {\bibfield
  {journal} {\bibinfo  {journal} {Nature}\ }\textbf {\bibinfo {volume} {539}},\
  \bibinfo {pages} {69} (\bibinfo {year} {2016})}\BibitemShut {NoStop}%
\bibitem [{\citenamefont {Klaer}\ and\ \citenamefont
  {Moore}(2017)}]{article:post_infl2}%
  \BibitemOpen
  \bibfield  {author} {\bibinfo {author} {\bibfnamefont {V.~B.}\ \bibnamefont
  {Klaer}}\ and\ \bibinfo {author} {\bibfnamefont {G.~D.}\ \bibnamefont
  {Moore}},\ }\bibfield  {title} {\bibinfo {title} {The dark-matter axion
  mass},\ }\href {https://doi.org/10.1088/1475-7516/2017/11/049} {\bibfield
  {journal} {\bibinfo  {journal} {J. of Cosmol. Astropart. Phys.}\ }\textbf
  {\bibinfo {volume} {11}},\ \bibinfo {pages} {49} (\bibinfo {year}
  {2017})}\BibitemShut {NoStop}%
\bibitem [{\citenamefont {Buschmann}\ \emph {et~al.}(2020)\citenamefont
  {Buschmann}, \citenamefont {Foster},\ and\ \citenamefont
  {Safdi}}]{article:post_infl3}%
  \BibitemOpen
  \bibfield  {author} {\bibinfo {author} {\bibfnamefont {M.}~\bibnamefont
  {Buschmann}}, \bibinfo {author} {\bibfnamefont {J.~W.}\ \bibnamefont
  {Foster}},\ and\ \bibinfo {author} {\bibfnamefont {B.~R.}\ \bibnamefont
  {Safdi}},\ }\bibfield  {title} {\bibinfo {title} {Early-universe simulations
  of the cosmological axion},\ }\href
  {https://doi.org/10.1103/PhysRevLett.124.161103} {\bibfield  {journal}
  {\bibinfo  {journal} {Phys. Rev. Lett.}\ }\textbf {\bibinfo {volume} {124}},\
  \bibinfo {pages} {161103} (\bibinfo {year} {2020})}\BibitemShut {NoStop}%
\bibitem [{\citenamefont {Turner}(1990)}]{article:Turner}%
  \BibitemOpen
  \bibfield  {author} {\bibinfo {author} {\bibfnamefont {M.~S.}\ \bibnamefont
  {Turner}},\ }\bibfield  {title} {\bibinfo {title} {Periodic signatures for
  the detection of cosmic axions},\ }\href
  {https://doi.org/10.1103/PhysRevD.42.3572} {\bibfield  {journal} {\bibinfo
  {journal} {Phys. Rev. D}\ }\textbf {\bibinfo {volume} {42}},\ \bibinfo
  {pages} {3572} (\bibinfo {year} {1990})}\BibitemShut {NoStop}%
\bibitem [{\citenamefont {Kim}(1979)}]{article:KSVZ1}%
  \BibitemOpen
  \bibfield  {author} {\bibinfo {author} {\bibfnamefont {J.~E.}\ \bibnamefont
  {Kim}},\ }\bibfield  {title} {\bibinfo {title} {Weak-interaction singlet and
  strong $\mathrm{CP}$ invariance},\ }\href
  {https://doi.org/10.1103/PhysRevLett.43.103} {\bibfield  {journal} {\bibinfo
  {journal} {Phys. Rev. Lett.}\ }\textbf {\bibinfo {volume} {43}},\ \bibinfo
  {pages} {103} (\bibinfo {year} {1979})}\BibitemShut {NoStop}%
\bibitem [{\citenamefont {Shifman}\ \emph {et~al.}(1980)\citenamefont
  {Shifman}, \citenamefont {Vainshtein},\ and\ \citenamefont
  {Zakharov}}]{article:KSVZ2}%
  \BibitemOpen
  \bibfield  {author} {\bibinfo {author} {\bibfnamefont {M.~A.}\ \bibnamefont
  {Shifman}}, \bibinfo {author} {\bibfnamefont {A.~I.}\ \bibnamefont
  {Vainshtein}},\ and\ \bibinfo {author} {\bibfnamefont {V.~I.}\ \bibnamefont
  {Zakharov}},\ }\bibfield  {title} {\bibinfo {title} {Can confinement ensure
  natural {CP} invariance of strong interactions?},\ }\href
  {https://doi.org/https://doi.org/10.1016/0550-3213(80)90209-6} {\bibfield
  {journal} {\bibinfo  {journal} {Nucl. Phys. B}\ }\textbf {\bibinfo {volume}
  {166}},\ \bibinfo {pages} {4933} (\bibinfo {year} {1980})}\BibitemShut
  {NoStop}%
\bibitem [{\citenamefont {Zhitnitsky}(1980)}]{article:DFSZ1}%
  \BibitemOpen
  \bibfield  {author} {\bibinfo {author} {\bibfnamefont {A.~P.}\ \bibnamefont
  {Zhitnitsky}},\ }\bibfield  {title} {\bibinfo {title} {Possible suppression
  of axion-hadron interactions},\ }\href {https://www.osti.gov/biblio/7063072}
  {\bibfield  {journal} {\bibinfo  {journal} {Sov. J. Nucl. Phys.}\ }\textbf
  {\bibinfo {volume} {31:2}},\ \bibinfo {pages} {260} (\bibinfo {year}
  {1980})}\BibitemShut {NoStop}%
\bibitem [{\citenamefont {Dine}\ \emph {et~al.}(1981)\citenamefont {Dine},
  \citenamefont {Fischler},\ and\ \citenamefont {Srednicki}}]{article:DFSZ2}%
  \BibitemOpen
  \bibfield  {author} {\bibinfo {author} {\bibfnamefont {M.}~\bibnamefont
  {Dine}}, \bibinfo {author} {\bibfnamefont {W.}~\bibnamefont {Fischler}},\
  and\ \bibinfo {author} {\bibfnamefont {M.}~\bibnamefont {Srednicki}},\
  }\bibfield  {title} {\bibinfo {title} {A simple solution to the strong {CP}
  problem with a harmless axion},\ }\href
  {https://doi.org/https://doi.org/10.1016/0370-2693(81)90590-6} {\bibfield
  {journal} {\bibinfo  {journal} {Physics Letters B}\ }\textbf {\bibinfo
  {volume} {104}},\ \bibinfo {pages} {199} (\bibinfo {year}
  {1981})}\BibitemShut {NoStop}%
\bibitem [{\citenamefont {Sikivie}(1983)}]{article:Sikivie83}%
  \BibitemOpen
  \bibfield  {author} {\bibinfo {author} {\bibfnamefont {P.}~\bibnamefont
  {Sikivie}},\ }\bibfield  {title} {\bibinfo {title} {Experimental tests of the
  ``invisible'' axion},\ }\href {https://doi.org/10.1103/PhysRevLett.51.1415}
  {\bibfield  {journal} {\bibinfo  {journal} {Phys. Rev. Lett.}\ }\textbf
  {\bibinfo {volume} {51}},\ \bibinfo {pages} {1415} (\bibinfo {year}
  {1983})}\BibitemShut {NoStop}%
\bibitem [{\citenamefont {Lee}\ \emph {et~al.}(2020)\citenamefont {Lee},
  \citenamefont {Ahn}, \citenamefont {Choi}, \citenamefont {Ko},\ and\
  \citenamefont {Semertzidis}}]{article:CAPP-8TB}%
  \BibitemOpen
  \bibfield  {author} {\bibinfo {author} {\bibfnamefont {S.}~\bibnamefont
  {Lee}}, \bibinfo {author} {\bibfnamefont {S.}~\bibnamefont {Ahn}}, \bibinfo
  {author} {\bibfnamefont {J.}~\bibnamefont {Choi}}, \bibinfo {author}
  {\bibfnamefont {B.~R.}\ \bibnamefont {Ko}},\ and\ \bibinfo {author}
  {\bibfnamefont {Y.~K.}\ \bibnamefont {Semertzidis}},\ }\bibfield  {title}
  {\bibinfo {title} {Axion dark matter search around $6.7\text{}\text{
  }\ensuremath{\mu}\mathrm{eV}$},\ }\href
  {https://doi.org/10.1103/PhysRevLett.124.101802} {\bibfield  {journal}
  {\bibinfo  {journal} {Phys. Rev. Lett.}\ }\textbf {\bibinfo {volume} {124}},\
  \bibinfo {pages} {101802} (\bibinfo {year} {2020})}\BibitemShut {NoStop}%
\bibitem [{\citenamefont {Jeong}\ \emph {et~al.}(2020)\citenamefont {Jeong}
  \emph {et~al.}}]{article:CAPP-9T}%
  \BibitemOpen
  \bibfield  {author} {\bibinfo {author} {\bibfnamefont {J.}~\bibnamefont
  {Jeong}} \emph {et~al.},\ }\bibfield  {title} {\bibinfo {title} {Search for
  invisible axion dark matter with a multiple-cell haloscope},\ }\href
  {https://doi.org/10.1103/PhysRevLett.125.221302} {\bibfield  {journal}
  {\bibinfo  {journal} {Phys. Rev. Lett.}\ }\textbf {\bibinfo {volume} {125}},\
  \bibinfo {pages} {221302} (\bibinfo {year} {2020})}\BibitemShut {NoStop}%
\bibitem [{\citenamefont {Kwon}\ \emph {et~al.}(2021)\citenamefont {Kwon} \emph
  {et~al.}}]{article:CAPP-PACE}%
  \BibitemOpen
  \bibfield  {author} {\bibinfo {author} {\bibfnamefont {O.}~\bibnamefont
  {Kwon}} \emph {et~al.},\ }\bibfield  {title} {\bibinfo {title} {First results
  from an axion haloscope at {CAPP} around $10.7\text{}\text{
  }\ensuremath{\mu}\mathrm{eV}$},\ }\href
  {https://doi.org/10.1103/PhysRevLett.126.191802} {\bibfield  {journal}
  {\bibinfo  {journal} {Phys. Rev. Lett.}\ }\textbf {\bibinfo {volume} {126}},\
  \bibinfo {pages} {191802} (\bibinfo {year} {2021})}\BibitemShut {NoStop}%
\bibitem [{\citenamefont {Kim}\ \emph {et~al.}(2023)\citenamefont {Kim} \emph
  {et~al.}}]{article:CAPP-PACE-JPA}%
  \BibitemOpen
  \bibfield  {author} {\bibinfo {author} {\bibfnamefont {J.}~\bibnamefont
  {Kim}} \emph {et~al.},\ }\bibfield  {title} {\bibinfo {title}
  {Near-quantum-noise axion dark matter search at {CAPP} around
  $9.5\text{}\text{ }\mathrm{\ensuremath{\mu}}\mathrm{eV}$},\ }\href
  {https://doi.org/10.1103/PhysRevLett.130.091602} {\bibfield  {journal}
  {\bibinfo  {journal} {Phys. Rev. Lett.}\ }\textbf {\bibinfo {volume} {130}},\
  \bibinfo {pages} {091602} (\bibinfo {year} {2023})}\BibitemShut {NoStop}%
\bibitem [{\citenamefont {Yi}\ \emph {et~al.}(2023{\natexlab{a}})\citenamefont
  {Yi} \emph {et~al.}}]{12TB-PRL}%
  \BibitemOpen
  \bibfield  {author} {\bibinfo {author} {\bibfnamefont {A.~K.}\ \bibnamefont
  {Yi}} \emph {et~al.},\ }\bibfield  {title} {\bibinfo {title} {Axion dark
  matter search around $4.55\text{}\text{
  }\mathrm{\ensuremath{\mu}}\mathrm{eV}$ with
  {Dine-Fischler-Srednicki-Zhitnitskii} sensitivity},\ }\href
  {https://doi.org/10.1103/PhysRevLett.130.071002} {\bibfield  {journal}
  {\bibinfo  {journal} {Phys. Rev. Lett.}\ }\textbf {\bibinfo {volume} {130}},\
  \bibinfo {pages} {071002} (\bibinfo {year} {2023}{\natexlab{a}})}\BibitemShut
  {NoStop}%
\bibitem [{\citenamefont {Du}\ \emph {et~al.}(2018)\citenamefont {Du} \emph
  {et~al.}}]{article:ADMX-1}%
  \BibitemOpen
  \bibfield  {author} {\bibinfo {author} {\bibfnamefont {N.}~\bibnamefont {Du}}
  \emph {et~al.} (\bibinfo {collaboration} {ADMX}),\ }\bibfield  {title}
  {\bibinfo {title} {Search for invisible axion dark matter with the axion dark
  matter experiment},\ }\href {https://doi.org/10.1103/PhysRevLett.120.151301}
  {\bibfield  {journal} {\bibinfo  {journal} {Phys. Rev. Lett.}\ }\textbf
  {\bibinfo {volume} {120}},\ \bibinfo {pages} {151301} (\bibinfo {year}
  {2018})}\BibitemShut {NoStop}%
\bibitem [{\citenamefont {Braine}\ \emph {et~al.}(2020)\citenamefont {Braine}
  \emph {et~al.}}]{article:ADMX-2}%
  \BibitemOpen
  \bibfield  {author} {\bibinfo {author} {\bibfnamefont {T.}~\bibnamefont
  {Braine}} \emph {et~al.} (\bibinfo {collaboration} {ADMX}),\ }\bibfield
  {title} {\bibinfo {title} {Extended search for the invisible axion with the
  axion dark matter experiment},\ }\href
  {https://doi.org/10.1103/PhysRevLett.124.101303} {\bibfield  {journal}
  {\bibinfo  {journal} {Phys. Rev. Lett.}\ }\textbf {\bibinfo {volume} {124}},\
  \bibinfo {pages} {101303} (\bibinfo {year} {2020})}\BibitemShut {NoStop}%
\bibitem [{\citenamefont {Bartram}\ \emph
  {et~al.}(2021{\natexlab{a}})\citenamefont {Bartram} \emph
  {et~al.}}]{article:ADMX-3}%
  \BibitemOpen
  \bibfield  {author} {\bibinfo {author} {\bibfnamefont {C.}~\bibnamefont
  {Bartram}} \emph {et~al.} (\bibinfo {collaboration} {ADMX}),\ }\bibfield
  {title} {\bibinfo {title} {Search for invisible axion dark matter in the
  $3.3-4.2\text{}\text{ }\ensuremath{\mu}\mathrm{eV}$ mass range},\ }\href
  {https://doi.org/10.1103/PhysRevLett.127.261803} {\bibfield  {journal}
  {\bibinfo  {journal} {Phys. Rev. Lett.}\ }\textbf {\bibinfo {volume} {127}},\
  \bibinfo {pages} {261803} (\bibinfo {year} {2021}{\natexlab{a}})}\BibitemShut
  {NoStop}%
\bibitem [{\citenamefont {Brubaker}\ \emph
  {et~al.}(2017{\natexlab{a}})\citenamefont {Brubaker} \emph
  {et~al.}}]{article:HAYSTAC-1}%
  \BibitemOpen
  \bibfield  {author} {\bibinfo {author} {\bibfnamefont {B.~M.}\ \bibnamefont
  {Brubaker}} \emph {et~al.},\ }\bibfield  {title} {\bibinfo {title} {First
  results from a microwave cavity axion search at $24\text{}\text{
  }\ensuremath{\mu}\mathrm{eV}$},\ }\href
  {https://doi.org/10.1103/PhysRevLett.118.061302} {\bibfield  {journal}
  {\bibinfo  {journal} {Phys. Rev. Lett.}\ }\textbf {\bibinfo {volume} {118}},\
  \bibinfo {pages} {061302} (\bibinfo {year} {2017}{\natexlab{a}})}\BibitemShut
  {NoStop}%
\bibitem [{\citenamefont {Backes}\ \emph
  {et~al.}(2021{\natexlab{a}})\citenamefont {Backes} \emph
  {et~al.}}]{article:HAYSTAC-2}%
  \BibitemOpen
  \bibfield  {author} {\bibinfo {author} {\bibfnamefont {K.}~\bibnamefont
  {Backes}} \emph {et~al.},\ }\bibfield  {title} {\bibinfo {title} {A quantum
  enhanced search for dark matter axions},\ }\href
  {https://doi.org/10.1038/s41586-021-03226-7} {\bibfield  {journal} {\bibinfo
  {journal} {Nature}\ }\textbf {\bibinfo {volume} {590}},\ \bibinfo {pages}
  {238} (\bibinfo {year} {2021}{\natexlab{a}})}\BibitemShut {NoStop}%
\bibitem [{\citenamefont {Jewell}\ \emph {et~al.}(2023)\citenamefont {Jewell}
  \emph {et~al.}}]{article:HAYSTAC-3}%
  \BibitemOpen
  \bibfield  {author} {\bibinfo {author} {\bibfnamefont {M.~J.}\ \bibnamefont
  {Jewell}} \emph {et~al.} (\bibinfo {collaboration} {HAYSTAC}),\ }\bibfield
  {title} {\bibinfo {title} {New results from {HAYSTAC}'s phase {II} operation
  with a squeezed state receiver},\ }\href
  {https://doi.org/10.1103/PhysRevD.107.072007} {\bibfield  {journal} {\bibinfo
   {journal} {Phys. Rev. D}\ }\textbf {\bibinfo {volume} {107}},\ \bibinfo
  {pages} {072007} (\bibinfo {year} {2023})}\BibitemShut {NoStop}%
\bibitem [{\citenamefont {Alesini}\ \emph {et~al.}(2019)\citenamefont {Alesini}
  \emph {et~al.}}]{article:QUAX-1}%
  \BibitemOpen
  \bibfield  {author} {\bibinfo {author} {\bibfnamefont {D.}~\bibnamefont
  {Alesini}} \emph {et~al.},\ }\bibfield  {title} {\bibinfo {title} {Galactic
  axions search with a superconducting resonant cavity},\ }\href
  {https://doi.org/10.1103/PhysRevD.99.101101} {\bibfield  {journal} {\bibinfo
  {journal} {Phys. Rev. D}\ }\textbf {\bibinfo {volume} {99}},\ \bibinfo
  {pages} {101101} (\bibinfo {year} {2019})}\BibitemShut {NoStop}%
\bibitem [{\citenamefont {Alesini}\ \emph {et~al.}(2021)\citenamefont {Alesini}
  \emph {et~al.}}]{article:QUAX-2}%
  \BibitemOpen
  \bibfield  {author} {\bibinfo {author} {\bibfnamefont {D.}~\bibnamefont
  {Alesini}} \emph {et~al.},\ }\bibfield  {title} {\bibinfo {title} {Search for
  invisible axion dark matter of mass ${\mathrm{m}}_{a}=43\text{}\text{
  }\ensuremath{\mu}\mathrm{eV}$ with the {QUAX}--$a\ensuremath{\gamma}$
  experiment},\ }\href {https://doi.org/10.1103/PhysRevD.103.102004} {\bibfield
   {journal} {\bibinfo  {journal} {Phys. Rev. D}\ }\textbf {\bibinfo {volume}
  {103}},\ \bibinfo {pages} {102004} (\bibinfo {year} {2021})}\BibitemShut
  {NoStop}%
\bibitem [{\citenamefont {McAllister}\ \emph {et~al.}(2017)\citenamefont
  {McAllister} \emph {et~al.}}]{article:ORGAN-1}%
  \BibitemOpen
  \bibfield  {author} {\bibinfo {author} {\bibfnamefont {B.~T.}\ \bibnamefont
  {McAllister}} \emph {et~al.},\ }\bibfield  {title} {\bibinfo {title} {The
  organ experiment: An axion haloscope above 15~{GHz}},\ }\href
  {https://doi.org/https://doi.org/10.1016/j.dark.2017.09.010} {\bibfield
  {journal} {\bibinfo  {journal} {Physics of the Dark Universe}\ }\textbf
  {\bibinfo {volume} {18}},\ \bibinfo {pages} {67} (\bibinfo {year}
  {2017})}\BibitemShut {NoStop}%
\bibitem [{\citenamefont {Quiskamp}\ \emph {et~al.}(2022)\citenamefont
  {Quiskamp} \emph {et~al.}}]{article:ORGAN-2}%
  \BibitemOpen
  \bibfield  {author} {\bibinfo {author} {\bibfnamefont {A.}~\bibnamefont
  {Quiskamp}} \emph {et~al.},\ }\bibfield  {title} {\bibinfo {title} {Direct
  search for dark matter axions excluding alp cogenesis in the $63\text{
  }\mathrm{to}\text{ }67\text{}\text{ }\ensuremath{\mu}\mathrm{eV}$ range with
  the organ experiment},\ }\href {https://doi.org/10.1126/sciadv.abq3765}
  {\bibfield  {journal} {\bibinfo  {journal} {Sci. Adv.}\ }\textbf {\bibinfo
  {volume} {8}},\ \bibinfo {pages} {27} (\bibinfo {year} {2022})}\BibitemShut
  {NoStop}%
\bibitem [{\citenamefont {Caldwell}\ \emph {et~al.}(2017)\citenamefont
  {Caldwell} \emph {et~al.}}]{article:MADMAX}%
  \BibitemOpen
  \bibfield  {author} {\bibinfo {author} {\bibfnamefont {A.}~\bibnamefont
  {Caldwell}} \emph {et~al.},\ }\bibfield  {title} {\bibinfo {title}
  {Dielectric haloscopes: A new way to detect axion dark matter},\ }\href
  {https://doi.org/10.1103/PhysRevLett.118.091801} {\bibfield  {journal}
  {\bibinfo  {journal} {Phys. Rev. Lett.}\ }\textbf {\bibinfo {volume} {118}},\
  \bibinfo {pages} {091801} (\bibinfo {year} {2017})}\BibitemShut {NoStop}%
\bibitem [{\citenamefont {Brouwer}\ \emph {et~al.}(2022)\citenamefont {Brouwer}
  \emph {et~al.}}]{article:DMRadio}%
  \BibitemOpen
  \bibfield  {author} {\bibinfo {author} {\bibfnamefont {L.}~\bibnamefont
  {Brouwer}} \emph {et~al.} (\bibinfo {collaboration} {DMRadio}),\ }\bibfield
  {title} {\bibinfo {title} {Proposal for a definitive search for {GUT}-scale
  {QCD} axions},\ }\href {https://doi.org/10.1103/PhysRevD.106.112003}
  {\bibfield  {journal} {\bibinfo  {journal} {Phys. Rev. D}\ }\textbf {\bibinfo
  {volume} {106}},\ \bibinfo {pages} {112003} (\bibinfo {year}
  {2022})}\BibitemShut {NoStop}%
\bibitem [{\citenamefont {Adair}\ \emph {et~al.}(2022)\citenamefont {Adair}
  \emph {et~al.}}]{article:CAST-CAPP}%
  \BibitemOpen
  \bibfield  {author} {\bibinfo {author} {\bibfnamefont {C.}~\bibnamefont
  {Adair}} \emph {et~al.} (\bibinfo {collaboration} {CAST-CAPP}),\ }\bibfield
  {title} {\bibinfo {title} {Search for dark matter axions with {CAST-CAPP}},\
  }\href {https://doi.org/10.1038/s41467-022-33913-6} {\bibfield  {journal}
  {\bibinfo  {journal} {Nature Communications}\ }\textbf {\bibinfo {volume}
  {13}},\ \bibinfo {pages} {6180} (\bibinfo {year} {2022})}\BibitemShut
  {NoStop}%
\bibitem [{\citenamefont {Anastassopoulos}\ \emph
  {et~al.}(2017{\natexlab{a}})\citenamefont {Anastassopoulos} \emph
  {et~al.}}]{article:CAST}%
  \BibitemOpen
  \bibfield  {author} {\bibinfo {author} {\bibfnamefont {V.}~\bibnamefont
  {Anastassopoulos}} \emph {et~al.} (\bibinfo {collaboration} {CAST}),\
  }\bibfield  {title} {\bibinfo {title} {New {CAST} limit on the axion–photon
  interaction},\ }\href {https://doi.org/10.1038/nphys4109} {\bibfield
  {journal} {\bibinfo  {journal} {Nature Phys.}\ }\textbf {\bibinfo {volume}
  {13}},\ \bibinfo {pages} {584} (\bibinfo {year}
  {2017}{\natexlab{a}})}\BibitemShut {NoStop}%
\bibitem [{\citenamefont {Armengaud}\ \emph {et~al.}(2019)\citenamefont
  {Armengaud} \emph {et~al.}}]{article:IAXO}%
  \BibitemOpen
  \bibfield  {author} {\bibinfo {author} {\bibfnamefont {E.}~\bibnamefont
  {Armengaud}} \emph {et~al.},\ }\bibfield  {title} {\bibinfo {title} {Physics
  potential of the {International Axion Observatory (IAXO)}},\ }\href
  {https://doi.org/10.1088/1475-7516/2019/06/047} {\bibfield  {journal}
  {\bibinfo  {journal} {J. Cosom. Astropart. Phys.}\ }\textbf {\bibinfo
  {volume} {06}},\ \bibinfo {pages} {047} (\bibinfo {year} {2019})}\BibitemShut
  {NoStop}%
\bibitem [{\citenamefont {Ballou}\ \emph {et~al.}(2015)\citenamefont {Ballou}
  \emph {et~al.}}]{article:OSQAR}%
  \BibitemOpen
  \bibfield  {author} {\bibinfo {author} {\bibfnamefont {R.}~\bibnamefont
  {Ballou}} \emph {et~al.} (\bibinfo {collaboration} {OSQAR}),\ }\bibfield
  {title} {\bibinfo {title} {New exclusion limits on scalar and pseudoscalar
  axionlike particles from light shining through a wall},\ }\href
  {https://doi.org/10.1103/PhysRevD.92.092002} {\bibfield  {journal} {\bibinfo
  {journal} {Phys. Rev. D}\ }\textbf {\bibinfo {volume} {92}},\ \bibinfo
  {pages} {092002} (\bibinfo {year} {2015})}\BibitemShut {NoStop}%
\bibitem [{\citenamefont {Ehret}\ \emph {et~al.}(2010)\citenamefont {Ehret}
  \emph {et~al.}}]{article:APLS}%
  \BibitemOpen
  \bibfield  {author} {\bibinfo {author} {\bibfnamefont {K.}~\bibnamefont
  {Ehret}} \emph {et~al.},\ }\bibfield  {title} {\bibinfo {title} {New {ALPS}
  results on hidden-sector lightweights},\ }\href
  {https://doi.org/https://doi.org/10.1016/j.physletb.2010.04.066} {\bibfield
  {journal} {\bibinfo  {journal} {Phys. Lett. B}\ }\textbf {\bibinfo {volume}
  {689}},\ \bibinfo {pages} {149} (\bibinfo {year} {2010})}\BibitemShut
  {NoStop}%
\bibitem [{\citenamefont {Baehre}\ \emph {et~al.}(2013)\citenamefont {Baehre}
  \emph {et~al.}}]{article:ALPSII}%
  \BibitemOpen
  \bibfield  {author} {\bibinfo {author} {\bibfnamefont {R.}~\bibnamefont
  {Baehre}} \emph {et~al.},\ }\bibfield  {title} {\bibinfo {title} {Any light
  particle search {II} — technical design report},\ }\href
  {https://doi.org/10.1088/1748-0221/8/09/T09001} {\bibfield  {journal}
  {\bibinfo  {journal} {Journal of Instrumentation}\ }\textbf {\bibinfo
  {volume} {8}},\ \bibinfo {pages} {T09001}}\BibitemShut {NoStop}%
\bibitem [{\citenamefont {Ma}\ \emph {et~al.}(2019)\citenamefont {Ma} \emph
  {et~al.}}]{OI-PAPER}%
  \BibitemOpen
  \bibfield  {author} {\bibinfo {author} {\bibfnamefont {W.}~\bibnamefont {Ma}}
  \emph {et~al.},\ }\bibfield  {title} {\bibinfo {title} {A new member of high
  field large bore superconducting research magnets family},\ }\href
  {https://doi.org/10.1088/1757-899X/502/1/012104} {\bibfield  {journal}
  {\bibinfo  {journal} {IOP Conf. Ser.}\ }\textbf {\bibinfo {volume} {502}},\
  \bibinfo {pages} {012104} (\bibinfo {year} {2019})}\BibitemShut {NoStop}%
\bibitem [{\citenamefont {Çağlar Kutlu}\ \emph {et~al.}(2021)\citenamefont
  {Çağlar Kutlu} \emph {et~al.}}]{article:Kutlu21}%
  \BibitemOpen
  \bibfield  {author} {\bibinfo {author} {\bibnamefont {Çağlar Kutlu}} \emph
  {et~al.},\ }\bibfield  {title} {\bibinfo {title} {Characterization of a
  flux-driven {Josephson} parametric amplifier with near quantum-limited added
  noise for axion search experiments},\ }\href
  {https://doi.org/10.1088/1361-6668/abf23b} {\bibfield  {journal} {\bibinfo
  {journal} {Supercond. Sci. Technol.}\ }\textbf {\bibinfo {volume} {34}},\
  \bibinfo {pages} {085013} (\bibinfo {year} {2021})}\BibitemShut {NoStop}%
\bibitem [{\citenamefont {Yi}\ \emph {et~al.}(2023{\natexlab{b}})\citenamefont
  {Yi} \emph {et~al.}}]{Sagitarius}%
  \BibitemOpen
  \bibfield  {author} {\bibinfo {author} {\bibfnamefont {A.~K.}\ \bibnamefont
  {Yi}} \emph {et~al.},\ }\bibfield  {title} {\bibinfo {title} {Search for the
  sagittarius tidal stream of axion dark matter around $4.55\text{}\text{
  }\mathrm{\ensuremath{\mu}e{v}}$},\ }\href
  {https://doi.org/10.1103/PhysRevD.108.L021304} {\bibfield  {journal}
  {\bibinfo  {journal} {Phys. Rev. D}\ }\textbf {\bibinfo {volume} {108}},\
  \bibinfo {pages} {L021304} (\bibinfo {year}
  {2023}{\natexlab{b}})}\BibitemShut {NoStop}%
\bibitem [{OI()}]{OI}%
  \BibitemOpen
  \href@noop {} {}\bibinfo {howpublished}
  {\url{https://www.oxinst.com}}\BibitemShut {NoStop}%
\bibitem [{\citenamefont {Schleeh}\ \emph {et~al.}(2012)\citenamefont
  {Schleeh}, \citenamefont {Alestig}, \citenamefont {Halonen}, \citenamefont
  {Malmros}, \citenamefont {Nilsson}, \citenamefont {Nilsson}, \citenamefont
  {Starski}, \citenamefont {Wadefalk}, \citenamefont {Zirath},\ and\
  \citenamefont {Grahn}}]{Shleeh2012}%
  \BibitemOpen
  \bibfield  {author} {\bibinfo {author} {\bibfnamefont {J.}~\bibnamefont
  {Schleeh}}, \bibinfo {author} {\bibfnamefont {G.}~\bibnamefont {Alestig}},
  \bibinfo {author} {\bibfnamefont {J.}~\bibnamefont {Halonen}}, \bibinfo
  {author} {\bibfnamefont {A.}~\bibnamefont {Malmros}}, \bibinfo {author}
  {\bibfnamefont {B.}~\bibnamefont {Nilsson}}, \bibinfo {author} {\bibfnamefont
  {P.~A.}\ \bibnamefont {Nilsson}}, \bibinfo {author} {\bibfnamefont {J.~P.}\
  \bibnamefont {Starski}}, \bibinfo {author} {\bibfnamefont {N.}~\bibnamefont
  {Wadefalk}}, \bibinfo {author} {\bibfnamefont {H.}~\bibnamefont {Zirath}},\
  and\ \bibinfo {author} {\bibfnamefont {J.}~\bibnamefont {Grahn}},\ }\bibfield
   {title} {\bibinfo {title} {Ultralow-power cryogenic {InP HEMT} with minimum
  noise temperature of 1 \uppercase{K} at 6 \uppercase{GH}z},\ }\href
  {https://doi.org/10.1109/LED.2012.2187422} {\bibfield  {journal} {\bibinfo
  {journal} {IEEE Electron Device Letters}\ }\textbf {\bibinfo {volume} {33}},\
  \bibinfo {pages} {664} (\bibinfo {year} {2012})}\BibitemShut {NoStop}%
\bibitem [{\citenamefont {Ivanov}\ \emph {et~al.}(2020)\citenamefont {Ivanov},
  \citenamefont {Volkhin}, \citenamefont {Novikov} \emph
  {et~al.}}]{Ivanov2020}%
  \BibitemOpen
  \bibfield  {author} {\bibinfo {author} {\bibfnamefont {B.~I.}\ \bibnamefont
  {Ivanov}}, \bibinfo {author} {\bibfnamefont {D.~I.}\ \bibnamefont {Volkhin}},
  \bibinfo {author} {\bibfnamefont {I.~L.}\ \bibnamefont {Novikov}}, \emph
  {et~al.},\ }\bibfield  {title} {\bibinfo {title} {A wideband cryogenic
  microwave low-noise amplifier},\ }\href
  {https://doi.org/10.3762/bjnano.11.131} {\bibfield  {journal} {\bibinfo
  {journal} {Beilstein Journal of Nanotechnology}\ }\textbf {\bibinfo {volume}
  {11}},\ \bibinfo {pages} {1484} (\bibinfo {year} {2020})}\BibitemShut
  {NoStop}%
\bibitem [{DRS()}]{DRS1000M}%
  \BibitemOpen
  \href@noop {} {}\bibinfo {howpublished}
  {\url{https://www.tokyoinst.co.jp/product_file/file/LCG02_cat01_ja.pdf}}\BibitemShut
  {NoStop}%
\bibitem [{Fis()}]{Fischer}%
  \BibitemOpen
  \href@noop {} {}\bibinfo {howpublished}
  {\url{https://fischerconnectors.com/en/circular-connectors/}}\BibitemShut
  {NoStop}%
\bibitem [{Api()}]{Apiezon}%
  \BibitemOpen
  \href@noop {} {}\bibinfo {howpublished}
  {\url{https://apiezon.com/products/vacuum-greases/apiezon-n-grease/}}\BibitemShut
  {NoStop}%
\bibitem [{LHe()}]{LHeP60}%
  \BibitemOpen
  \href@noop {} {}\bibinfo {howpublished}
  {\url{https://bluefors.com/products/liquid-helium-management-products/liquid-helium-plants/}}\BibitemShut
  {NoStop}%
\bibitem [{HeR()}]{HeRL60}%
  \BibitemOpen
  \href@noop {} {}\bibinfo {howpublished}
  {\url{https://bluefors.com/products/liquid-helium-management-products/helium-reliquefiers/}}\BibitemShut
  {NoStop}%
\bibitem [{COM()}]{COMSOL}%
  \BibitemOpen
  \href {https://www.comsol.com/} {\bibinfo {title} {Comsol multiphysics® v.
  6.1}}\BibitemShut {NoStop}%
\bibitem [{ANC()}]{ANC350}%
  \BibitemOpen
  \href@noop {} {}\bibinfo {howpublished}
  {\url{https://www.attocube.com}}\BibitemShut {NoStop}%
\bibitem [{\citenamefont {Ekin}(2006)}]{Ekin_cryo_book}%
  \BibitemOpen
  \bibfield  {author} {\bibinfo {author} {\bibfnamefont {J.}~\bibnamefont
  {Ekin}},\ }\href {https://doi.org/10.1093/acprof:oso/9780198570547.001.0001}
  {\emph {\bibinfo {title} {Experimental Techniques for Low-Temperature
  Measurements: Cryostat Design, Material Properties and Superconductor
  Critical-Current Testing}}}\ (\bibinfo  {publisher} {Oxford University
  Press},\ \bibinfo {year} {2006})\BibitemShut {NoStop}%
\bibitem [{\citenamefont {Zhong}\ \emph {et~al.}(2013)\citenamefont {Zhong},
  \citenamefont {Menzel}, \citenamefont {Candia} \emph {et~al.}}]{Zhong2013}%
  \BibitemOpen
  \bibfield  {author} {\bibinfo {author} {\bibfnamefont {L.}~\bibnamefont
  {Zhong}}, \bibinfo {author} {\bibfnamefont {E.}~\bibnamefont {Menzel}},
  \bibinfo {author} {\bibfnamefont {R.~D.}\ \bibnamefont {Candia}}, \emph
  {et~al.},\ }\bibfield  {title} {\bibinfo {title} {Squeezing with a
  flux-driven {Josephson} parametric amplifier},\ }\href
  {https://doi.org/10.1088/1367-2630/15/12/125013} {\bibfield  {journal}
  {\bibinfo  {journal} {New Journal Of Physics}\ }\textbf {\bibinfo {volume}
  {15}},\ \bibinfo {pages} {125013} (\bibinfo {year} {2013})}\BibitemShut
  {NoStop}%
\bibitem [{\citenamefont {Clerk}\ \emph {et~al.}(2010)\citenamefont {Clerk},
  \citenamefont {Devoret},\ and\ \citenamefont {Girvin}}]{Clerk2010}%
  \BibitemOpen
  \bibfield  {author} {\bibinfo {author} {\bibfnamefont {A.}~\bibnamefont
  {Clerk}}, \bibinfo {author} {\bibfnamefont {M.}~\bibnamefont {Devoret}},\
  and\ \bibinfo {author} {\bibfnamefont {S.}~\bibnamefont {Girvin}},\
  }\bibfield  {title} {\bibinfo {title} {Introduction to quantum noise,
  measurement, and amplification},\ }\href
  {https://doi.org/10.1103/RevModPhys.82.1155} {\bibfield  {journal} {\bibinfo
  {journal} {Rev. Mod. Phys.}\ }\textbf {\bibinfo {volume} {82}},\ \bibinfo
  {pages} {1155} (\bibinfo {year} {2010})}\BibitemShut {NoStop}%
\bibitem [{\citenamefont {Backes}\ \emph
  {et~al.}(2021{\natexlab{b}})\citenamefont {Backes}, \citenamefont {Palken},
  \citenamefont {Kenany} \emph {et~al.}}]{Backes2021}%
  \BibitemOpen
  \bibfield  {author} {\bibinfo {author} {\bibfnamefont {K.~M.}\ \bibnamefont
  {Backes}}, \bibinfo {author} {\bibfnamefont {D.~A.}\ \bibnamefont {Palken}},
  \bibinfo {author} {\bibfnamefont {S.~A.}\ \bibnamefont {Kenany}}, \emph
  {et~al.},\ }\bibfield  {title} {\bibinfo {title} {A quantum enhanced search
  for dark matter axions},\ }\href {https://doi.org/10.1038/s41586-021-03226-7}
  {\bibfield  {journal} {\bibinfo  {journal} {Nature}\ }\textbf {\bibinfo
  {volume} {590}},\ \bibinfo {pages} {238} (\bibinfo {year}
  {2021}{\natexlab{b}})}\BibitemShut {NoStop}%
\bibitem [{\citenamefont {Yamamoto}\ \emph {et~al.}(2008)\citenamefont
  {Yamamoto}, \citenamefont {Inomata}, \citenamefont {Watanabe} \emph
  {et~al.}}]{Yamamoto08}%
  \BibitemOpen
  \bibfield  {author} {\bibinfo {author} {\bibfnamefont {T.}~\bibnamefont
  {Yamamoto}}, \bibinfo {author} {\bibfnamefont {K.}~\bibnamefont {Inomata}},
  \bibinfo {author} {\bibfnamefont {M.}~\bibnamefont {Watanabe}}, \emph
  {et~al.},\ }\bibfield  {title} {\bibinfo {title} {Flux-driven {Josephson}
  parametric amplifier},\ }\href {https://doi.org/10.1063/1.2964182} {\bibfield
   {journal} {\bibinfo  {journal} {Applied Physics Letters}\ }\textbf {\bibinfo
  {volume} {93}},\ \bibinfo {pages} {042510} (\bibinfo {year}
  {2008})}\BibitemShut {NoStop}%
\bibitem [{\citenamefont {Roy}\ and\ \citenamefont {Devoret}(2016)}]{Roy2016}%
  \BibitemOpen
  \bibfield  {author} {\bibinfo {author} {\bibfnamefont {A.}~\bibnamefont
  {Roy}}\ and\ \bibinfo {author} {\bibfnamefont {M.}~\bibnamefont {Devoret}},\
  }\bibfield  {title} {\bibinfo {title} {Introduction to parametric
  amplification of quantum signals with {Josephson} circuits},\ }\href
  {https://doi.org/10.1016/j.crhy.2016.07.012} {\bibfield  {journal} {\bibinfo
  {journal} {Comptes Rendus Physique}\ }\textbf {\bibinfo {volume} {17}},\
  \bibinfo {pages} {740} (\bibinfo {year} {2016})}\BibitemShut {NoStop}%
\bibitem [{\citenamefont {Uchaikin}\ \emph
  {et~al.}(2023{\natexlab{a}})\citenamefont {Uchaikin}, \citenamefont {Ivanov},
  \citenamefont {Kim} \emph {et~al.}}]{Uchaikin23-LT29}%
  \BibitemOpen
  \bibfield  {author} {\bibinfo {author} {\bibfnamefont {S.~V.}\ \bibnamefont
  {Uchaikin}}, \bibinfo {author} {\bibfnamefont {B.~I.}\ \bibnamefont
  {Ivanov}}, \bibinfo {author} {\bibfnamefont {J.}~\bibnamefont {Kim}}, \emph
  {et~al.},\ }\bibinfo {title} {Josephson parametric amplifier in axion
  experiments},\ in\ \href {https://doi.org/10.7566/JPSCP.38.011201} {\emph
  {\bibinfo {booktitle} {Proceedings of the 29th International Conference on
  Low Temperature Physics (LT29)}}},\ Vol.~\bibinfo {volume} {38}\ (\bibinfo
  {year} {2023})\ p.\ \bibinfo {pages} {011201}\BibitemShut {NoStop}%
\bibitem [{ANS()}]{ANSYS}%
  \BibitemOpen
  \href {https://www.ansys.com/} {\bibinfo {title} {Ansys simulation
  software}}\BibitemShut {NoStop}%
\bibitem [{\citenamefont {Uchaikin}\ \emph
  {et~al.}(2023{\natexlab{b}})\citenamefont {Uchaikin} \emph
  {et~al.}}]{Uchaikin-LTD20}%
  \BibitemOpen
  \bibfield  {author} {\bibinfo {author} {\bibfnamefont {S.~V.}\ \bibnamefont
  {Uchaikin}} \emph {et~al.},\ }\bibfield  {title} {\bibinfo {title} {will be
  published soon},\ }\href@noop {} {\bibfield  {journal} {\bibinfo  {journal}
  {Proceedings of LTD20}\ }\textbf {\bibinfo {volume} {0}},\ \bibinfo {pages}
  {0} (\bibinfo {year} {2023}{\natexlab{b}})}\BibitemShut {NoStop}%
\bibitem [{LNF()}]{LNF-LNC0.6_2A}%
  \BibitemOpen
  \href@noop {} {}\bibinfo {howpublished}
  {\url{https://lownoisefactory.com/product/lnf-lnc0-6_2a/}}\BibitemShut
  {NoStop}%
\bibitem [{\citenamefont {Engen}(1970)}]{Engen70}%
  \BibitemOpen
  \bibfield  {author} {\bibinfo {author} {\bibfnamefont {G.~F.}\ \bibnamefont
  {Engen}},\ }\bibfield  {title} {\bibinfo {title} {A new method of
  characterizing amplifier noise performance},\ }\href
  {https://doi.org/10.1109/TIM.1970.4313925} {\bibfield  {journal} {\bibinfo
  {journal} {IEEE Trans. Instrum. Meas.}\ }\textbf {\bibinfo {volume} {19}},\
  \bibinfo {pages} {344} (\bibinfo {year} {1970})}\BibitemShut {NoStop}%
\bibitem [{\citenamefont {Ivanov}\ \emph {et~al.}(2023)\citenamefont {Ivanov},
  \citenamefont {Kim}, \citenamefont {\c{C}. Kutlu} \emph
  {et~al.}}]{Ivanov23-LT29}%
  \BibitemOpen
  \bibfield  {author} {\bibinfo {author} {\bibfnamefont {B.~I.}\ \bibnamefont
  {Ivanov}}, \bibinfo {author} {\bibfnamefont {J.}~\bibnamefont {Kim}},
  \bibinfo {author} {\bibnamefont {\c{C}. Kutlu}}, \emph {et~al.},\ }\bibinfo
  {title} {Four-channel system for characterization of {Josephson} parametric
  amplifiers},\ in\ \href {https://doi.org/10.7566/JPSCP.38.011200} {\emph
  {\bibinfo {booktitle} {Proceedings of the 29th International Conference on
  Low Temperature Physics (LT29)}}},\ Vol.~\bibinfo {volume} {38}\ (\bibinfo
  {year} {2023})\ p.\ \bibinfo {pages} {011200}\BibitemShut {NoStop}%
\bibitem [{\citenamefont {Kapitza}(1941)}]{Kapitza41}%
  \BibitemOpen
  \bibfield  {author} {\bibinfo {author} {\bibfnamefont {P.~L.}\ \bibnamefont
  {Kapitza}},\ }\bibfield  {title} {\bibinfo {title} {Heat transfer and
  superfluidity of {Helium II}},\ }\href
  {https://doi.org/10.1103/PhysRev.60.354} {\bibfield  {journal} {\bibinfo
  {journal} {Phys.Rev.}\ }\textbf {\bibinfo {volume} {60}},\ \bibinfo {pages}
  {354} (\bibinfo {year} {1941})}\BibitemShut {NoStop}%
\bibitem [{\citenamefont {Little}(1959)}]{Little59}%
  \BibitemOpen
  \bibfield  {author} {\bibinfo {author} {\bibfnamefont {W.~A.}\ \bibnamefont
  {Little}},\ }\bibfield  {title} {\bibinfo {title} {The transport of heat
  between disimilar solids at low tempeartures},\ }\href
  {https://doi.org/https://doi.org/10.1139/p59-037} {\bibfield  {journal}
  {\bibinfo  {journal} {Canadian Journal of Physics}\ }\textbf {\bibinfo
  {volume} {37}},\ \bibinfo {pages} {334} (\bibinfo {year} {1959})}\BibitemShut
  {NoStop}%
\bibitem [{\citenamefont {Lee}(2017)}]{jphysconfser_898_032035_2017}%
  \BibitemOpen
  \bibfield  {author} {\bibinfo {author} {\bibfnamefont {S.}~\bibnamefont
  {Lee}},\ }\bibfield  {title} {\bibinfo {title} {Development of a data
  acquisition software for the {CULTASK} experiment},\ }\href
  {https://doi.org/10.1088/1742-6596/898/3/032035} {\bibfield  {journal}
  {\bibinfo  {journal} {J. Phys.: Conf. Ser.}\ }\textbf {\bibinfo {volume}
  {898}},\ \bibinfo {pages} {032035} (\bibinfo {year} {2017})}\BibitemShut
  {NoStop}%
\bibitem [{\citenamefont {Brun}\ and\ \citenamefont {Rademakers}(1997)}]{ROOT}%
  \BibitemOpen
  \bibfield  {author} {\bibinfo {author} {\bibfnamefont {R.}~\bibnamefont
  {Brun}}\ and\ \bibinfo {author} {\bibfnamefont {F.}~\bibnamefont
  {Rademakers}},\ }\bibfield  {title} {\bibinfo {title} {{ROOT} — an object
  oriented data analysis framework},\ }\href
  {https://doi.org/10.1016/S0168-9002(97)00048-X} {\bibfield  {journal}
  {\bibinfo  {journal} {Nucl. Instrum. Meth. Phys. Res. A}\ }\textbf {\bibinfo
  {volume} {389}},\ \bibinfo {pages} {81} (\bibinfo {year} {1997})}\BibitemShut
  {NoStop}%
\bibitem [{spe()}]{spectrum_instrumentation}%
  \BibitemOpen
  \href@noop {} {}\bibinfo {howpublished}
  {\url{https://spectrum-instrumentation.com}}\BibitemShut {NoStop}%
\bibitem [{\citenamefont {Ahn}\ \emph {et~al.}(2022{\natexlab{a}})\citenamefont
  {Ahn}, \citenamefont {Lee}, \citenamefont {Yi}, \citenamefont {Yeo},
  \citenamefont {Ko},\ and\ \citenamefont
  {Semertzidis}}]{jinst_17_p05025_2022}%
  \BibitemOpen
  \bibfield  {author} {\bibinfo {author} {\bibfnamefont {S.}~\bibnamefont
  {Ahn}}, \bibinfo {author} {\bibfnamefont {M.~J.}\ \bibnamefont {Lee}},
  \bibinfo {author} {\bibfnamefont {A.~K.}\ \bibnamefont {Yi}}, \bibinfo
  {author} {\bibfnamefont {B.}~\bibnamefont {Yeo}}, \bibinfo {author}
  {\bibfnamefont {B.~R.}\ \bibnamefont {Ko}},\ and\ \bibinfo {author}
  {\bibfnamefont {Y.~K.}\ \bibnamefont {Semertzidis}},\ }\bibfield  {title}
  {\bibinfo {title} {Fast {DAQ} system with image rejection for axion dark
  matter searches},\ }\href {https://doi.org/10.1088/1748-0221/17/05/P05025}
  {\bibfield  {journal} {\bibinfo  {journal} {J. Inst.}\ }\textbf {\bibinfo
  {volume} {17}},\ \bibinfo {pages} {P05025} (\bibinfo {year}
  {2022}{\natexlab{a}})}\BibitemShut {NoStop}%
\bibitem [{\citenamefont {Caspers}()}]{Fritz}%
  \BibitemOpen
  \bibfield  {author} {\bibinfo {author} {\bibfnamefont {F.}~\bibnamefont
  {Caspers}},\ }\href@noop {} {}\bibinfo {howpublished} {private communication,
  2015.}\BibitemShut {Stop}%
\bibitem [{\citenamefont {Vogelsberger}\ and\ \citenamefont
  {White}(2011)}]{article:stream1}%
  \BibitemOpen
  \bibfield  {author} {\bibinfo {author} {\bibfnamefont {M.}~\bibnamefont
  {Vogelsberger}}\ and\ \bibinfo {author} {\bibfnamefont {S.~D.~M.}\
  \bibnamefont {White}},\ }\bibfield  {title} {\bibinfo {title} {Streams and
  caustics: the fine-grained structure of $\ensuremath{\Lambda}$ cold dark
  matter haloes},\ }\href {https://doi.org/10.1111/j.1365-2966.2011.18224.x}
  {\bibfield  {journal} {\bibinfo  {journal} {Monthly Notices of the Royal
  Astronomical Society}\ }\textbf {\bibinfo {volume} {413}},\ \bibinfo {pages}
  {1419} (\bibinfo {year} {2011})},\ \Eprint
  {https://arxiv.org/abs/https://academic.oup.com/mnras/article-pdf/413/2/1419/18595473/mnras0413-1419.pdf}
  {https://academic.oup.com/mnras/article-pdf/413/2/1419/18595473/mnras0413-1419.pdf}
  \BibitemShut {NoStop}%
\bibitem [{\citenamefont {Hoffmann}\ \emph {et~al.}(2003)\citenamefont
  {Hoffmann}, \citenamefont {Jacoby},\ and\ \citenamefont
  {Zioutas}}]{article:KZ1}%
  \BibitemOpen
  \bibfield  {author} {\bibinfo {author} {\bibfnamefont {D.~H.~H.}\
  \bibnamefont {Hoffmann}}, \bibinfo {author} {\bibfnamefont {J.}~\bibnamefont
  {Jacoby}},\ and\ \bibinfo {author} {\bibfnamefont {K.}~\bibnamefont
  {Zioutas}},\ }\bibfield  {title} {\bibinfo {title} {Gravitational lensing by
  the sun of non-relativistic penetrating particles},\ }\href
  {https://doi.org/https://doi.org/10.1016/S0927-6505(03)00138-5} {\bibfield
  {journal} {\bibinfo  {journal} {Astroparticle Physics}\ }\textbf {\bibinfo
  {volume} {20}},\ \bibinfo {pages} {73} (\bibinfo {year} {2003})}\BibitemShut
  {NoStop}%
\bibitem [{\citenamefont {Patla}\ \emph {et~al.}(2013)\citenamefont {Patla}
  \emph {et~al.}}]{article:KZ2}%
  \BibitemOpen
  \bibfield  {author} {\bibinfo {author} {\bibfnamefont {B.~R.}\ \bibnamefont
  {Patla}} \emph {et~al.},\ }\bibfield  {title} {\bibinfo {title} {Flux
  enhancement of slow-moving particles by sun or {Jupyter}: Can they be
  detected on {Earth}?},\ }\href {https://doi.org/10.1088/0004-637X/780/2/158}
  {\bibfield  {journal} {\bibinfo  {journal} {The Astrophysical Journal}\
  }\textbf {\bibinfo {volume} {780}},\ \bibinfo {pages} {158} (\bibinfo {year}
  {2013})}\BibitemShut {NoStop}%
\bibitem [{\citenamefont {Zioutas}\ \emph {et~al.}(2017)\citenamefont {Zioutas}
  \emph {et~al.}}]{article:zioutas2017}%
  \BibitemOpen
  \bibfield  {author} {\bibinfo {author} {\bibfnamefont {K.}~\bibnamefont
  {Zioutas}} \emph {et~al.},\ }\href
  {https://doi.org/https://doi.org/10.48550/arXiv.1703.01436} {\bibinfo {title}
  {Search for axions in streaming dark matter}} (\bibinfo {year} {2017}),\
  \Eprint {https://arxiv.org/abs/1703.01436} {arXiv:1703.01436
  [physics.ins-det]} \BibitemShut {NoStop}%
\bibitem [{\citenamefont {Fischer}\ \emph {et~al.}(2017)\citenamefont
  {Fischer}, \citenamefont {Semertzidis},\ and\ \citenamefont
  {Zioutas}}]{Fischer2017}%
  \BibitemOpen
  \bibfield  {author} {\bibinfo {author} {\bibfnamefont {H.}~\bibnamefont
  {Fischer}}, \bibinfo {author} {\bibfnamefont {Y.}~\bibnamefont
  {Semertzidis}},\ and\ \bibinfo {author} {\bibfnamefont {K.}~\bibnamefont
  {Zioutas}},\ }\bibfield  {title} {\bibinfo {title} {Search for axions in
  streaming dark matter},\ }\href
  {https://ep-news.web.cern.ch/content/search-axions-streaming-dark-matter} {\
  (\bibinfo {year} {2017})},\ \bibinfo {note}
  {\url{https://ep-news.web.cern.ch/content/search-axions-streaming-dark-matter}}\BibitemShut
  {NoStop}%
\bibitem [{Fre()}]{FrequencyDistribution}%
  \BibitemOpen
  \href@noop {} {}\bibinfo {howpublished}
  {\url{https://www.msit.go.kr/bbs/view.do?bbsSeqNo=83&nttSeqNo=3175660}}\BibitemShut
  {NoStop}%
\bibitem [{\citenamefont {Ahn}\ \emph {et~al.}(2021)\citenamefont {Ahn} \emph
  {et~al.}}]{article:Ahn-anal}%
  \BibitemOpen
  \bibfield  {author} {\bibinfo {author} {\bibfnamefont {S.}~\bibnamefont
  {Ahn}} \emph {et~al.},\ }\bibfield  {title} {\bibinfo {title} {Improved axion
  haloscope search analysis},\ }\href {https://doi.org/10.1007/JHEP04(2021)297}
  {\bibfield  {journal} {\bibinfo  {journal} {Journal of High Energy Physics}\
  }\textbf {\bibinfo {volume} {2021}},\ \bibinfo {pages} {297} (\bibinfo {year}
  {2021})}\BibitemShut {NoStop}%
\bibitem [{\citenamefont {Bartram}\ \emph
  {et~al.}(2021{\natexlab{b}})\citenamefont {Bartram} \emph
  {et~al.}}]{article:ADMX-anal}%
  \BibitemOpen
  \bibfield  {author} {\bibinfo {author} {\bibfnamefont {C.}~\bibnamefont
  {Bartram}} \emph {et~al.} (\bibinfo {collaboration} {ADMX Collaboration}),\
  }\bibfield  {title} {\bibinfo {title} {Axion dark matter experiment: Run {1B}
  analysis details},\ }\href {https://doi.org/10.1103/PhysRevD.103.032002}
  {\bibfield  {journal} {\bibinfo  {journal} {Phys. Rev. D}\ }\textbf {\bibinfo
  {volume} {103}},\ \bibinfo {pages} {032002} (\bibinfo {year}
  {2021}{\natexlab{b}})}\BibitemShut {NoStop}%
\bibitem [{\citenamefont {Brubaker}\ \emph
  {et~al.}(2017{\natexlab{b}})\citenamefont {Brubaker}, \citenamefont {Zhong},
  \citenamefont {Lamoreaux}, \citenamefont {Lehnert},\ and\ \citenamefont {van
  Bibber}}]{article:HAYSTAC-anal}%
  \BibitemOpen
  \bibfield  {author} {\bibinfo {author} {\bibfnamefont {B.~M.}\ \bibnamefont
  {Brubaker}}, \bibinfo {author} {\bibfnamefont {L.}~\bibnamefont {Zhong}},
  \bibinfo {author} {\bibfnamefont {S.~K.}\ \bibnamefont {Lamoreaux}}, \bibinfo
  {author} {\bibfnamefont {K.~W.}\ \bibnamefont {Lehnert}},\ and\ \bibinfo
  {author} {\bibfnamefont {K.~A.}\ \bibnamefont {van Bibber}},\ }\bibfield
  {title} {\bibinfo {title} {{HAYSTAC} axion search analysis procedure},\
  }\href {https://doi.org/10.1103/PhysRevD.96.123008} {\bibfield  {journal}
  {\bibinfo  {journal} {Phys. Rev. D}\ }\textbf {\bibinfo {volume} {96}},\
  \bibinfo {pages} {123008} (\bibinfo {year} {2017}{\natexlab{b}})}\BibitemShut
  {NoStop}%
\bibitem [{\citenamefont {Chang}\ \emph
  {et~al.}(2022{\natexlab{a}})\citenamefont {Chang} \emph
  {et~al.}}]{article:TASEH-anal}%
  \BibitemOpen
  \bibfield  {author} {\bibinfo {author} {\bibfnamefont {H.}~\bibnamefont
  {Chang}} \emph {et~al.} (\bibinfo {collaboration} {TASEH Collaboration}),\
  }\bibfield  {title} {\bibinfo {title} {Taiwan axion search experiment with
  haloscope: {CD102} analysis details},\ }\href
  {https://doi.org/10.1103/PhysRevD.106.052002} {\bibfield  {journal} {\bibinfo
   {journal} {Phys. Rev. D}\ }\textbf {\bibinfo {volume} {106}},\ \bibinfo
  {pages} {052002} (\bibinfo {year} {2022}{\natexlab{a}})}\BibitemShut
  {NoStop}%
\bibitem [{\citenamefont {Palken}\ \emph {et~al.}(2020)\citenamefont {Palken},
  \citenamefont {Brubaker}, \citenamefont {Malnou} \emph
  {et~al.}}]{article:bayesian_method}%
  \BibitemOpen
  \bibfield  {author} {\bibinfo {author} {\bibfnamefont {D.~A.}\ \bibnamefont
  {Palken}}, \bibinfo {author} {\bibfnamefont {B.~M.}\ \bibnamefont
  {Brubaker}}, \bibinfo {author} {\bibfnamefont {M.}~\bibnamefont {Malnou}},
  \emph {et~al.},\ }\bibfield  {title} {\bibinfo {title} {Improved analysis
  framework for axion dark matter searches},\ }\href
  {https://doi.org/10.1103/PhysRevD.101.123011} {\bibfield  {journal} {\bibinfo
   {journal} {Phys. Rev. D}\ }\textbf {\bibinfo {volume} {101}},\ \bibinfo
  {pages} {123011} (\bibinfo {year} {2020})}\BibitemShut {NoStop}%
\bibitem [{\citenamefont {Dolan}\ \emph {et~al.}(2022)\citenamefont {Dolan},
  \citenamefont {Hiskens},\ and\ \citenamefont {Volkas}}]{Dolan_2022}%
  \BibitemOpen
  \bibfield  {author} {\bibinfo {author} {\bibfnamefont {M.~J.}\ \bibnamefont
  {Dolan}}, \bibinfo {author} {\bibfnamefont {F.~J.}\ \bibnamefont {Hiskens}},\
  and\ \bibinfo {author} {\bibfnamefont {R.~R.}\ \bibnamefont {Volkas}},\
  }\bibfield  {title} {\bibinfo {title} {Advancing globular cluster constraints
  on the axion-photon coupling},\ }\href
  {https://doi.org/10.1088/1475-7516/2022/10/096} {\bibfield  {journal}
  {\bibinfo  {journal} {Journal of Cosmology and Astroparticle Physics}\
  }\textbf {\bibinfo {volume} {2022}}\bibinfo  {number} { (10)},\ \bibinfo
  {pages} {096}}\BibitemShut {NoStop}%
\bibitem [{\citenamefont {Foster}\ \emph {et~al.}(2020)\citenamefont {Foster},
  \citenamefont {Kahn}, \citenamefont {Macias}, \citenamefont {Sun},
  \citenamefont {Eatough}, \citenamefont {Kondratiev}, \citenamefont {Peters},
  \citenamefont {Weniger},\ and\ \citenamefont
  {Safdi}}]{PhysRevLett.125.171301}%
  \BibitemOpen
\bibfield  {number} {  }\bibfield  {author} {\bibinfo {author} {\bibfnamefont
  {J.~W.}\ \bibnamefont {Foster}}, \bibinfo {author} {\bibfnamefont
  {Y.}~\bibnamefont {Kahn}}, \bibinfo {author} {\bibfnamefont {O.}~\bibnamefont
  {Macias}}, \bibinfo {author} {\bibfnamefont {Z.}~\bibnamefont {Sun}},
  \bibinfo {author} {\bibfnamefont {R.~P.}\ \bibnamefont {Eatough}}, \bibinfo
  {author} {\bibfnamefont {V.~I.}\ \bibnamefont {Kondratiev}}, \bibinfo
  {author} {\bibfnamefont {W.~M.}\ \bibnamefont {Peters}}, \bibinfo {author}
  {\bibfnamefont {C.}~\bibnamefont {Weniger}},\ and\ \bibinfo {author}
  {\bibfnamefont {B.~R.}\ \bibnamefont {Safdi}},\ }\bibfield  {title} {\bibinfo
  {title} {Green bank and effelsberg radio telescope searches for axion dark
  matter conversion in neutron star magnetospheres},\ }\href
  {https://doi.org/10.1103/PhysRevLett.125.171301} {\bibfield  {journal}
  {\bibinfo  {journal} {Phys. Rev. Lett.}\ }\textbf {\bibinfo {volume} {125}},\
  \bibinfo {pages} {171301} (\bibinfo {year} {2020})}\BibitemShut {NoStop}%
\bibitem [{\citenamefont {Darling}(2020)}]{Darling_2020}%
  \BibitemOpen
  \bibfield  {author} {\bibinfo {author} {\bibfnamefont {J.}~\bibnamefont
  {Darling}},\ }\bibfield  {title} {\bibinfo {title} {New limits on axionic
  dark matter from the magnetar {PSR J1745-2900}},\ }\href
  {https://doi.org/10.3847/2041-8213/abb23f} {\bibfield  {journal} {\bibinfo
  {journal} {The Astrophysical Journal Letters}\ }\textbf {\bibinfo {volume}
  {900}},\ \bibinfo {pages} {L28} (\bibinfo {year} {2020})}\BibitemShut
  {NoStop}%
\bibitem [{\citenamefont {Foster}\ \emph {et~al.}(2022)\citenamefont {Foster},
  \citenamefont {Witte}, \citenamefont {Lawson}, \citenamefont {Linden},
  \citenamefont {Gajjar}, \citenamefont {Weniger},\ and\ \citenamefont
  {Safdi}}]{PhysRevLett.129.251102}%
  \BibitemOpen
  \bibfield  {author} {\bibinfo {author} {\bibfnamefont {J.~W.}\ \bibnamefont
  {Foster}}, \bibinfo {author} {\bibfnamefont {S.~J.}\ \bibnamefont {Witte}},
  \bibinfo {author} {\bibfnamefont {M.}~\bibnamefont {Lawson}}, \bibinfo
  {author} {\bibfnamefont {T.}~\bibnamefont {Linden}}, \bibinfo {author}
  {\bibfnamefont {V.}~\bibnamefont {Gajjar}}, \bibinfo {author} {\bibfnamefont
  {C.}~\bibnamefont {Weniger}},\ and\ \bibinfo {author} {\bibfnamefont {B.~R.}\
  \bibnamefont {Safdi}},\ }\bibfield  {title} {\bibinfo {title}
  {Extraterrestrial axion search with the breakthrough listen galactic center
  survey},\ }\href {https://doi.org/10.1103/PhysRevLett.129.251102} {\bibfield
  {journal} {\bibinfo  {journal} {Phys. Rev. Lett.}\ }\textbf {\bibinfo
  {volume} {129}},\ \bibinfo {pages} {251102} (\bibinfo {year}
  {2022})}\BibitemShut {NoStop}%
\bibitem [{\citenamefont {Battye}\ \emph {et~al.}(2023)\citenamefont {Battye},
  \citenamefont {Keith}, \citenamefont {McDonald}, \citenamefont {Srinivasan},
  \citenamefont {Stappers},\ and\ \citenamefont
  {Weltevrede}}]{battye2023searching}%
  \BibitemOpen
  \bibfield  {author} {\bibinfo {author} {\bibfnamefont {R.~A.}\ \bibnamefont
  {Battye}}, \bibinfo {author} {\bibfnamefont {M.~J.}\ \bibnamefont {Keith}},
  \bibinfo {author} {\bibfnamefont {J.~I.}\ \bibnamefont {McDonald}}, \bibinfo
  {author} {\bibfnamefont {S.}~\bibnamefont {Srinivasan}}, \bibinfo {author}
  {\bibfnamefont {B.~W.}\ \bibnamefont {Stappers}},\ and\ \bibinfo {author}
  {\bibfnamefont {P.}~\bibnamefont {Weltevrede}},\ }\href@noop {} {\bibinfo
  {title} {Searching for time-dependent axion dark matter signals in pulsars}}
  (\bibinfo {year} {2023}),\ \Eprint {https://arxiv.org/abs/2303.11792}
  {arXiv:2303.11792 [astro-ph.CO]} \BibitemShut {NoStop}%
\bibitem [{\citenamefont {DePanfilis}\ \emph {et~al.}(1987)\citenamefont
  {DePanfilis}, \citenamefont {Melissinos}, \citenamefont {Moskowitz} \emph
  {et~al.}}]{article:Panfilis1987}%
  \BibitemOpen
  \bibfield  {author} {\bibinfo {author} {\bibfnamefont {S.}~\bibnamefont
  {DePanfilis}}, \bibinfo {author} {\bibfnamefont {A.~C.}\ \bibnamefont
  {Melissinos}}, \bibinfo {author} {\bibfnamefont {B.~E.}\ \bibnamefont
  {Moskowitz}}, \emph {et~al.},\ }\bibfield  {title} {\bibinfo {title} {Limits
  on the abundance and coupling of cosmic axions at
  $4.5<\ensuremath{m_a}<5.0\text{}\text{ }\ensuremath{\mu}\mathrm{eV}$},\
  }\href {https://doi.org/10.1103/PhysRevLett.59.839} {\bibfield  {journal}
  {\bibinfo  {journal} {Phys. Rev. Lett.}\ }\textbf {\bibinfo {volume} {59}},\
  \bibinfo {pages} {839} (\bibinfo {year} {1987})}\BibitemShut {NoStop}%
\bibitem [{\citenamefont {Hagmann}\ \emph {et~al.}(1990)\citenamefont
  {Hagmann}, \citenamefont {Sikivie}, \citenamefont {Sullivan},\ and\
  \citenamefont {Tanner}}]{article:Hagmann1990}%
  \BibitemOpen
  \bibfield  {author} {\bibinfo {author} {\bibfnamefont {C.}~\bibnamefont
  {Hagmann}}, \bibinfo {author} {\bibfnamefont {P.}~\bibnamefont {Sikivie}},
  \bibinfo {author} {\bibfnamefont {N.~S.}\ \bibnamefont {Sullivan}},\ and\
  \bibinfo {author} {\bibfnamefont {D.~B.}\ \bibnamefont {Tanner}},\ }\bibfield
   {title} {\bibinfo {title} {Results from a search for cosmic axions},\ }\href
  {https://doi.org/10.1103/PhysRevD.42.1297} {\bibfield  {journal} {\bibinfo
  {journal} {Phys. Rev. D}\ }\textbf {\bibinfo {volume} {42}},\ \bibinfo
  {pages} {1297} (\bibinfo {year} {1990})}\BibitemShut {NoStop}%
\bibitem [{\citenamefont {Anastassopoulos}\ \emph
  {et~al.}(2017{\natexlab{b}})\citenamefont {Anastassopoulos}, \citenamefont
  {Aune}, \citenamefont {Barth} \emph {et~al.}}]{Anastassopoulos2017}%
  \BibitemOpen
  \bibfield  {author} {\bibinfo {author} {\bibfnamefont {V.}~\bibnamefont
  {Anastassopoulos}}, \bibinfo {author} {\bibfnamefont {S.}~\bibnamefont
  {Aune}}, \bibinfo {author} {\bibfnamefont {K.}~\bibnamefont {Barth}}, \emph
  {et~al.},\ }\bibfield  {title} {\bibinfo {title} {New cast limit on the
  axion--photon interaction},\ }\href {https://doi.org/10.1038/nphys4109}
  {\bibfield  {journal} {\bibinfo  {journal} {Nature Physics}\ }\textbf
  {\bibinfo {volume} {13}},\ \bibinfo {pages} {584} (\bibinfo {year}
  {2017}{\natexlab{b}})}\BibitemShut {NoStop}%
\bibitem [{\citenamefont {Yang}\ \emph {et~al.}(2023)\citenamefont {Yang},
  \citenamefont {Yoon}, \citenamefont {Ahn}, \citenamefont {Lee},\ and\
  \citenamefont {Yoo}}]{article:CAPP18T2023}%
  \BibitemOpen
  \bibfield  {author} {\bibinfo {author} {\bibfnamefont {B.}~\bibnamefont
  {Yang}}, \bibinfo {author} {\bibfnamefont {H.}~\bibnamefont {Yoon}}, \bibinfo
  {author} {\bibfnamefont {M.}~\bibnamefont {Ahn}}, \bibinfo {author}
  {\bibfnamefont {Y.}~\bibnamefont {Lee}},\ and\ \bibinfo {author}
  {\bibfnamefont {J.}~\bibnamefont {Yoo}},\ }\bibfield  {title} {\bibinfo
  {title} {Extended axion dark matter search using the {CAPP18T} haloscope},\
  }\href {https://doi.org/10.1103/PhysRevLett.131.081801} {\bibfield  {journal}
  {\bibinfo  {journal} {Phys. Rev. Lett.}\ }\textbf {\bibinfo {volume} {131}},\
  \bibinfo {pages} {081801} (\bibinfo {year} {2023})}\BibitemShut {NoStop}%
\bibitem [{\citenamefont {Lee}\ \emph {et~al.}(2022)\citenamefont {Lee},
  \citenamefont {Yang}, \citenamefont {Yoon} \emph {et~al.}}]{CAPP-18T}%
  \BibitemOpen
  \bibfield  {author} {\bibinfo {author} {\bibfnamefont {Y.}~\bibnamefont
  {Lee}}, \bibinfo {author} {\bibfnamefont {B.}~\bibnamefont {Yang}}, \bibinfo
  {author} {\bibfnamefont {H.}~\bibnamefont {Yoon}}, \emph {et~al.},\
  }\bibfield  {title} {\bibinfo {title} {Searching for invisible axion dark
  matter with an {18 T} magnet haloscope},\ }\href
  {https://doi.org/10.1103/PhysRevLett.128.241805} {\bibfield  {journal}
  {\bibinfo  {journal} {Phys. Rev. Lett.}\ }\textbf {\bibinfo {volume} {128}},\
  \bibinfo {pages} {241805} (\bibinfo {year} {2022})}\BibitemShut {NoStop}%
\bibitem [{\citenamefont {Chang}\ \emph
  {et~al.}(2022{\natexlab{b}})\citenamefont {Chang} \emph {et~al.}}]{TASEH-1}%
  \BibitemOpen
  \bibfield  {author} {\bibinfo {author} {\bibfnamefont {H.}~\bibnamefont
  {Chang}} \emph {et~al.} (\bibinfo {collaboration} {TASEH Collaboration}),\
  }\bibfield  {title} {\bibinfo {title} {First results from the taiwan axion
  search experiment with a haloscope at $19.6\text{}\text{
  }\ensuremath{\mu}\mathrm{eV}$},\ }\href
  {https://doi.org/10.1103/PhysRevLett.129.111802} {\bibfield  {journal}
  {\bibinfo  {journal} {Phys. Rev. Lett.}\ }\textbf {\bibinfo {volume} {129}},\
  \bibinfo {pages} {111802} (\bibinfo {year} {2022}{\natexlab{b}})}\BibitemShut
  {NoStop}%
\bibitem [{\citenamefont {Grenet}\ \emph {et~al.}(2021)\citenamefont {Grenet},
  \citenamefont {Ballou}, \citenamefont {Basto} \emph
  {et~al.}}]{grenet2021grenoble}%
  \BibitemOpen
  \bibfield  {author} {\bibinfo {author} {\bibfnamefont {T.}~\bibnamefont
  {Grenet}}, \bibinfo {author} {\bibfnamefont {R.}~\bibnamefont {Ballou}},
  \bibinfo {author} {\bibfnamefont {Q.}~\bibnamefont {Basto}}, \emph {et~al.},\
  }\href@noop {} {\bibinfo {title} {The {Grenoble} axion haloscope platform
  ({GrAHal}): development plan and first results}} (\bibinfo {year} {2021}),\
  \Eprint {https://arxiv.org/abs/2110.14406} {arXiv:2110.14406 [hep-ex]}
  \BibitemShut {NoStop}%
\bibitem [{\citenamefont {O'Hare}(2020)}]{AxionLimits}%
  \BibitemOpen
  \bibfield  {author} {\bibinfo {author} {\bibfnamefont {C.}~\bibnamefont
  {O'Hare}},\ }\href {https://doi.org/10.5281/zenodo.3932430} {\bibinfo {title}
  {cajohare/axionlimits: Axionlimits}},\ \bibinfo {howpublished}
  {\url{https://cajohare.github.io/AxionLimits/}} (\bibinfo {year}
  {2020})\BibitemShut {NoStop}%
\bibitem [{\citenamefont {Ahn}\ \emph {et~al.}(2019{\natexlab{a}})\citenamefont
  {Ahn} \emph {et~al.}}]{Danho19_1}%
  \BibitemOpen
  \bibfield  {author} {\bibinfo {author} {\bibfnamefont {D.}~\bibnamefont
  {Ahn}} \emph {et~al.},\ }\bibfield  {title} {\bibinfo {title} {High quality
  factor high-temperature superconducting microwave cavity development for the
  dark matter axion search in a strong magnetic field},\ }\bibfield  {journal}
  {\bibinfo  {journal} {ArXiv}\ }\href
  {https://doi.org/10.48550/arXiv.1902.04551} {10.48550/arXiv.1902.04551}
  (\bibinfo {year} {2019}{\natexlab{a}}),\ \Eprint
  {https://arxiv.org/abs/1902.04551} {1902.04551} \BibitemShut {NoStop}%
\bibitem [{\citenamefont {Ahn}\ \emph {et~al.}(2019{\natexlab{b}})\citenamefont
  {Ahn}, \citenamefont {Kwon}, \citenamefont {Chung}, \citenamefont {Jang},
  \citenamefont {Lee}, \citenamefont {Lee}, \citenamefont {Youn}, \citenamefont
  {Youm},\ and\ \citenamefont {Semertzidis}}]{Danho19_2}%
  \BibitemOpen
  \bibfield  {author} {\bibinfo {author} {\bibfnamefont {D.}~\bibnamefont
  {Ahn}}, \bibinfo {author} {\bibfnamefont {O.}~\bibnamefont {Kwon}}, \bibinfo
  {author} {\bibfnamefont {W.}~\bibnamefont {Chung}}, \bibinfo {author}
  {\bibfnamefont {W.}~\bibnamefont {Jang}}, \bibinfo {author} {\bibfnamefont
  {D.}~\bibnamefont {Lee}}, \bibinfo {author} {\bibfnamefont {J.}~\bibnamefont
  {Lee}}, \bibinfo {author} {\bibfnamefont {S.~W.}\ \bibnamefont {Youn}},
  \bibinfo {author} {\bibfnamefont {D.}~\bibnamefont {Youm}},\ and\ \bibinfo
  {author} {\bibfnamefont {Y.~K.}\ \bibnamefont {Semertzidis}},\ }\href@noop {}
  {\bibinfo {title} {Maintaining high {Q}-factor of superconducting
  {YBa$_2$Cu$_3$O$_{7-x}$} microwave cavity in a high magnetic field}}
  (\bibinfo {year} {2019}{\natexlab{b}}),\ \Eprint
  {https://arxiv.org/abs/1904.05111} {arXiv:1904.05111 [physics.ins-det]}
  \BibitemShut {NoStop}%
\bibitem [{\citenamefont {Ahn}\ \emph {et~al.}(2022{\natexlab{b}})\citenamefont
  {Ahn}, \citenamefont {Kwon}, \citenamefont {Chung} \emph {et~al.}}]{Danho22}%
  \BibitemOpen
  \bibfield  {author} {\bibinfo {author} {\bibfnamefont {D.}~\bibnamefont
  {Ahn}}, \bibinfo {author} {\bibfnamefont {O.}~\bibnamefont {Kwon}}, \bibinfo
  {author} {\bibfnamefont {W.}~\bibnamefont {Chung}}, \emph {et~al.},\
  }\bibfield  {title} {\bibinfo {title} {Biaxially textured
  {${\mathrm{YBa}}_{2}{\mathrm{Cu}}_{3}{\mathrm{O}}_{7\ensuremath{-}x}$}
  microwave cavity in a high magnetic field for a dark-matter axion search},\
  }\href {https://doi.org/10.1103/PhysRevApplied.17.L061005} {\bibfield
  {journal} {\bibinfo  {journal} {Phys. Rev. Appl.}\ }\textbf {\bibinfo
  {volume} {17}},\ \bibinfo {pages} {L061005} (\bibinfo {year}
  {2022}{\natexlab{b}})}\BibitemShut {NoStop}%
\bibitem [{\citenamefont {Kim}(2023)}]{Jinsu_Kim2023}%
  \BibitemOpen
  \bibfield  {author} {\bibinfo {author} {\bibfnamefont {J.}~\bibnamefont
  {Kim}},\ }\bibfield  {title} {\bibinfo {title} {The first axion quark nugget
  experiment using a haloscope at {CAPP}},\ }\href
  {https://agenda.infn.it/event/34455/timetable/#20230706.detailed} {\bibfield
  {journal} {\bibinfo  {journal} {Private communication, presentation at
  Patras18}\ } (\bibinfo {year} {2023})}\BibitemShut {NoStop}%
\bibitem [{\citenamefont {Cervantes}\ \emph {et~al.}(2022)\citenamefont
  {Cervantes}, \citenamefont {Carosi}, \citenamefont {Hanretty}, \citenamefont
  {Kimes}, \citenamefont {LaRoque}, \citenamefont {Leum}, \citenamefont
  {Mohapatra}, \citenamefont {Oblath}, \citenamefont {Ottens}, \citenamefont
  {Park}, \citenamefont {Rybka}, \citenamefont {Sinnis},\ and\ \citenamefont
  {Yang}}]{article:Cervantes2022}%
  \BibitemOpen
  \bibfield  {author} {\bibinfo {author} {\bibfnamefont {R.}~\bibnamefont
  {Cervantes}}, \bibinfo {author} {\bibfnamefont {G.}~\bibnamefont {Carosi}},
  \bibinfo {author} {\bibfnamefont {C.}~\bibnamefont {Hanretty}}, \bibinfo
  {author} {\bibfnamefont {S.}~\bibnamefont {Kimes}}, \bibinfo {author}
  {\bibfnamefont {B.~H.}\ \bibnamefont {LaRoque}}, \bibinfo {author}
  {\bibfnamefont {G.}~\bibnamefont {Leum}}, \bibinfo {author} {\bibfnamefont
  {P.}~\bibnamefont {Mohapatra}}, \bibinfo {author} {\bibfnamefont {N.~S.}\
  \bibnamefont {Oblath}}, \bibinfo {author} {\bibfnamefont {R.}~\bibnamefont
  {Ottens}}, \bibinfo {author} {\bibfnamefont {Y.}~\bibnamefont {Park}},
  \bibinfo {author} {\bibfnamefont {G.}~\bibnamefont {Rybka}}, \bibinfo
  {author} {\bibfnamefont {J.}~\bibnamefont {Sinnis}},\ and\ \bibinfo {author}
  {\bibfnamefont {J.}~\bibnamefont {Yang}},\ }\bibfield  {title} {\bibinfo
  {title} {Search for $70\text{}\text{ }\ensuremath{\mu}\mathrm{eV}$ dark
  photon dark matter with a dielectrically loaded multiwavelength microwave
  cavity},\ }\href {https://doi.org/10.1103/PhysRevLett.129.201301} {\bibfield
  {journal} {\bibinfo  {journal} {Phys. Rev. Lett.}\ }\textbf {\bibinfo
  {volume} {129}},\ \bibinfo {pages} {201301} (\bibinfo {year}
  {2022})}\BibitemShut {NoStop}%
\bibitem [{\citenamefont {Jeong}\ \emph {et~al.}(2018)\citenamefont {Jeong},
  \citenamefont {Youn}, \citenamefont {Ahn}, \citenamefont {Kim},\ and\
  \citenamefont {Semertzidis}}]{article:pizza_cavity}%
  \BibitemOpen
  \bibfield  {author} {\bibinfo {author} {\bibfnamefont {J.}~\bibnamefont
  {Jeong}}, \bibinfo {author} {\bibfnamefont {S.}~\bibnamefont {Youn}},
  \bibinfo {author} {\bibfnamefont {S.}~\bibnamefont {Ahn}}, \bibinfo {author}
  {\bibfnamefont {J.~E.}\ \bibnamefont {Kim}},\ and\ \bibinfo {author}
  {\bibfnamefont {Y.~K.}\ \bibnamefont {Semertzidis}},\ }\bibfield  {title}
  {\bibinfo {title} {Concept of multiple-cell cavity for axion dark matter
  search},\ }\href
  {https://doi.org/https://doi.org/10.1016/j.physletb.2017.12.066} {\bibfield
  {journal} {\bibinfo  {journal} {Physics Letters B}\ }\textbf {\bibinfo
  {volume} {777}},\ \bibinfo {pages} {412} (\bibinfo {year}
  {2018})}\BibitemShut {NoStop}%
\bibitem [{\citenamefont {Kim}\ \emph {et~al.}(2020)\citenamefont {Kim} \emph
  {et~al.}}]{article:wheel_mechanism}%
  \BibitemOpen
  \bibfield  {author} {\bibinfo {author} {\bibfnamefont {J.}~\bibnamefont
  {Kim}} \emph {et~al.},\ }\bibfield  {title} {\bibinfo {title} {Exploiting
  higher-order resonant modes for axion haloscopes},\ }\href
  {https://doi.org/10.1088/1361-6471/ab5ace} {\bibfield  {journal} {\bibinfo
  {journal} {J. Phys. G}\ }\textbf {\bibinfo {volume} {47}},\ \bibinfo {pages}
  {035203} (\bibinfo {year} {2020})}\BibitemShut {NoStop}%
\bibitem [{\citenamefont {Bae}\ \emph {et~al.}(2023)\citenamefont {Bae},
  \citenamefont {Youn},\ and\ \citenamefont
  {Jeong}}]{article:photonic_crystal}%
  \BibitemOpen
  \bibfield  {author} {\bibinfo {author} {\bibfnamefont {S.}~\bibnamefont
  {Bae}}, \bibinfo {author} {\bibfnamefont {S.}~\bibnamefont {Youn}},\ and\
  \bibinfo {author} {\bibfnamefont {J.}~\bibnamefont {Jeong}},\ }\bibfield
  {title} {\bibinfo {title} {Tunable photonic crystal haloscope for high-mass
  axion searches},\ }\href {https://doi.org/10.1103/PhysRevD.107.015012}
  {\bibfield  {journal} {\bibinfo  {journal} {Phys. Rev. D}\ }\textbf {\bibinfo
  {volume} {107}},\ \bibinfo {pages} {015012} (\bibinfo {year}
  {2023})}\BibitemShut {NoStop}%
\bibitem [{\citenamefont {Omarov}\ \emph {et~al.}(2023)\citenamefont {Omarov},
  \citenamefont {Jeong},\ and\ \citenamefont {Semertzidis}}]{article:variance}%
  \BibitemOpen
  \bibfield  {author} {\bibinfo {author} {\bibfnamefont {Z.}~\bibnamefont
  {Omarov}}, \bibinfo {author} {\bibfnamefont {J.}~\bibnamefont {Jeong}},\ and\
  \bibinfo {author} {\bibfnamefont {Y.~K.}\ \bibnamefont {Semertzidis}},\
  }\bibfield  {title} {\bibinfo {title} {Speeding axion haloscope experiments
  using heterodyne-variance-based detection with a power meter},\ }\href
  {https://doi.org/10.1103/PhysRevD.107.103005} {\bibfield  {journal} {\bibinfo
   {journal} {Phys. Rev. D}\ }\textbf {\bibinfo {volume} {107}},\ \bibinfo
  {pages} {103005} (\bibinfo {year} {2023})}\BibitemShut {NoStop}%
\bibitem [{\citenamefont {Lamoreaux}\ \emph {et~al.}(2013)\citenamefont
  {Lamoreaux}, \citenamefont {van Bibber}, \citenamefont {Lehnert},\ and\
  \citenamefont {Carosi}}]{article:SPC_QLA}%
  \BibitemOpen
  \bibfield  {author} {\bibinfo {author} {\bibfnamefont {S.~K.}\ \bibnamefont
  {Lamoreaux}}, \bibinfo {author} {\bibfnamefont {K.~A.}\ \bibnamefont {van
  Bibber}}, \bibinfo {author} {\bibfnamefont {K.~W.}\ \bibnamefont {Lehnert}},\
  and\ \bibinfo {author} {\bibfnamefont {G.}~\bibnamefont {Carosi}},\
  }\bibfield  {title} {\bibinfo {title} {Analysis of single-photon and linear
  amplifier detectors for microwave cavity dark matter axion searches},\ }\href
  {https://doi.org/10.1103/PhysRevD.88.035020} {\bibfield  {journal} {\bibinfo
  {journal} {Phys. Rev. D}\ }\textbf {\bibinfo {volume} {88}},\ \bibinfo
  {pages} {035020} (\bibinfo {year} {2013})}\BibitemShut {NoStop}%
\bibitem [{\citenamefont {Oelsner}\ \emph {et~al.}(2017)\citenamefont
  {Oelsner}, \citenamefont {Andersen}, \citenamefont {Reh\'ak}, \citenamefont
  {Schmelz}, \citenamefont {Anders}, \citenamefont {Grajcar}, \citenamefont
  {H\"ubner}, \citenamefont {M\o{}lmer},\ and\ \citenamefont
  {Il'ichev}}]{article:SPD}%
  \BibitemOpen
  \bibfield  {author} {\bibinfo {author} {\bibfnamefont {G.}~\bibnamefont
  {Oelsner}}, \bibinfo {author} {\bibfnamefont {C.~K.}\ \bibnamefont
  {Andersen}}, \bibinfo {author} {\bibfnamefont {M.}~\bibnamefont {Reh\'ak}},
  \bibinfo {author} {\bibfnamefont {M.}~\bibnamefont {Schmelz}}, \bibinfo
  {author} {\bibfnamefont {S.}~\bibnamefont {Anders}}, \bibinfo {author}
  {\bibfnamefont {M.}~\bibnamefont {Grajcar}}, \bibinfo {author} {\bibfnamefont
  {U.}~\bibnamefont {H\"ubner}}, \bibinfo {author} {\bibfnamefont
  {K.}~\bibnamefont {M\o{}lmer}},\ and\ \bibinfo {author} {\bibfnamefont
  {E.}~\bibnamefont {Il'ichev}},\ }\bibfield  {title} {\bibinfo {title}
  {Detection of weak microwave fields with an underdamped {Josephson}
  junction},\ }\href {https://doi.org/10.1103/PhysRevApplied.7.014012}
  {\bibfield  {journal} {\bibinfo  {journal} {Phys. Rev. Appl.}\ }\textbf
  {\bibinfo {volume} {7}},\ \bibinfo {pages} {014012} (\bibinfo {year}
  {2017})}\BibitemShut {NoStop}%
\bibitem [{\citenamefont {Revin}\ \emph {et~al.}(2020)\citenamefont {Revin},
  \citenamefont {Pankratov}, \citenamefont {Gordeeva}, \citenamefont
  {Yablokov}, \citenamefont {Rakut}, \citenamefont {Zbrozhek},\ and\
  \citenamefont {Kuzmin}}]{article:Kuzmin2020}%
  \BibitemOpen
  \bibfield  {author} {\bibinfo {author} {\bibfnamefont {L.~S.}\ \bibnamefont
  {Revin}}, \bibinfo {author} {\bibfnamefont {A.~L.}\ \bibnamefont
  {Pankratov}}, \bibinfo {author} {\bibfnamefont {A.~V.}\ \bibnamefont
  {Gordeeva}}, \bibinfo {author} {\bibfnamefont {A.~A.}\ \bibnamefont
  {Yablokov}}, \bibinfo {author} {\bibfnamefont {I.~V.}\ \bibnamefont {Rakut}},
  \bibinfo {author} {\bibfnamefont {V.~O.}\ \bibnamefont {Zbrozhek}},\ and\
  \bibinfo {author} {\bibfnamefont {L.~S.}\ \bibnamefont {Kuzmin}},\ }\bibfield
   {title} {\bibinfo {title} {Microwave photon detection by an al {Josephson}
  junction},\ }\href {https://doi.org/10.3762/bjnano.11.80} {\bibfield
  {journal} {\bibinfo  {journal} {Beilstein Journal of Nanotechnology}\
  }\textbf {\bibinfo {volume} {11}},\ \bibinfo {pages} {960} (\bibinfo {year}
  {2020})}\BibitemShut {NoStop}%
\bibitem [{\citenamefont {D'Elia}\ \emph {et~al.}(2023)\citenamefont {D'Elia},
  \citenamefont {Rettaroli}, \citenamefont {Tocci}, \citenamefont {Babusci},
  \citenamefont {Barone}, \citenamefont {Beretta}, \citenamefont {Buonomo},
  \citenamefont {Chiarello}, \citenamefont {Chikhi}, \citenamefont
  {Di~Gioacchino}, \citenamefont {Felici}, \citenamefont {Filatrella},
  \citenamefont {Fistul}, \citenamefont {Foggetta}, \citenamefont {Gatti},
  \citenamefont {Il'ichev}, \citenamefont {Ligi}, \citenamefont {Lisitskiy},
  \citenamefont {Maccarrone}, \citenamefont {Mattioli}, \citenamefont
  {Oelsner}, \citenamefont {Pagano}, \citenamefont {Piersanti}, \citenamefont
  {Ruggiero}, \citenamefont {Torrioli},\ and\ \citenamefont
  {Zagoskin}}]{article:Gatti2023}%
  \BibitemOpen
  \bibfield  {author} {\bibinfo {author} {\bibfnamefont {A.}~\bibnamefont
  {D'Elia}}, \bibinfo {author} {\bibfnamefont {A.}~\bibnamefont {Rettaroli}},
  \bibinfo {author} {\bibfnamefont {S.}~\bibnamefont {Tocci}}, \bibinfo
  {author} {\bibfnamefont {D.}~\bibnamefont {Babusci}}, \bibinfo {author}
  {\bibfnamefont {C.}~\bibnamefont {Barone}}, \bibinfo {author} {\bibfnamefont
  {M.}~\bibnamefont {Beretta}}, \bibinfo {author} {\bibfnamefont
  {B.}~\bibnamefont {Buonomo}}, \bibinfo {author} {\bibfnamefont
  {F.}~\bibnamefont {Chiarello}}, \bibinfo {author} {\bibfnamefont
  {N.}~\bibnamefont {Chikhi}}, \bibinfo {author} {\bibfnamefont
  {D.}~\bibnamefont {Di~Gioacchino}}, \bibinfo {author} {\bibfnamefont
  {G.}~\bibnamefont {Felici}}, \bibinfo {author} {\bibfnamefont
  {G.}~\bibnamefont {Filatrella}}, \bibinfo {author} {\bibfnamefont
  {M.}~\bibnamefont {Fistul}}, \bibinfo {author} {\bibfnamefont
  {L.}~\bibnamefont {Foggetta}}, \bibinfo {author} {\bibfnamefont
  {C.}~\bibnamefont {Gatti}}, \bibinfo {author} {\bibfnamefont
  {E.}~\bibnamefont {Il'ichev}}, \bibinfo {author} {\bibfnamefont
  {C.}~\bibnamefont {Ligi}}, \bibinfo {author} {\bibfnamefont {M.}~\bibnamefont
  {Lisitskiy}}, \bibinfo {author} {\bibfnamefont {G.}~\bibnamefont
  {Maccarrone}}, \bibinfo {author} {\bibfnamefont {F.}~\bibnamefont
  {Mattioli}}, \bibinfo {author} {\bibfnamefont {G.}~\bibnamefont {Oelsner}},
  \bibinfo {author} {\bibfnamefont {S.}~\bibnamefont {Pagano}}, \bibinfo
  {author} {\bibfnamefont {L.}~\bibnamefont {Piersanti}}, \bibinfo {author}
  {\bibfnamefont {B.}~\bibnamefont {Ruggiero}}, \bibinfo {author}
  {\bibfnamefont {G.}~\bibnamefont {Torrioli}},\ and\ \bibinfo {author}
  {\bibfnamefont {A.}~\bibnamefont {Zagoskin}},\ }\bibfield  {title} {\bibinfo
  {title} {Stepping closer to pulsed single microwave photon detectors for
  axions search},\ }\href {https://doi.org/10.1109/TASC.2022.3218072}
  {\bibfield  {journal} {\bibinfo  {journal} {IEEE Transactions on Applied
  Superconductivity}\ }\textbf {\bibinfo {volume} {33}},\ \bibinfo {pages} {1}
  (\bibinfo {year} {2023})}\BibitemShut {NoStop}%
\bibitem [{\citenamefont {Govenius}\ \emph {et~al.}(2016)\citenamefont
  {Govenius}, \citenamefont {Lake}, \citenamefont {Tan},\ and\ \citenamefont
  {M\"ott\"onen}}]{article:Mottonen2016}%
  \BibitemOpen
  \bibfield  {author} {\bibinfo {author} {\bibfnamefont {J.}~\bibnamefont
  {Govenius}}, \bibinfo {author} {\bibfnamefont {R.~E.}\ \bibnamefont {Lake}},
  \bibinfo {author} {\bibfnamefont {K.~Y.}\ \bibnamefont {Tan}},\ and\ \bibinfo
  {author} {\bibfnamefont {M.}~\bibnamefont {M\"ott\"onen}},\ }\bibfield
  {title} {\bibinfo {title} {Detection of zeptojoule microwave pulses using
  electrothermal feedback in proximity-induced {Josephson} junctions},\ }\href
  {https://doi.org/10.1103/PhysRevLett.117.030802} {\bibfield  {journal}
  {\bibinfo  {journal} {Phys. Rev. Lett.}\ }\textbf {\bibinfo {volume} {117}},\
  \bibinfo {pages} {030802} (\bibinfo {year} {2016})}\BibitemShut {NoStop}%
\bibitem [{\citenamefont {Matsuki}\ \emph {et~al.}(1996)\citenamefont
  {Matsuki}, \citenamefont {Ogawa}, \citenamefont {Nakamura}, \citenamefont
  {Tada}, \citenamefont {Yamamoto},\ and\ \citenamefont
  {Masaike}}]{article:CARRACK}%
  \BibitemOpen
  \bibfield  {author} {\bibinfo {author} {\bibfnamefont {S.}~\bibnamefont
  {Matsuki}}, \bibinfo {author} {\bibfnamefont {I.}~\bibnamefont {Ogawa}},
  \bibinfo {author} {\bibfnamefont {S.}~\bibnamefont {Nakamura}}, \bibinfo
  {author} {\bibfnamefont {M.}~\bibnamefont {Tada}}, \bibinfo {author}
  {\bibfnamefont {K.}~\bibnamefont {Yamamoto}},\ and\ \bibinfo {author}
  {\bibfnamefont {A.}~\bibnamefont {Masaike}},\ }\bibfield  {title} {\bibinfo
  {title} {Rydberg-atom cavity detector for dark matter axion search in
  {Kyoto}},\ }\href
  {https://doi.org/https://doi.org/10.1016/S0920-5632(96)00518-X} {\bibfield
  {journal} {\bibinfo  {journal} {Nuclear Physics B - Proceedings Supplements}\
  }\textbf {\bibinfo {volume} {51}},\ \bibinfo {pages} {213} (\bibinfo {year}
  {1996})},\ \bibinfo {note} {proceedings of the International Symposium on
  Sources and Detection of Dark Matter in the Universe}\BibitemShut {NoStop}%
\bibitem [{\citenamefont {Kim}\ and\ \citenamefont
  {Ahn}(2023)}]{article:Ahn23}%
  \BibitemOpen
  \bibfield  {author} {\bibinfo {author} {\bibfnamefont {K.}~\bibnamefont
  {Kim}}\ and\ \bibinfo {author} {\bibfnamefont {J.}~\bibnamefont {Ahn}},\
  }\bibfield  {title} {\bibinfo {title} {Quantum tomography of rydberg atom
  graphs by configurable ancillas},\ }\href
  {https://doi.org/10.1103/PRXQuantum.4.020316} {\bibfield  {journal} {\bibinfo
   {journal} {PRX Quantum}\ }\textbf {\bibinfo {volume} {4}},\ \bibinfo {pages}
  {020316} (\bibinfo {year} {2023})}\BibitemShut {NoStop}%
\end{thebibliography}%

\end{document}